\documentclass[a4paper,11pt]{article}
\pdfoutput=1 

\usepackage{jheppub} 
\usepackage{graphicx,color}
\usepackage{enumitem}
\usepackage{braket}
\usepackage{slashed}
\usepackage[utf8]{inputenc}
\usepackage{hyperref}
\usepackage{float}
\usepackage{caption}
\usepackage{subfigure}
\usepackage{makecell}
\usepackage{epsfig}
\usepackage{amsmath,amssymb}
\usepackage{hyperref}
\hypersetup{
    colorlinks=flase,
    linkcolor=black,
    filecolor=magenta,      
    urlcolor=blue,
}

\hypersetup{
	colorlinks,
	citecolor=black,
	filecolor=black,
	linkcolor=blue,
	urlcolor=black
}

\newcommand{\bea}{\begin{eqnarray}}
\newcommand{\eea}{\end{eqnarray}}
\newcommand{\be}{\begin{equation}}
\newcommand{\ee}{\end{equation}}
\newcommand{\ba}{\begin{align}}
\newcommand{\ea}{\end{align}}
\newcommand{\ben}{\begin{enumerate}}
\newcommand{\een}{\end{enumerate}}
\newcommand{\bi}{\begin{itemize}}
\newcommand{\ei}{\end{itemize}}
\newcommand{\comments}[1]{}
\def\nn{\nonumber}

\def\bel#1{\begin{equation} \label{#1}}

\def\vo{\mathcal{V}}

\def\KK{{\scriptscriptstyle KK}}

\def\ED3{{\scriptscriptstyle ED3}}

\newcommand{\beq}{\begin{equation}}  \newcommand{\eeq}{\end{equation}}
\newcommand{\bal}{\begin{aligned}}   \newcommand{\eal}{\end{aligned}}
\def\ov{\overline}
\newcommand{\bmat}{\left(\begin{array}}
\newcommand{\emat}{\end{array}\right)}

\newcommand{\cO}{\mathcal{O}}

\newcommand{\cP}{\mathcal{P}}

\newcommand{\cN}{\mathcal{N}}

\newcommand{\cR}{\mathcal{R}}

\newcommand{\cV}{\mathcal{V}}

\def\Tr{{\rm Tr \,}}
\def\bZ{\mathbb{Z}}
\newcommand{\I}{\text{i}}

\newcommand{\kom}{\, ,\quad }
\newcommand*{\raw}{\rightarrow}
\newcommand*{\dif}{{\,\rm d}}
\newcommand*{\tr}{{\text{Tr}}}
\newcommand*{\p}{\mathop{}\!\mathrm \partial}

\newcommand{\mc}{\mathcal}
\newcommand{\ap}{\alpha'}
\newcommand{\bC}{\mathbb{C}}
\newcommand{\bP}{\mathbb{P}}

\usepackage{tikz}
\usepgflibrary{arrows}  
\usetikzlibrary{arrows,shapes,positioning} 
\usetikzlibrary{decorations.markings}
\tikzstyle arrowstyle=[scale=1]
\tikzstyle directed=[postaction={decorate,decoration={markings,
    mark=at position .5 with {\pgftransformscale{1.6}\arrow[arrowstyle]{stealth}}}}]
    
    \tikzstyle directedThree=[postaction={decorate,decoration={markings,
    mark=at position .5 with {\pgftransformscale{1.6}\arrow[arrowstyle]{stealth}},mark=at position .4 with {\pgftransformscale{1.6}\arrow[arrowstyle]{stealth}},mark=at position .6 with {\pgftransformscale{1.6}\arrow[arrowstyle]{stealth}}}}]

\tikzstyle reverse directed=[postaction={decorate,decoration={markings,
    mark=at position .5 with {\pgftransformscale{1.6}\arrowreversed[arrowstyle]{stealth};}}}]
    
    \tikzstyle directedR=[postaction={decorate,decoration={markings,
    mark=at position .45 with {\pgftransformscale{1.6}\arrow[arrowstyle]{stealth}},mark=at position .55 with {\pgftransformscale{1.6}\arrowreversed[arrowstyle]{stealth}}}}]

\title{Systematics of the $\alpha'$ Expansion in F-Theory}

\author[a]{M. Cicoli,}
\author[b]{F. Quevedo,}
\author[c]{R. Savelli,}
\author[b]{A. Schachner,}
\author[d]{R. Valandro}

\affiliation[a]{\footnotesize Dipartimento di Fisica e Astronomia, Universitá di Bologna, via Irnerio 46, 40126 Bologna, Italy and INFN, Sezione di Bologna, viale Berti Pichat 6/2, 40127 Bologna, Italy}
\affiliation[b]{\footnotesize DAMTP, University of Cambridge, Wilberforce Road, Cambridge, CB3 0WA, UK}
\affiliation[c]{\footnotesize Dipartimento di Fisica, Universitá di Roma Tor Vergata and  INFN - Sezione di Roma2, via della Ricerca Scientifica, I-00133 Roma, Italy}
\affiliation[d]{\footnotesize Dipartimento di Fisica, Universitá di Trieste, Strada Costiera 11, I-34151 Trieste, Italy and INFN, Sezione di Trieste, via Valerio 2, I-34127 Trieste, Italy}

\emailAdd{michele.cicoli@unibo.it}
\emailAdd{F.Quevedo@damtp.cam.ac.uk}
\emailAdd{raffaele.savelli@uniroma2.it}
\emailAdd{as2673@maths.cam.ac.uk}
\emailAdd{roberto.valandro@ts.infn.it}

\abstract{
Extracting reliable low-energy information from string compactifications notoriously requires a detailed understanding of the UV sensitivity of the corresponding effective field theories. Despite  past efforts in computing perturbative string corrections to the tree-level action, neither a systematic approach nor a unified framework has emerged yet. We make progress in this direction, focusing on the moduli dependence of perturbative corrections to the 4D scalar potential of type IIB Calabi-Yau orientifold compactifications. We proceed by employing two strategies. First, we use two rescaling symmetries of type IIB string theory to infer the dependence of any perturbative correction  on both the dilaton and the Calabi-Yau volume. Second, we use F/M-theory duality to conclude that KK reductions on elliptically-fibred Calabi-Yau fourfolds of the M-theory action at any order in the derivative expansion can only generate $(\alpha')^{\rm even}$ corrections to the 4D scalar potential, which, moreover, all vanish for trivial fibrations. We finally give evidence that $(\alpha')^{\rm odd}$ effects arise from integrating out KK and winding modes on the elliptic fibration and argue that the leading no-scale breaking effects at string tree-level arise from $(\alpha')^3$ effects, modulo potential logarithmic corrections.
}

\begin{document} 
\maketitle
\flushbottom

\section{Introduction}
\label{Intro}

Effective field theories (EFT) have been the subject of recent debates regarding their relative importance for a UV complete theory of gravity. On the one hand, based on the outstanding success of EFTs to describe all kinds of physical phenomena \cite{Burgess:2020tbq}, a common bottom-up attitude is to fully concentrate on EFTs at low-energies assuming that their self-consistency is enough to expect that they can be completed in the UV. On the other hand, the swampland programme argues that most EFTs cannot be UV completed, and concentrates on conjectures that could eliminate general classes of EFTs \cite{Vafa:2005ui}. In this paper we take an alternative, more traditional, top-down approach where we perform a systematic study of $\alpha'$ corrections to the 4D effective action of compactified string theories which automatically provide a UV completion.

From the topological understanding of Calabi-Yau (CY) compactifications, direct dimensional reduction, supersymmetry and scaling symmetries, we have a very good control over tree-level effective actions for $\mc{N}=1$ supersymmetric compactifications in terms of the K\"ahler potential $K$ and superpotential $W$ for moduli and matter fields. Focussing on type IIB compactifications, these EFTs are of the no-scale type in the sense that the corresponding 4D scalar potential for the K\"ahler moduli vanishes identically when the other moduli are fixed supersymmetrically, since $K^{i\bar{j}} K_i K_{\bar{j}}=3$. Given that this is a tree-level property, $\alpha'$ and string loop corrections are in general expected to lift these flat directions and to play a crucial role in stabilising moduli.

The challenge is further complicated by the fact that string theory does not have free parameters since the higher derivative and string loop expansions are controlled respectively by the vacuum expectation values of the CY volume modulus $\vo$ and the imaginary part of the axio-dilaton $\tau$. Hence it is only after determining their value that we can assess if the expansion parameters are small enough to trust the calculations. Furthermore, the fact that free 10D string theory is always a solution already indicates that the determination of other vacua will never be under full computational control since the scalar potential for $\vo$ and $\tau$ runs away towards their value at infinity. This is the well-known Dine-Seiberg problem \cite{Dine:1985he}. It is essentially the prize string theory has to pay for not having free parameters and is a fully general situation independent of any scenario of moduli stabilisation.

Not having arbitrary good control of perturbative expansions is not a string theory disease but a condition we have to live with. Fortunately there are extra parameters appearing from the nature of the compactification which can play an important role to allow non-trivial moduli stabilisation at couplings which are weak enough to trust the perturbative expansions. These are usually discrete parameters such as the CY Euler number, the rank of condensing gauge groups and the many integer fluxes which are ubiquitous in string compactifications. The derivation of non-trivial vacua necessarily involves a combination between these discrete parameters as well as perturbative and non-perturbative corrections to the 4D scalar potential. 

Together with the dilaton, every K\"ahler modulus, which measures the size of a 4-cycle, can be considered as an expansion parameter, since it determines the gauge coupling of the EFT on D7-branes wrapped on the corresponding 4-cycle. Thus the 4D EFT has many expansion parameters which on the one hand make the calculations more involved since each of them has to be stabilised within the regime of validity of the approximations. On the other hand, however, they allow to stabilise the moduli at weak coupling since, as it happens in the Large Volume Scenario (LVS) \cite{Balasubramanian:2005zx, Conlon:2005ki}, a vacuum can arise from balancing terms of two different expansions without causing a breakdown of perturbation theory. 

On top of moduli stabilisation, identifying the leading no-scale breaking effects beyond the tree-level approximation is crucial to shed light on several important implications of string vacua for cosmology and particle phenomenology. Promising inflationary models based on K\"ahler moduli \cite{Burgess:2001vr, Conlon:2005jm, Cicoli:2008gp, Cicoli:2011ct, Burgess:2013sla, Broy:2015zba, Cicoli:2016chb} feature a shallow potential which is protected by approximate non-compact rescaling shift symmetries~\cite{Burgess:2014tja,Burgess:2016owb} that are broken by no-scale breaking effects. As shown in \cite{Burgess:2020qsc}, leading-order perturbative corrections to the K\"ahler potential are in general also crucial to determine the mass spectrum of the K\"ahler moduli. Moreover, in sequestered models with D3-branes at singularities, the mass scale of the soft terms is set by the dominant no-scale breaking effect at perturbative level \cite{Aparicio:2014wxa, Cicoli:2012sz, Reece:2015qbf}.

In this article we present a systematic analysis of $\alpha'$ corrections to the 4D scalar potential of type IIB string compactifications. It is well-known that in 10D type IIB string theory the leading higher derivative corrections arise only at order $(\alpha')^3$. These include the $\mc{R}^4$ correction to the Einstein-Hilbert action plus its supersymmetric extensions. This property is inherited by $\mc{N}=2$ CY compactifications where the corresponding $(\alpha')^3$ correction to the K\"ahler potential has been computed in \cite{Becker:2002nn}. Additional $\mc{N}=2$ string loop corrections to $K$ at $\mc{O}((\alpha')^2)$ and $\mc{O}((\alpha')^4)$ have been computed in \cite{Berg:2005ja, Berg:2007wt, vonGersdorff:2005bf}, and in the F-theory context in \cite{GarciaEtxebarria:2012zm}, but they yield subdominant contributions to the scalar potential due to a cancellation of $\mc{O}((\alpha')^2)$ terms named `extended no-scale' in \cite{Cicoli:2007xp}. Backreaction of $(\alpha')^3$ effects on the internal geometry have been considered in \cite{Bonetti:2016dqh} which however found only moduli redefinitions. Further $\mc{O}((\alpha')^3)$ terms have been shown in \cite{Ciupke:2015msa, Grimm:2017okk} to give rise to contributions to the 4D scalar potential at $F^4$ order, where $F$ denotes the F-term of the moduli fields. 

Genuine $\mc{N}=1$ corrections are less understood. Different papers found shifts of the CY Euler number induced by $\mc{O}((\alpha')^3)$ corrections at tree- \cite{Minasian:2015bxa} and loop-level \cite{Berg:2014ama, Haack:2015pbv,Haack:2018ufg,Antoniadis:2018hqy}. Using M/F-theory duality, novel $\mc{O}((\alpha')^2)$ effects were found in \cite{Grimm:2013gma,Grimm:2013bha}, which can however be affected by field redefinitions of the $11$D fields \cite{Junghans:2014zla}.
More corrections in the $\mc{N}=1$ 4D effective action of F-theory were discussed in \cite{Weissenbacher:2019mef}, which were further constrained recently in \cite{Klaewer:2020lfg} by studying infinite distance limits. A full understanding of $\alpha'$ corrections to the type IIB $\mc{N}=1$ effective action is not available yet. In particular any correction that would dominate over the $(\alpha')^3$ ones may play an important role in moduli stabilisation scenarios as LVS \cite{Balasubramanian:2005zx, Conlon:2005ki} and KKLT \cite{Kachru:2003aw}. They may shift the minimum, provide a potential de Sitter uplift or destabilise the original vacuum. Moreover, subdominant higher derivative or string loop corrections can still be relevant for lifting leading order flat directions in scenarios with more than one K\"ahler modulus \cite{Cicoli:2016chb,Cicoli:2007xp,Cicoli:2008va,AbdusSalam:2020ywo}. 

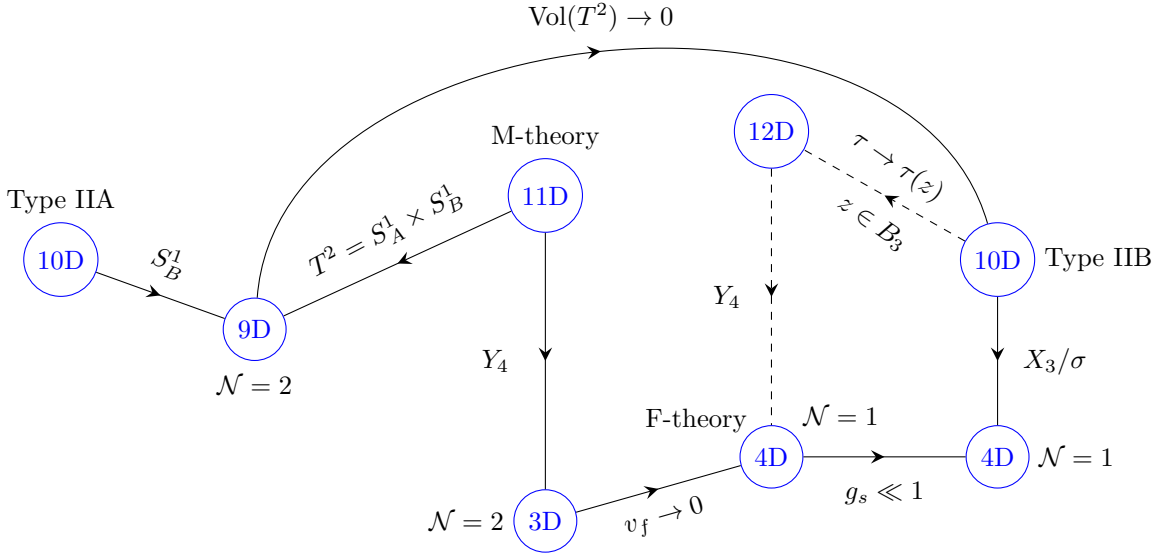
\begin{figure}[t!]
\centering
{\small
\begin{tikzpicture}[node distance=0.6cm, scale=0.85]
\draw (-4.5,3) node(7)[anchor=south,circle,draw,blue]{{\small 9\text{D}}};
\draw (-7.5,4) node(8)[anchor=south,circle,draw,blue]{{\small 10\text{D}}};
\draw (0,5) node(1)[anchor=south,circle,draw,blue]{{\small 11\text{D}}};
\draw (3.5,6) node(2)[anchor=south,circle,draw,blue]{{\small 12\text{D}}};
\draw (7,4) node(3)[anchor=south,circle,draw,blue]{{\small 10\text{D}}};
\draw (0,1) node(4)[anchor=north,circle,draw,blue]{{\small 3\text{D}}};
\draw (3.5,2) node(5)[anchor=north,circle,draw,blue]{{\small 4\text{D}}};
\draw (7,2) node(6)[anchor=north,circle,draw,blue]{{\small 4\text{D}}};
\draw[directed,black] (1) --node[left=10pt,rotate=0]{{\small $Y_4$}}  (4)node[left=12pt]{{\small $\cN=2$}};
\draw[directed,dashed,black] (2) --node[left=8pt]{$Y_4$} (5)node[above right=12pt]{{\small $\cN=1$}};
\draw[directed,black] (3)node[right=14pt]{{\small Type IIB}} --node[right=6pt]{$X_3/\sigma$} (6)node[right=12pt]{{\small $\cN=1$}};
\draw[directed,black] (4) --node[below=3pt,rotate=16]{$v_f\raw 0$} (5);
\draw[directed,black] (5)node[above left=8pt]{{\small F-theory}}  --node[below=6pt]{$g_s \ll 1$} (6);
\draw[directed,dashed,black] (3) --node[above=4pt,rotate=-35]{$\tau\raw \tau(z)$}node[below=4pt,rotate=-35]{$z\in B_3$} (2);
\draw[directed,black] (1)node[above=14pt]{{\small M-theory}}  --node[above=4pt,rotate=28]{{\small $T^2=S^1_A\times S_B^1$}} (7);
\draw[directed,black] (8)node[above=14pt]{{\small Type IIA}} --node[above=4pt,rotate=-20]{$S_B^1$} (7)node[below=14pt]{{\small $\cN=2$}} ;
\draw[directed,black] (7) to[bend left=80pt]node[above=4pt]{{\small $\mathrm{Vol}(T^2)\raw 0$}} (3);
\end{tikzpicture}
}
\caption{Schematic picture of dualities and limits of the various theories in different dimensions employed throughout this paper.}
\label{fig:MFIIB}
\end{figure}

A complete analysis of $\alpha'$ corrections is too ambitious to be achievable. Here we will extract information on the moduli dependence of the low-energy scalar potential by combining techniques that rely on either symmetries of 10D type IIB string theory or on dimensional analysis in M/F-theory compactifications which come along with a rich web of dualities summarised pictorially in Fig.~\ref{fig:MFIIB}. Our analysis is simplified by concentrating only on the dilaton and overall volume dependence of perturbative corrections. Even if a full dependence of arbitrary $\alpha'$ and $g_s$ effects on all the K\"ahler moduli is beyond our reach, it is the $\vo$-dependence that is the most relevant information for moduli stabilisation.

In practice, investigations of F-theory compactifications start from M-theory by reducing the 11D action on a CY fourfold $Y_4$ which leads to 3D gauged $\cN=2$ supergravity \cite{Haack:1999zv,Haack:2001jz,Berg:2002es,deWit:2004yr,Grimm:2011tb}. Under the assumption that $Y_4$ is elliptically fibred over a 6D K\"ahler base manifold $B_3$, one takes the point-wise limit of vanishing fibre volume, $v_f\raw 0$, thereby decompactifying a single direction giving rise to F-theory in 4D, that is type IIB compactified on the base manifold $B_3$ \cite{Vafa:1996xn,Morrison:1996na,Morrison:1996pp,Denef:2008wq,Grimm:2010ks}. The elliptic curve in $Y_4$ is effectively keeping track of the dynamics of the axio-dilaton. The singular loci of the fibre are associated with 7-branes on the base whose precise realisation within $Y_4$ specifies the gauge algebra (see \cite{Weigand:2018rez} for a recent review). In the weak coupling limit $g_s \ll 1$ of F-theory, the so-called \emph{Sen limit} \cite{Sen:1997gv}, one recovers perturbative type IIB orientifold compactifications on the double cover $X_3$ of the base $B_3$ \cite{Denef:2008wq}. The existence of an elliptic fibre leads to an $\mathrm{SL}(2,\bZ)$ symmetry acting on the axio-dilaton $\tau$ in type IIB. More precisely, the strong coupling dynamics of string theory is accessible via dualities even if the microscopic origin remains elusive. The most basic duality is M-theory compactified on an $S_A^1$ giving rise to 10D type IIA supergravity. After a subsequent reduction on another circle $S_B^1$, we can use T-duality \cite{Giveon:1994fu,Sen:2001di} to obtain type IIB supergravity on a circle. This is equivalent to compactifying M-theory on a torus $T^2=S_A^1\times S_B^1$ and taking the ${\rm Vol}(T^2)\raw 0$ limit \cite{Dijkgraaf:1997vv}, as we said above. Recently an effective 12D approach (indicated by dashed lines in Fig.~\ref{fig:MFIIB}) has been put forward in \cite{Minasian:2015bxa} which in principle allows for a new access to $\alpha'$ effects in F-theory. 
In this paper we will take advantage of these dualities together with scaling symmetries to extract direct information on $\alpha'$ corrections to the 4D scalar potential. 

This paper is structured as follows. In Sec.~\ref{Sec2} we follow \cite{Burgess:2020qsc} and use the symmetries of the 10D type IIB action to organise different perturbative corrections to the $\mc{N}=1$ 4D effective action concentrating on the dilaton and volume dependence of each order in the $\alpha'$ and string loop expansions. We make use of the fact that each of the two expansions is directly related to the existence of two scaling symmetries of the 10D action. In particular we present the general expression of the 4D scalar potential including each order in the $\alpha'$ and $g_s$ expansions as well as the number of powers of F-terms of the low-energy moduli which corresponds to an expansion in terms of inverse powers of the Kaluza-Klein (KK) scale \cite{Cicoli:2013swa}, as typical of KK compactifications. We recover all known corrections that have been computed so far as particular cases of our general expression. 

In Sec.~\ref{AbsenceOfAPOne} we see how the absence of $(\alpha')^1$ corrections at string tree-level to either the 10D bulk action or to the 8D action of localised sources, combined with our symmetry considerations and the extended no-scale structure, allow us to infer that the leading no-scale breaking effects at tree-level in $g_s$ should arise from $(\alpha')^3$ effects. We confirm this claim by dimensional reduction and dimensional analysis considering all potential sources for these corrections: bulk terms, brane effects and backreaction. 

Sec.~\ref{sec:MF} is the core of the paper, where we utilise F/M-theory duality techniques as well as a dimensional analysis to extract systematic information on the $\alpha'$ expansion of the 4D scalar potential. We present the rules to perform the F-theory limit by first considering the 3D EFT obtained via a fourfold compactification of 11D M-theory and then taking the vanishing fibre volume limit to extract information on 4D compactifications of F-theory. In particular, using a very general ansatz for the metric of an elliptically fibred fourfold, we constrain the moduli dependence of higher derivative corrections to the 4D scalar potential. We find that conventional KK reductions on elliptically fibred CY fourfolds of the 11D supergravity action, corrected at arbitrary order in the derivative expansion, can generate only $(\alpha')^{\rm even}$ corrections to the 4D scalar potential. We come to the conclusion that only a certain class of higher-order terms in the 11D Planck length $\ell_M$ gives rise to a finite contribution in the F-theory limit. Remarkably, this class of 11D higher derivative structures precisely falls into the general pattern of the M-theory higher derivative expansion as conjectured by \cite{Damour:2005zb}, using symmetry constraints from the Kac-Moody algebra E$_{10}$.\footnote{More precisely, we show that higher derivative corrections in M-theory should appear only at order $\ell_M^{6\mathfrak{p}}$ with $\ell_M$ the 11D fundamental length and $\mathfrak{p}\in\mathbb{N}$, assuming that they contribute to the effective action in the F-theory limit.} Furthermore, for the case of trivial fibrations, we find that all such higher derivative corrections give vanishing contribution in 4D. 

In contrast, we argue that $(\alpha')^{\rm odd}$ effects arise from a proper process of integrating out KK and winding states on the elliptic fibration, which we outline in Sec.~\ref{sec:QuantumReductions}. Here we provide evidence in favour of this claim by focusing on the simple case of trivial fibrations where we manage to show that our approach based on dimensional analysis allows to reproduce, from 11D loops, known $(\alpha')^3$ corrections at different orders in the low-energy F-term expansion.

We present our conclusions and outlook in Sec.~\ref{sec:Con} and leave some technical aspects to the appendices. In App.~\ref{app:TensorsEllFibGen} we collect some results on higher curvature terms for elliptic fibrations. For completeness, in App.~\ref{AppB} we explore the effects that potential loop corrections at order $(\alpha')^1$ could have, if they existed, on moduli stabilisation. Interestingly we find that they could give rise to new dS vacua in a regime where the EFT is under relatively good control.

\section{Perturbative corrections from symmetries in type IIB}
\label{Sec2}

In this section we show how perturbative corrections to the 4D EFT of type IIB string theory can be constrained using the symmetries of the underlying 10D theory. 

\subsection{Tree-level effective action}

\subsubsection*{10D perspective}

The low energy description of string theory can be obtained by computing scattering amplitudes of massless string excitations. This gives rise to a 10D EFT whose action can be written as $S_{\rm IIB} = S_{\rm bulk} + S_{\rm loc}$, where $S_{\rm bulk}$ describes the dynamics of the bulk degrees of freedom while $S_{\rm loc}$ is associated to objects localised in the extra dimensions, like D-branes and O-planes. The bosonic bulk action at tree-level and in Einstein frame reads \cite{Polchinski:1998rr}:
\be
S_{\rm bulk}^{(0)} = \frac{1}{2\kappa_{10}^2}\,\int\, \sqrt{-\tilde{g}}\left( \mc{R}-\frac{|\nabla\tau|^2}{2\, ({\rm Im} \tau)^2}-\frac{|G_3|^2}{12\, {\rm Im}\tau}-\dfrac{ |\tilde{F}_5|^2}{4\cdot 5!}\right) +\frac{1}{8\I\kappa_{10}^2}\int\, \frac{{C}_4\wedge G_3\wedge \overline{{G}}_3}{{\rm Im} \tau},
\label{Sbulktree} 
\ee
where $\tilde{g}_{MN}$ is the 10D Einstein frame metric, $\tau= C_0 +{\rm i} \,e^{-\phi}$ is the axio-dilaton whose imaginary part controls the string coupling ($e^\phi = g_s$), and $G_3=F_3-\tau H_3$ is the 3-form background flux with: 
\be
H_3=\dif B_2\,, \qquad F_{p+1}=\dif C_p\,, \qquad \tilde{F}_5=F_5-\frac12 C_2\wedge H_3 + \frac12 B_2\wedge F_3\,.
\ee
In addition to the equations of motion, the 5-form flux must satisfy the self-duality condition $\tilde{F}_5=\star_{10} \tilde{F}_5$.  Beyond general coordinate invariance, $\mc{N}=2$ supersymmetry and the gauge symmetries of the $p$-forms, the tree-level bulk action (\ref{Sbulktree}) enjoys the following accidental symmetries:
\begin{itemize}
\item $SL(2,\mathbb{R})$
\be
\tau\rightarrow \frac{a \tau+b}{c\tau + d}\,, \qquad G_3\rightarrow \frac{G_3}{c\tau+d} \qquad\text{with}\quad ad-bc=1\,.
\ee
This symmetry is broken by $\alpha'$ and $g_s$ corrections. However two subgroups survive at higher order: the axionic shift symmetry of $C_0$ is unbroken at perturbative level, while $SL(2,\mathbb{Z})$ is an exact symmetry of the whole non-perturbative theory. 

\item {\it Scale invariance} 

Scaling the bosonic fields with two arbitrary weights $\omega$ and $\nu$ as \cite{Burgess:2020qsc}:
\be
\tilde{g}_{MN}\rightarrow \lambda^\nu \tilde{g}_{MN}, \quad \tau\rightarrow \lambda^{2(\omega-\nu)}\tau, \quad B_2\rightarrow \lambda^{2\nu-\omega} B_2, \quad C_2\rightarrow \lambda^\omega C_2, \quad C_4 \rightarrow \lambda^{2\nu} C_4\,,
\label{10Drescalings}
\ee
the bulk action (\ref{Sbulktree}) transforms as:
\be
S_{\rm bulk}^{(0)}\rightarrow \lambda^{4\nu}\, S_{\rm bulk}^{(0)}\,,
\ee
showing that it enjoys two families of classical scale invariance that are expected to be broken by corrections beyond tree-level. Notice that for $\nu\neq 0$ the equations of motion are still invariant even if $S_{\rm bulk}^{(0)}$ is not, while the case with $\nu=0$ reproduces the scale invariance included in $SL(2,\mathbb{R})$ for $b=c=0$ and $a=1/d$.

Let us stress that the existence of two scaling symmetries is closely related to the fact that the EFT features two independent perturbative expansions: in terms of $g_s$ controlled by the dilaton (corresponding to worldsheet topologies/loops in the spacetime theory), and $\alpha'$ controlled by the metric (associated to loops in the worldsheet theory/higher derivative terms from the spacetime point of view). This property is shared by all five different 10D superstring theories but it does not hold for the effective action of 11D supergravity since its massless spectrum does not include a dilaton field. This implies that in this case there is just a single perturbative expansion which is reflected in the existence of a single scaling symmetry. In fact, all terms of the 11D supergravity action: 
\be
S_{11}=\frac{1}{2\kappa_{11}^2}\int d^{11}x\left[\sqrt{-g^{(11)}}\left(R-\frac{1}{48}G_4^2\right) +\frac{1}{6}G_4\wedge G_4\wedge C_3\right], 
\ee
scale homogeneously as $S_{11}\rightarrow \lambda^{9\omega} S_{11}$ under the 1-family rescalings:
\be
g^{(11)}_{MN}\rightarrow \lambda^{2\omega} g^{(11)}_{MN} \qquad\text{and}\qquad C_3\rightarrow \lambda^{3\omega} C_3\,. 
\ee

\end{itemize}

Coming back to the type IIB action, let us now include localised sources in 10D. The action of a D$p$-brane contains a DBI and a Wess-Zumino (WZ) contribution. It can be shown that, under the rescalings (\ref{10Drescalings}), both of them scale as \cite{Burgess:2020qsc}:
\be
S_{\rm loc}^{(0)}\rightarrow \lambda^{\rho}\, S_{\rm loc}^{(0)}\qquad\text{with}\quad \rho=(p-1)\nu-\frac12 (p-3)\omega\,.
\ee
Given that $\rho\neq 4\nu$ $\forall p$, we realise that the D$p$-brane action breaks the 2-family scale invariance of the bulk action down to a 1-family scaling symmetry parametrised by the relation $2(p-5)\nu = (p-3)\omega$. This can be easily understood from noticing that the 10D string frame metric $\hat{g}_{MN} = \tilde{g}_{MN}/\sqrt{{\rm Im}\,\tau}$ scales with weight $2\nu-\omega$. Hence choosing $\omega=2\nu$, the 10D string frame metric does not rescale and $\rho = 2\nu$ $\forall p$. In this case $S_{\rm bulk}^{(0)}\propto e^{-2\phi} = g_s^{-2}$ while $S_{\rm loc}^{(0)}\propto e^{-\phi}= g_s^{-1}$, showing that $S_{\rm loc}$ can be seen as a higher order effect in the expansion of the action in powers of $g_s$ that breaks one of the two scaling symmetries enjoyed by the leading expression. This remaining scale invariance is then expected to be broken by additional $g_s$ and $\alpha'$ corrections. 

\subsubsection*{4D perspective}

Type IIB string theory compactified on a CY threefold $X_3$ yields an $\mc{N}=2$ 4D EFT which can be broken down to $\mc{N}=1$ by the inclusion of O-planes and D-branes. The scaling properties of the 4D fields inherited from the higher dimensional theory can be understood from looking at the decomposition of the 10D metric:
\be
\dif \tilde{s}^2_{(10)} = \tilde{g}_{\mu\nu}\dif x^\mu \dif x^\nu + \tilde{g}_{mn} \dif z^m \dif z^n\,,
\ee
where we ignored the warp factor since it does not scale. Thus we realise that (\ref{10Drescalings}) implies that the 4D metric scales as the 10D one, $\tilde{g}_{\mu\nu}\to \lambda^\nu \tilde{g}_{\mu\nu}$, while the Einstein frame CY volume scales as:
\be
\vo = \frac{1}{\ell_s^6}\int_{X_3} \dif^6 z \sqrt{-\tilde{g}^{(6)}} \to \lambda^{3\nu} \,\vo\,,
\label{voscaling}
\ee
where we measured $\vo$ in units of the string length $\ell_s = 2\pi\sqrt{\alpha'}$. Therefore the 4D Einstein frame metric and Lagrangian scale as:
\be
g_{\mu\nu} = \vo \,\tilde{g}_{\mu\nu}\to \lambda^{4\nu}\, g_{\mu\nu}\qquad \text{and}\qquad \mc{L}\to \lambda^{4\nu}\,\mc{L}\,,
\label{Lscal}
\ee
where the scaling of $\mc{L}$ is fixed by the Einstein-Hilbert term just by knowing the scaling of $g_{\mu\nu}$. The scaling of the overall volume (\ref{voscaling}) implies that also the K\"ahler moduli rescale since $\vo$ can be rewritten as:
\be
\vo = \frac16 \int_{X_3}\, J\wedge J\wedge J = \frac16\, k_{\alpha\beta\gamma}\, t^\alpha t^{\beta}t^{\gamma}\,,
\label{eq:VolCY} 
\ee
where $J$ is the K\"ahler form which we expanded in a basis $\hat{D}_\alpha$ of $H_+^{(1,1)}(X_3,\bZ)$ as $J=\sum_{\alpha=1}^{h_+^{1,1}}\, t^\alpha \hat{D}_\alpha$, the $t^\alpha$'s, $\alpha=1,...,h_+^{1,1}$, are 2-cycle volumes and $k_{\alpha\beta\gamma}$ are the triple intersection numbers given by:
\be
k_{\alpha\beta\gamma}=\int_{X_3}\, \hat{D}_\alpha \wedge \hat{D}_\beta \wedge \hat{D}_\gamma\,.
\ee
Given that the K\"ahler moduli are defined as $T_\alpha=b_\alpha+\I \,\tau_\alpha$ with $b_\alpha=\int\,C_4\wedge\hat{D}_\alpha$ and $\tau_\alpha=\frac12 k_{\alpha\beta\gamma} t^\beta t^\gamma$, (\ref{10Drescalings}) and (\ref{voscaling}) imply $T_\alpha \to \lambda^{2\nu}\, T_\alpha$ $\forall \alpha$. Since the K\"ahler moduli are the scalar components of $h^{1,1}_+$ chiral superfields $\mc{T}_\alpha = T_\alpha + \sqrt{2}\theta \psi_\alpha + F_\alpha$, the superspace coordinate $\theta$ has to rescale as $\theta \to \lambda^\nu \theta$ together with $\psi_\alpha \to \lambda^\nu\,\psi_\alpha$ to ensure that the fermionic kinetic term $\bar\psi e^\mu_a \gamma^a \partial_\mu \psi$ scales as the bosonic one $g^{\mu\nu}\partial_\mu T\partial_\nu T$. The $h^{1,2}_-$ complex structure moduli $Z_i$ instead do not rescale. 

The implications of these scaling symmetries can be easily understood by using the superconformal formalism which allows to write the Lagrangian in terms of the chiral compensator $\Phi$ as (ignoring the contribution from the gauge kinetic function):
\be
\frac{\mc{L}}{\sqrt{-g}} = -3 \int \dif^4\theta\, e^{-K/3}\, \ov{\Phi}\Phi + \int \dif^2 \theta W \Phi^3 + {\rm h.c.}\,, 
\label{Lag}
\ee
where $K$ and $W$ denote respectively the K\"ahler potential and the superpotential. Using (\ref{Lscal}) together with ${\rm d}\theta \to \lambda^{-\nu} {\rm d}\theta$, we obtain:
\bea
e^{-K/3}\, \ov{\Phi}\Phi&\to&\lambda^{\omega_{\mc{L}}-4\nu}\,e^{-K/3}\, \ov{\Phi}\Phi = e^{-K/3}\, \ov{\Phi}\Phi\,, \label{Krescal} \\
W \Phi^3 &\to& \lambda^{\omega_{\mc{L}}-6\nu} \,W \Phi^3 = \lambda^{-2\nu} \,W \Phi^3\,, \label{Wrescal}
\eea
where $\omega_{\mc{L}}$ is the weight of the Lagrangian, with $\omega_{\mc{L}}=4\nu$ at tree-level. These two relations can be used to derive the dependence of $K$ on two combinations of rescaling fields (due to the presence of two scaling symmetries) once the weight of $W$ is known. This can be deduced from direct dimensional reduction which yields the tree-level flux superpotential $W_0 =\int_{X_3} G_3\wedge \Omega$ where $\Omega$ is the CY holomorphic $(3,0)$-form. Since $\Omega$ is a function of the complex structure moduli which do not rescale, $W_0$ scales as $G_3$ whose weight is $\omega$ (see the weight of $C_2$ in (\ref{10Drescalings})). Thus (\ref{Wrescal}) can be used to fix the weight of the chiral compensator which in turn determines the weight of the tree-level K\"ahler potential $K_0$ from (\ref{Krescal}) as:
\be
\Phi\to \lambda^{-\frac13 (\omega+2\nu)}\, \Phi   \qquad\Rightarrow\qquad e^{-K_0/3}\to \lambda^{\frac23(\omega+2\nu)}\,e^{-K_0/3}\,.
\label{Kscal}
\ee
Using the scaling properties of the dilaton and the volume mode together with the fact that axionic shift symmetries forbid a dependence of the tree-level $K_0$ on $C_0$ and $C_4$-axions, the relation (\ref{Kscal}) allows us to fix:
\be
e^{-K_0/3} = \mc{A}\,({\rm Im}\,\tau)^{1/3}\,\vo^{2/3}\,,
\ee
where $\mc{A}$ is a scale invariant combination of all other 4D fields. Notice that this expression reproduces the one obtained by direct dimensional reduction:
\be
K_0 =-2\ln \vo -\ln(-\I(\tau-\bar{\tau}))-\ln\left (\I\int_{X_3}\,\Omega\wedge\overline{\Omega}\right )\qquad\text{for}\qquad 
\mc{A} = \left(\int_{X_3}\,\Omega\wedge\overline{\Omega}\right)^{1/3}\,.   \nn
\ee
Thus we have seen the dependence of $K_0$ on $\phi$ and $\vo$ can be fixed  without the need to perform any computation but just by symmetry arguments via a combination of supersymmetry, scale invariance and shift symmetries. As shown in \cite{Burgess:2020qsc}, these symmetry considerations are also enough to infer that the 4D EFT enjoys a no-scale cancellation where the associated flat direction, the volume mode, corresponds to the Goldstone boson of one of the two scaling symmetries which is spontaneously broken by the vacuum expectation value of the metric. The other scaling symmetry is also spontaneously broken by the vacuum expectation value of the dilaton. However the corresponding would-be Goldstone mode, the dilaton, would become massive in the presence of non-zero 3-form flux quanta which would break the rescaling symmetry explicitly. In fact, as can be seen from (\ref{10Drescalings}), $G_3$ rescales with weight $\omega$, and so any 4D EFT with $G_3$ fixed at a non-zero background value would necessarily break this symmetry explicitly. 

\subsection{Perturbative corrections}
\label{TypeIIBscalings}

\subsubsection*{10D perspective}

As already stressed above, the two rescaling symmetries of the bulk tree-level action are expected to be broken by higher order $g_s$ and $\alpha'$ effects (we have already seen that any D$p$-brane action already breaks one of these two scale invariances). However these breaking effects arise in a controllable manner since the parameters which control these two perturbative expansions are two fields, $\phi$ and $\vo$, which rescale with a non-trivial weight. We thus expect to be able to infer the dependence on $\phi$ and $\vo$ of any perturbative correction to $K$ at all orders in $g_s$ and $\alpha'$. This can be achieved by exploiting again the superconformal chiral compensator formalism together with the scaling properties of the 10D and 4D EFT. 

Before seeing how this works, let us remind the reader that the $10$D type IIB supergravity action can in general be expanded as:
\be
S_{\rm IIB} = S_{\rm bulk}^{(0)}+\sum_{m=3}^\infty \sum_{n=0}^\infty\, (\alpha')^m g_s^n\, S_{\rm bulk}^{(m,n)}
+S_{\rm loc}^{(0)}+\sum_{m=2}^\infty\sum_{n=0}^\infty\, (\alpha')^m g_s^n\,S_{\rm loc}^{(m,n)}\,.
\label{eq:ch1:ActionIIB10DFullExpAP} 
\ee
Notice that, because of $\mc{N}=2$ supersymmetry, the first higher derivative corrections to the bulk action arise only at order $(\alpha')^3$. Corrections to the action of localised sources are instead expected to emerge only at $(\alpha')^2$ order. The higher derivative corrections in \eqref{eq:ch1:ActionIIB10DFullExpAP} can be obtained from string amplitudes \cite{Peeters:2001ub,Richards:2008jg,Richards:2008sa,Liu:2013dna,Liu:2019ses}, the pure spinor formalism \cite{Policastro:2006vt,Policastro:2008hg}, via duality to M-theory \cite{Green:1997tv,Green:1997di,Green:1997as,Green:1997tn,Green:1998yf,Green:1999by,Grimm:2017okk} or supersymmetry \cite{Green:1998by,Peeters:2000qj,Howe:1983sra,Green:1999qt,deHaro:2002vk,Rajaraman:2005up,Green:2003an,Paulos:2008tn} (see also \cite{Bakhtiarizadeh:2017bpl,Blaback:2019zig,Garousi:2020mqn,Codina:2020yma,Mayer:2020lpa}). While $\mc{R}^2$ corrections can arise in the heterotic string, in type II theories the greater degree of supersymmetry forbids $\mc{R}^2$, $\mc{R}^3$ as well as all other terms at order $\ap$ and $(\ap)^2$ \cite{Polchinski:1998rr}. At order $(\alpha')^3$, one finds schematically \cite{Conlon:2005ki}:
\bea
S_{\rm bulk}^{(3,0)}&\sim& \int\dif^{10}x\sqrt{-\tilde{g}}\,\left[\mc{R}^4+\mc{R}^3\left (\tilde{G}_3^2 + |\tilde{G}_3|^2 + \overline{\tilde{G}}_3^2+\tilde{F}_5^2+|\cP|^2\right ) \right. 
\label{AP3Action} \\
&+& \left. \mc{R}^2 \left (|\nabla{\tilde{G}}_3|^2+(\nabla\tilde{F}_5)^2 + \tilde{G}_3^4 + \ldots\right )+\mc{R}\left (\tilde{G}_3^6+\ldots\right )+\left (\tilde{G}_3^8+(|\nabla{\tilde{G}}_3|^2)^{2}+\ldots\right )\right] \nn \,,
\eea
where:
\be
\tilde{G}_3\equiv \frac{G_3}{\sqrt{{\rm Im}\,\tau}}\qquad\text{and}\qquad \cP\equiv \frac{\I\,\nabla\tau}{{\rm Im}\,\tau} \,.
\ee
In general, an arbitrary correction to the 10D bulk action in string frame at order $(\alpha')^m g_s^n$ involving the dilaton, the curvature and the 3-form flux $H_3$ can be written as:
\be
S_{\rm bulk}^{(m,n)} 
\sim \int \dif^{10}x\, \sqrt{- \hat{g}} \left( \frac{1}{{\rm Im}\,\tau} \right)^{(n-2)} \left(  \hat{g}^{\circ \circ} \hat{R}^\circ_{\circ \circ\circ} \right)^{p+1} \left[ \hat{g}^{\circ \circ}  \hat{g}^{\circ \circ}  \hat{g}^{\circ \circ} H_{\circ\circ\circ} H_{\circ\circ\circ} \right]^r\qquad m=p+r \,,
\label{Scorrstring}
\ee
where $\circ$ denotes the appropriate index structure and $m=p+r$ since each power of $\hat{R}$ and $H_3^2$ contains two derivatives. In (\ref{Scorrstring}) we ignored potential contributions from gradients of the dilaton since $\phi$ is set to be constant by the equations of motion (except in the
vicinity of localised sources). Higher derivative corrections are expected to introduce a dependence of the dilaton on the internal coordinates but, given that explicit computations have shown that this dependence can be rewritten in terms of the curvature \cite{Becker:2002nn}, we expect this effect to be captured by (\ref{Scorrstring}). Notice that (\ref{Scorrstring}) is generic enough to describe also contributions of the form $\mc{R}^{p+1} (\nabla{G}_3)^{2r}$ since they would scale as $\mc{R}^{p+r+1} G_3^{2r}$. Moreover in (\ref{Scorrstring}) we neglected potential $\tilde{F}_5$-dependent higher derivative corrections since $\tilde{F}_5=0$ in the absence of warping. Writing $H_3$ in terms of $G_3$ and converting (\ref{Scorrstring}) to Einstein frame via $\hat{g}_{MN} = \tilde{g}_{MN}/\sqrt{{\rm Im}\,\tau}$, we end up with:
\be
S_{\rm bulk}^{(m,n)} 
\sim \int \dif^{10}x\, \sqrt{- \tilde{g}} \left( \frac{1}{\hbox{Im}\, \tau} \right)^{(2n-p+r)/2} \left(\tilde{g}^{\circ \circ} \tilde{R}^\circ_{\circ \circ\circ} \right)^{p+1}  \left[ \tilde{g}^{\circ \circ}  \tilde{g}^{\circ \circ}  \tilde{g}^{\circ \circ} G_{\circ\circ\circ} G_{\circ\circ\circ} \right]^r \,.  
\label{ScorrEinst}
\ee
Notice that for $n=m=0$ (\ref{ScorrEinst}) reproduces the correct scaling of two terms in (\ref{Sbulktree}): the Einstein-Hilbert term for $p=r=0$, and the kinetic terms of $G_3$ for $p=-1$ and $r=1$. Using (\ref{10Drescalings}), we can easily infer that the generic $\mc{O}\left((\alpha')^m g_s^n\right)$ correction (\ref{ScorrEinst}) rescales as:
\be
S_{\rm bulk}^{(m,n)} \to \lambda^{4\nu-2n(w-\nu) +m (w-2\nu)} \, S_{\rm bulk}^{(m,n)}\,.
\label{Scorrscaling}
\ee

\subsubsection*{4D perspective}

Non-renormalisation theorems ensure that the superpotential receives only tree-level and non-perturbative contributions, whereas the K\"ahler potential can be corrected at all orders in $\ap$ and $g_s$. The 10D action (\ref{ScorrEinst}) is therefore expected to yield a perturbative correction to the 4D K\"ahler potential. Using again the two scaling symmetries of the classical action and the chiral compensator formalism, we can work out the dilaton and volume mode dependence of a generic $\mc{O}\left((\alpha')^m g_s^n\right)$ perturbative correction to $K$. Combining (\ref{Krescal}) with (\ref{Wrescal}) for $\omega_{\mc{L}} = 4\nu-2n(w-\nu) + m (w-2\nu)$ from (\ref{Scorrscaling}), we realise that:
\be
\left(e^{-K/3}\right)_{(m,n)} \to \lambda^{\frac23(\omega+2\nu)} \left[\lambda^{-2(w-\nu)}\right]^n \left[\lambda^{w-2\nu}\right]^m 
\left(e^{-K/3}\right)_{(m,n)}\,.
\ee
This rescaling property, together with $\tau \to \lambda^{2(\omega-\nu)}\,\tau$, $\vo\to\lambda^{3\nu}\,\vo$ and the axionic shift symmetries, implies that the perturbative K\"ahler potential has to take the form \cite{Burgess:2020qsc}:
\be
e^{-K/3} = ({\rm Im}\,\tau)^{1/3} \vo^{2/3}\sum_{n,m} \mc{A}_{(n,m)} \left( \frac{1}{{\rm Im}\,\tau} \right)^n \left[ \frac{({\rm Im}\,\tau)^{1/2}}{\vo^{1/3}} \right]^m  \,,
\label{Kcorrections}
\ee
where $\mc{A}_{(n,m)}$ are scale invariant combinations involving other non-axionic fields. Interestingly, supersymmetry dictates that the quantity which is corrected at a given order in $\alpha'$ and $g_s$ is $e^{-K/3}$ and not directly $K$. This observation explains why some corrections beyond tree-level which break scale invariance can still satisfy a generalised no-scale condition \cite{Burgess:2020qsc} which accounts for the presence of an extended no-scale structure \cite{Cicoli:2007xp}. The expression (\ref{Kcorrections}) is valid for $m=p+r$ where $p$ controls the number of curvature contributions while $r$ counts the factors of $G_3^2$ in 10D. Since a non-zero $G_3$ gives rise to the 4D superpotential $W_0 =\int_{X_3} G_3\wedge \Omega$, when $r\neq 0$ $\mc{A}_{(n,m)}$ should be proportional to $W_0^{2r}$. Knowing that the weight of $W_0$ is $\omega$, it is easy to deduce that the corresponding scale invariant combination has to be:
\be
\mc{A}_{(n,m)} = \hat{\mc{A}}_{(n,m)}\left(\frac{W_0^2}{\vo^{2/3}\,{\rm Im}\,\tau}\right)^r\qquad \text{with}\quad m=p+r\,,
\label{Spurion}
\ee
where $\hat{\mc{A}}_{(n,m)}$ is another scale invariant combination. As shown in \cite{Cicoli:2013swa}, the ratio appearing in (\ref{Spurion}) corresponds exactly to the parameter which controls the 4D superspace derivative expansion since:
\be
\left(\frac{g F}{M_\KK^2}\right)^2\sim \left(\frac{m_{3/2}}{M_\KK}\right)^2 \sim \frac{W_0^2}{\vo^{2/3}\,{\rm Im}\,\tau}\,,
\ee
where $F$ denotes the F-term of the light fields and $g$ is the coupling between heavy KK modes and light states. Thus in the regime where the superspace derivative expansion is under control, i.e. when $g F/M_\KK \ll 1$, the leading correction at fixed order in $\alpha'$ is expected to be the one corresponding to $r=0$. Notice that these higher F-term corrections might not be incorporated into $K$ but they might induce directly a correction to the scalar potential. This difference does not matter for our scaling arguments (which can be applied equally well by extending (\ref{Lag}) to the more general case of corrections to $\int \dif^4\theta D$), and so we shall consider them as `effective' corrections to $K$.

Perturbative $g_s$ and $\alpha'$ contributions to the scalar potential of the 4D EFT can be obtained by plugging $W=W_0$ and the K\"ahler potential given by (\ref{Kcorrections}) and (\ref{Spurion}) into the general expression:
\be
V = e^K\left ( K^{A\bar{B}}\, D_A W\, D_{\bar{B}}\overline{W}-3|W|^2\right ) = e^K\left( K^{\alpha\bar{\beta}}\, K_\alpha \, K_{\bar{\beta}}-3\right)|W_{0}|^2\,,
\label{eq:IIB:ScalarPotN1}
\ee
where $\alpha$ and $\beta$ run only over the K\"ahler moduli and, in the second equality, the dilaton and the complex structure moduli have been fixed supersymmetrically. This yields a generic $\mc{O}\left((\alpha')^m g_s^n\right)$ correction at $\mc{O}(F^{2r})$ of the form:
\bea
V_{(n,m,r)} &=& \hat{\mc{A}}_{(n,m,r)}\,\frac{W_0^2}{\vo^2\, {\rm Im}\,\tau}\left(\frac{W_0^2}{\vo^{2/3}\,{\rm Im}\,\tau}\right)^{r-1} \left( \frac{1}{{\rm Im}\,\tau} \right)^n \left[ \frac{({\rm Im}\,\tau)^{1/2}}{\vo^{1/3}} \right]^m 
\label{ExpParam} \\
&=& \hat{\mc{A}}_{(n,m,r)} \left( \frac{1}{{\rm Im}\,\tau} \right)^{n+r-m/2} \frac{W_0^{2r}}{\vo^{2+\frac{m}{3}+\frac23(r-1)}} \,.
\label{GenVcorr}
\eea
Notice that (\ref{ExpParam}) displays clearly the 3 expansion parameters of the EFT associated to higher F-terms, string loops and $\alpha'$ effects. The relation (\ref{GenVcorr}) implies also that an arbitrary contribution of the form $V \sim g_s^s\,W_0^{2r}/\vo^q$ corresponds to an $\mc{O}\left((\alpha')^m\, g_s^n\right)$ correction at $\mc{O}(F^{2r})$ with:
\be
m=3q-2(2+r)\qquad\text{and}\qquad n=s-2(r+1)+\frac32q\,.
\label{Important}
\ee
From $p\geq -1$ one finds also $r=m-p \leq m+1$ which implies that at a fixed $(\alpha')^m$ order, one can have higher F-term corrections up to $F^{2(m+1)}$. 

Moreover, the expression (\ref{GenVcorr}) reproduces several known perturbative effects:
\ben
\item $m=n=0$ and $r=1$ $\Rightarrow$ $p=-1$: This is the standard tree-level scalar potential arising from the 10D $\tilde{G}_3^2$ term:
\be
V_{(0,0,1)} \sim \frac{\hat{\mc{A}}_{(0,0,1)}}{{\rm Im}\,\tau}\, \frac{W_0^2}{\vo^2} \,.
\ee
The coefficient of this term is zero due to the no-scale cancellation: $\hat{\mc{A}}_{(0,0,1)}=0$.

\item $n=0$, $m=3$ and $r=1$ $\Rightarrow$ $p=2$: $(\alpha')^3$ correction at $\mc{O}(F^2)$ like the one computed by \cite{Becker:2002nn}, which should arise from 10D terms like $\mc{R}^3 \tilde{G}_3^2$ and $\mc{R}^2 |\nabla\tilde{G}_3|^2$:
\be
V_{(0,3,1)} \sim \hat{\mc{A}}_{(0,3,1)} \sqrt{{\rm Im}\,\tau}\,  \frac{W_0^2}{\vo^3} \,.
\label{VBBHL}
\ee
Notice that the dilaton dependence of this correction, when written in terms of the number of closed string loops $\ell_c$, reproduces the scaling expected from modular invariance:
\be
\left( \frac{1}{{\rm Im}\, \tau} \right)^{2\ell_c-1/2}= \frac{1}{{\rm Im}\,\tau}\left[({\rm Im}\,\tau)^{3/2} (\ell_c=0) + \frac{1}{\sqrt{{\rm Im}\,\tau}} (\ell_c=1) +\ldots \right]\sim \frac{f^{(0,0)}_{3/2}(\tau,\bar\tau)}{{\rm Im}\,\tau}\,. \nn
\ee

\item $n=0$, $m=3$ and $r=2$ $\Rightarrow$ $p=1$: $(\alpha')^3$ contribution at $\mc{O}(F^4)$, like those derived in~\cite{Ciupke:2015msa}, which should come from 10D terms like $\mc{R}^2 \tilde{G}_3^4$ and $(|\nabla\tilde{G}_3|^2)^2$:
\be
V_{(0,3,2)} \sim  \frac{\hat{\mc{A}}_{(0,3,2)}}{\sqrt{{\rm Im}\,\tau}}\, \frac{W_0^4}{\vo^{11/3}} \,.
\ee

\item $n=2$, $m=2$ and $r=1$ $\Rightarrow$ $p=1$: $(\alpha')^2$ open string 1-loop corrections (notice that $n=\ell_o+1$ for $\ell_o$ open string loops) at $\mc{O}(F^2)$, like those worked out in \cite{Berg:2005ja} which, from the closed string viewpoint, can be seen as due to the tree-level exchange of KK modes between parallel stacks of branes:
\be
V_{(2,2,1)} \sim  \frac{\hat{\mc{A}}_{(2,2,1)}}{({\rm Im}\,\tau)^2}\, \frac{W_0^2}{\vo^{8/3}} \,,
\ee
where however $\hat{\mc{A}}_{(2,2,1)}=0$ due to the extended no-scale cancellation. 

\item $n=2$, $m=4$ and $r=1$ $\Rightarrow$ $p=1$: $(\alpha')^4$ open string 1-loop effects at $\mc{O}(F^2)$, like those derived in \cite{Berg:2005ja}, which can be interpreted as the tree-level exchange of winding modes between intersecting stacks of branes:
\be
V_{(2,4,1)} \sim  \frac{\hat{\mc{A}}_{(2,4,1)}}{{\rm Im}\,\tau}\, \frac{W_0^2}{\vo^{10/3}} \,.
\ee
\een
This shows that the leading no-scale breaking effects in a large volume expansion seem to be $(\alpha')^3$ corrections at $\mc{O}(F^2)$, like the one derived in \cite{Becker:2002nn,Minasian:2015bxa}. Interestingly our scaling analysis combined with generalised no-scale relations is powerful enough to argue that $(\alpha')^2$ corrections should be absent at any order in $g_s$ \cite{Burgess:2020qsc} (unless they come along with $\ln\vo$-factors \cite{Weissenbacher:2019mef}). On the other hand, $(\alpha')^1$ effects, if they existed at some order in the $g_s$ expansion, would dominate over (\ref{VBBHL}) for $\vo\gg 1$ since they would scale as $\vo^{-7/3}$. However also these perturbative effects might not be generated. In fact, in Sec.~\ref{AbsenceOfAPOne} we provide evidence for the absence at tree-level in $g_s$ ($n=0$) of any correction to the 4D scalar potential which scales as $\vo^{-7/3}$. In App.~\ref{AppB} we discuss instead the effect on moduli stabilisation of potential $(\alpha')^1$ corrections arising at loop level.

\section{Leading no-scale breaking effects in type IIB}
\label{AbsenceOfAPOne} 

In this section we shall try to understand what is the leading order no-scale breaking contribution to the 4D scalar potential in the limit where the EFT is under control, i.e. for $\vo\gg 1$ and at tree-level in the string loop expansion. We shall first exploit the symmetry considerations of Sec.~\ref{TypeIIBscalings}, and we shall then confirm our findings with a combination of dimensional reduction and dimensional analysis.

\subsection{Symmetry considerations}

Symmetry arguments led us to the fundamental result (\ref{Important}) which implies:
\ben
\item Any contribution to the scalar potential at $\mc{O}(F^{2r})$ should feature $r\geq 1$. This implies $q\geq 2+m/3$. At tree-level, i.e. $m=0$, one has $q\geq 2$, and so the first dangerous higher derivative correction arises at order $(\alpha')^1$, i.e. $m=1$, corresponding to $q\geq 7/3$. 

\item A term as $\vo^{-7/3}$ can arise only at order $(\alpha')^m\, F^{2r}$ with $m = 3-2r$. For $r=0$ one has $m=3$, corresponding to the $\mc{O}((\alpha')^3) \mc{R}^4$ term. However $r=0$ implies $F^0$, and so no contribution to the 4D potential. This fits with the fact that the integral of $\mc{R}^4$ over a CY threefold gives zero. For $r=1$ one has instead $m=1$ at $\mc{O}(F^2)$, potentially at different orders in the string loop expansion counted by the powers of $g_s$. However, given that at tree-level there are no $(\alpha')^1$ corrections since the bulk action starts being corrected at $\mc{O}((\alpha')^3)$ while the brane action at $\mc{O}((\alpha')^2)$, no $\vo^{-7/3}$ term can be generated at tree-level in $g_s$. For $r=2$, $m$ becomes negative, leading to an absurd result.

\item A correction which scales as $\vo^{-8/3}$ would correspond to $m = 4-2r$. For $r=1$, we have $m=2$, and so an $(\alpha')^2\, F^2$ term which however should come with a zero coefficient due to the extended no-scale cancellation, regardless of the order in the $g_s$ expansion.\footnote{Modulo $\ln\vo$ corrections \cite{Grimm:2017pid,Weissenbacher:2019mef,Weissenbacher:2020cyf}, if present at all.} This can be easily seen from the fact that, in a supersymmetric theory, such a term should come from a $\vo$-independent correction $c$ to the K\"ahler potential of the form $e^{-K/3} = \vo^{2/3}+c$ that would however satisfy a generalised no-scale relation \cite{Burgess:2020qsc}. For $r=2$, $m=0$ which would be an $F^4$ term at tree-level. This would correspond to the $\vo^{-8/3}$ term used in T-brane uplifting scenarios \cite{Cicoli:2015ylx} since it is a tree-level effect that scales in terms of F-terms of matter fields as $(F^{\rm matter})^2$ where it can be easily seen that they are related to the F-terms of the K\"ahler moduli as $F^{\rm matter} \sim (F^T)^2$.

\item A perturbative correction which scales as $\vo^{-3}$ features $m = 5-2r$. For $r=1$, one has $m=3$, and so standard $(\alpha')^3$ corrections at $\mc{O}(F^2)$ \cite{Becker:2002nn,Minasian:2015bxa}. For $r=2$ one would have instead $m=1$ but we have just recalled that there are no $(\alpha')^1$ corrections in 10D at tree-level. The $r=3$ case can instead be safely ignored since the $\alpha'$ order would become negative.
\een
This analysis, just based on symmetries and the known absence of $(\alpha')^1$ corrections at tree-level, implies that the leading no-scale breaking effect in the 4D scalar potential at tree-level should arise from $(\alpha')^3$ effects and should scale as $\vo^{-3}$.

\subsection{Arguments from dimensional analysis}

Let us now provide further evidence in favour of this claim from arguments based on a dimensional analysis combined with dimensional reduction. As we have seen above, the order in $\alpha'$ and the number of F-terms is dictated by the $\vo$ and $W_0$ dependence of a generic perturbative correction. This has been derived in (\ref{GenVcorr}) using symmetry arguments and it agrees with the expectations from direct dimensional reduction. In fact, when all components of tensors and derivatives are taken along internal directions\footnote{Upon modifying the volume factor coming from Weyl rescaling, this dimensional analysis can in principle be applied to determining the volume behaviour of any other term in the $4$D effective action.}, the generic $\mc{O}\left((\alpha')^m g_s^n\right)$ 10D correction (\ref{ScorrEinst}) generates a contribution to the 4D scalar potential whose $\vo$ dependence can be inferred as follows \cite{Conlon:2005ki}:
\begin{enumerate}
\item[(i)] the Weyl rescaling to 4D Einstein frame yields a $\vo^{-2}$ factor;
\item[(ii)] the integration over $X_3$ brings a $\vo$ contribution;
\item[(iii)] as can be seen from (\ref{voscaling}), each inverse metric factor introduces a $\vo^{-1/3}$ dependence.
\end{enumerate}
\noindent The number of F-terms and the associated $W_0$ dependence can instead be easily deduced from the number of $G_3$ terms in 10D. Hence dimensional reduction is expected to produce:
\be
V_{(n,m,r)} \sim \frac{W_0^{2r}}{\vo^{1+\frac{\lambda}{3}}}\qquad\text{with}\qquad \lambda =3r + p + 1\,,
\label{GenVcorrDimRed}
\ee
where $\lambda$ counts the net number of inverse metric factors, and its expression in terms of $r$ and $p$ follows from (\ref{ScorrEinst}).
This formula can further be motivated as follows: flux quantisation implies $\tilde{G}_{3}\sim \cO(\alpha')$ as the leading order solution to the $10$D equations of motion. At a fixed order $m$ in the $10$D $\alpha'$ expansion, one has $\lambda=2r+m+1$ net factors of inverse metrics.
Hence, for dimensional reasons, each power of $(\tilde{G}_3)^2$ introduces an additional $\vo^{-2/3}$ power in \eqref{GenVcorrDimRed}.\footnote{\label{fn:F5}Similar arguments apply to $\tilde{F}_5$ which satisfies $\tilde{F}_5\sim\cO((\alpha')^2)$ due to the 5-form Bianchi identity. Hence, each power of $(\tilde{F}_5)^2$ comes with a volume factor of $\vo^{-4/3}$.}
Using $p = m-r$ it is straightforward to realise that the $\vo$ dependence in (\ref{GenVcorrDimRed}) agrees with the one in (\ref{GenVcorr}).
We summarise in Tab.~\ref{Tab1} the volume scaling and the F-term order of different $\alpha'$ contributions to the 4D scalar potential arising from various 10D terms. For completeness we include also higher derivative terms like $\mc{R}^2$ and $\mc{R}^3$ which are forbidden in the type IIB action due to supersymmetry \cite{Polchinski:1998rr} and $\mc{R}^4$ even if it does not contribute to the 4D scalar potential due to Ricci-flatness and K\"ahlerity of the underlying manifold \cite{Grisaru:1986px} (see App.~\ref{app:TensorsEllFibGen} for details).

\begin{table}[t!]
\centering
\begin{tabular}{|c|c|c|c|c|c|c|c|c|}
\hline 
&&&&&&&& \\ [-1.0em]
$(\ap)^m$ & $p$ & $r$ & 10D term & 10D $\cN=2$ &  $\lambda$ & $V(\vo)$  & 4D $V$ & $F^{2r}$ \\  [0.3em]
\hline 
\hline 
&&&&&&&& \\ [-1.em]
0 & 0 & 0 & $\mc{R}$ & $\surd$ & 1 & $\vo^{-4/3}$ & $\times$ & -- \\ [0.3em]
\hline 
&&&&&&&& \\ [-0.8em]
0 & -1 & 1 & $|\tilde{G}_3|^2$ & $\surd$ & 3  & $\vo^{-2}$ & $\surd$ & $F^2$ \\ [0.3em]
\hline 
&&&&&&&& \\ [-0.9em]
1 & 1 & 0 & $\mc{R}^2$ & $\times$ &  2 & $\vo^{-5/3}$ & $\times$ & --\\ [0.3em]
\hline 
&&&&&&&& \\ [-0.9em]
1 & 0 & 1 & $\mc{R} |\tilde{G}_3|^2$ & $\times$ &  4 & $\vo^{-7/3}$ & $\times$ & -- \\ [0.3em]
\hline 
&&&&&&&& \\ [-0.9em]
1 & -1 & 2 & $|\tilde{G}_3|^4$ & $\times$ &  6 & $\vo^{-3}$ & $\times$ & -- \\ [0.3em]
\hline 
&&&&&&&& \\ [-0.8em]
2 & 2 & 0 & $\mc{R}^3$ & $\times$ & 3 & $\vo^{-2}$ & $\times$ & --\\ [0.3em]
\hline 
&&&&&&&& \\ [-0.8em]
2 & 1 & 1 & $\mc{R}^2 |\tilde{G}_3|^2$ & $\times$ &  5 & $\vo^{-8/3}$ & $\times$ & --\\ [0.3em]
\hline 
&&&&&&&& \\ [-0.9em]
3 & 3 & 0 & $\mc{R}^4$ & $\surd$ & 4 & $\vo^{-7/3}$ & $\times$ & -- \\ [0.3em]
\hline 
&&&&&&&& \\ [-0.8em]
3 & 2 & 1 & $\mc{R}^3 |\tilde{G}_3|^2$ & $\surd$ &  6 & $\vo^{-3}$ & $\surd$ & $F^2$ \\ [0.3em]
\hline 
&&&&&&&& \\ [-0.8em]
3 & 1 & 2 & $\mc{R}^2 |\tilde{G}_3|^4$ & $\surd$ &  8 & $\vo^{-11/3}$ & $\surd$ & $F^4$ \\ [0.3em]
\hline 
&&&&&&&& \\ [-0.8em]
3 & 0 & 3 & $\mc{R} |\tilde{G}_3|^6$ & $\surd$ &  10 & $\vo^{-13/3}$ & $\surd$ & $F^6$ \\ [0.3em]
\hline 
&&&&&&&& \\ [-0.8em]
3 & -1 & 4 & $|\tilde{G}_3|^8$ & $\surd$ &  12 & $\vo^{-5}$ & $\surd$ & $F^8$ \\ [0.3em]
\hline 
\end{tabular} 
\caption{Volume scaling and F-term order of different $\alpha'$ corrections to the 4D scalar potential generated by various 10D terms. The $\vo$ dependence and the number of inverse metric factors $\lambda$ is obtained from (\ref{GenVcorr}) and (\ref{GenVcorrDimRed}). 10D terms of the form $\mc{R}^{p+1} (\nabla\tilde{G}_3)^{2r}$ can also be incorporated by noticing that they would scale as $\mc{R}^{p+r+1} \tilde{G}_3^{2r}$.}
\label{Tab1} 
\end{table}

Coming back to the leading order no-scale breaking effects, we now apply the dimensional analysis to argue against the presence of $\vo^{-7/3}$ corrections at string tree-level in the $4$D scalar potential.
The starting points are two higher dimensional actions: the 10D bulk type IIB action and the 8D D7/O7 DBI and WZ actions. In this type of compactifications, localised D5-brane sources are projected out by the orientifold. Localised D3-branes are instead relevant for the 4D 2-derivative effective action as far as their backreaction on the closed string background is concerned, but higher derivative couplings on their worldvolume can clearly be ignored. We will come back to D3-branes later. Upon dimensional reduction (on CY threefolds and on K\"ahler twofolds respectively), these two actions potentially give rise to a plethora of perturbative corrections to the 4D scalar potential which we now discuss schematically.

\subsubsection*{Bulk corrections}

The 32 supercharges characterising the 10D bulk theory force the first higher derivative corrections to arise only at order $(\ap)^3$. Schematically, one finds:
\be
\mc{L}_{\rm bulk}=\mc{R}+|\mc{P}|^2+|\tilde{G}_3|^2+|\tilde{F}_5|^2+(\ap)^3 \mc{L}'(\tau;\mc{R},\mc{P},\tilde{G}_3,\tilde{F}_5)+\mc{O}((\ap)^5)\,,
\label{10dBulk}
\ee
where $\mc{L}'$ collects all possible 8-derivative couplings, and we have neglected the classical Chern-Simons term since it is irrelevant for the present discussion. Notice that the axio-dilaton can appear either in $\mc{P}$ (with gradients involved) or in the modular functions multiplying the various kinematic structures. Moreover all terms in $\mc{L}'$ contain an even number of $\tilde{G}_3$'s and $\tilde{F}_5$'s due to parity invariance. 

Recall also that, due to Ricci flatness and K\"ahlerity, terms in \eqref{10dBulk} involving only powers of $\mc{R}$ give vanishing contributions to the 4D scalar potential when integrated on the internal manifold.\footnote{It is well-known that in flux-less $\cN=2$ compactifications, moduli remain massless to all orders in $\alpha'$ and $g_s$ \cite{Conlon:2005ki}. In App.~\ref{app:TensorsEllFibGen} we show explicitly that $\mc{R}^4$ does not contribute to the scalar potential.} The same holds true for CY fourfold compactifications of M-theory down to 3D where a 3D scalar potential can be generated only for a non-vanishing $G_4$ flux \cite{Haack:2001jz,Grimm:2015mua}. Considering an elliptically fibred fourfold and performing the F-theory limit, we therefore conclude that a 4D scalar potential can be generated only by turning on either $\tilde{G}_3$ or $\tilde{F}_5$ in the bulk, or $\mc{F}_2$ on D7-branes. This is a crucial statement since terms like $\mc{R}^4$ or $\mc{P}^{2n}(\nabla{\mc{P}})^m \mc{R}^{4-n-m}$ with $1\leq n\leq 4$, $0\leq m\leq4$, if they were contributing to the 4D scalar potential, would produce corrections which scale as $\vo^{-7/3}$. This is easy to see: each power of $\mc{R}$ and of $\nabla{\mc{P}}$, and each pair of $\mc{P}$'s need one net factor of inverse metric of the CY threefold to give rise to a Lorentz invariant. Hence $\lambda=4$ and (\ref{GenVcorrDimRed}) yields $V\sim \vo^{-7/3}$.  

As we have already seen, the leading corrections beyond the tree-level $\vo^{-2}$ term coming from $|\tilde{G}_3|^2$, originate from reductions of terms like $\mc{R}^3 \tilde{G}_3^2$ and $\mc{R}^2 |\nabla\tilde{G}_3|^2$ which scale like $\vo^{-3}$ with $\lambda=6$ in \eqref{GenVcorrDimRed} (corresponding to $(\ap)^3$ corrections at $F^2$ order). Every pair of $\tilde{G}_3$'s that replaces a power of $\mc{R}$ introduces an additional $\vo^{-2/3}$ suppression. Analogous considerations hold for higher derivative terms containing $\tilde{F}_5$, which start contributing at order $\vo^{-11/3}$ and acquire an additional $\vo^{-4/3}$ suppression each time a pair of $\tilde{F}_5$'s replaces an $\mc{R}$ (recall footnote \ref{fn:F5}). Notice that, contrary to the purely gravitational sector, in $\mc{N}=1$ compactifications there is no reason to exclude contributions from terms of the form $\mc{R}^{3-n} \tilde{G}_3^2 \mc{P}^{2n}$ with $1\leq n\leq3$ (or analogous terms involving also $\nabla{\mc{P}}$). This is because the presence of D7-branes induces non-trivial gradients for the axio-dilaton.\footnote{Corrections of this type are e.g.~those discussed in \cite{Minasian:2015bxa} from a 12D viewpoint.} 

To summarise, the classical KK reduction of the 8-derivative bulk 10D action down to 4D on a orientifolded CY threefold gives rise to only $(\ap)^{\rm odd}$ corrections to the scalar potential, starting from $(\ap)^3$ at tree-level in $g_s$ (sphere level) which yields $V\sim \vo^{-3}$. 

\subsubsection*{Brane corrections}

The 16 supercharges of the 8D worldvolume theory of a stack of D7-branes (or O7-planes) fix to $(\ap)^2$ the order of the leading higher derivative corrections. Schematically, this amounts to:
\be
\mc{L}_{\rm loc}=\frac{\sqrt{-g}}{(\ap)^2}+\Tr\left(|\mc{F}_2|^2+|D\Phi|^2+[\Phi,\Phi]^2\right)+(\ap)^2\mc{L}''(\tau; \mc{R}, \mc{F}_2, D\Phi,[\Phi,\Phi])+\mc{O}((\ap)^4)\,,
\label{10dBrane}
\ee
where again we have ignored the classical Chern-Simons couplings to RR forms since they are irrelevant for our discussion. All bulk quantities in \eqref{10dBrane} are meant to be pulled-back to the brane world-volume, $g$ denotes the determinant of the induced metric with $\mc{R}$ its curvature 2-form\footnote{Here it is not relevant to distinguish between curvature of the tangent and of the normal bundle.}, $\mc{F}_2$ is the gauge invariant world-volume field-strength and $\Phi$ collectively denotes worldvolume scalars (possibly non-Abelian). The $\tau$ dependence is again due to the modular functions multiplying the various kinematic structures and we have dropped all terms of the type $\mc{R}^2$ since there is no 4D scalar potential generated purely by geometry. It is a common convention (T-duality friendly) to take world-volume fields, like $\Phi$ and the gauge field $A$, to have mass dimension 1 (as opposed to bulk fields). Moreover, T-duality and gauge invariance force any possible correction to be written just in terms of the arguments of $\mc{L}''$.\footnote{With the only exception of the implicit dependence of bulk quantities on the normal coordinates $\sqrt{\alpha'}\,\Phi$, which is often used to encode backreaction effects of the branes on the closed string background.} 

D7-brane tadpole cancellation guarantees that the classical tension does not contribute to the 4D scalar potential which would have otherwise yielded a $\vo^{-4/3}$ dependence from integrating the first term in \eqref{10dBrane}. In fact, following the same logic which led to (\ref{GenVcorrDimRed}) with the only difference that now the internal integration gives a $\vo^{2/3}$ instead of a $\vo$ factor, we can easily infer that a generic term in the localised action (\ref{10dBrane}) can in principle generate a contribution to the 4D scalar potential which scales as $V \sim \vo^{-(4+\lambda)/3}$ where $\lambda$ counts again the number of inverse metric factors. The classical tension would correspond to $\lambda=0$.

The classical 4D scalar potential thus arises from integrating $|\mc{F}_2|^2$ over the internal 4-cycle.\footnote{More precisely, only the anti-self-dual part of $\mc{F}_2$ generates a potential since the self-dual part contributes to D3-brane tadpole cancellation.} If we have a non-Abelian stack and/or the brane has a non-trivial profile in the normal directions, further contributions to the 4D scalar potential come from integrating $|D\Phi|^2$ and $[\Phi,\Phi]^2$ \cite{Myers:1999ps}. Notice that all of these terms would produce a scalar potential which scales as $\vo^{-2}$. In fact they all have $\lambda=2$ since they involve 2 pairs of indices (both longitudinal for $|\mc{F}_2|^2$, both transverse for $[\Phi,\Phi]^2$, while one longitudinal and one transverse for $|D\Phi|^2$), and hence need 2 inverse metric factors to give rise to a Lorentz invariant. The term proportional to $\mc{F}_2$ is the well-known D-term scalar potential contribution from moduli-dependent Fayet-Iliopoulos (FI) terms \cite{Haack:2006cy}.

The leading higher derivative corrections to the brane action are all encoded in $\mc{L}''$ and are all \emph{quartic} in $\mc{R}$, $\mc{F}_2$, $D\Phi$ and $[\Phi,\Phi]$.\footnote{Again terms of the type $\mc{R}^4$ are not expected to contribute to the 4D potential. Moreover terms where a power of $\mc{F}_2$ is replaced by a pair of $D$'s give the same $\vo$ dependence.} Thus each term necessarily involves 4 pairs of indices (which may be all longitudinal, all transverse, or mixed) which implies $\lambda=4$ and $V \sim \vo^{-8/3}$, as expected for $(\ap)^2$ corrections. As stressed above, supersymmetry and generalised no-scale relations should imply the absence of these corrections \cite{Burgess:2020qsc} (in the sense that they might just induce moduli redefinitions \cite{Grimm:2013gma}). However if they arise with an additional logarithmic dependence on the K\"ahler moduli, they might still represent the leading no-scale breaking effect \cite{Weissenbacher:2019mef}. Symmetries and scaling arguments are clearly not enough to provide a definite answer to this important issue. 

To summarise, the classical KK reduction of the 8D higher derivative brane action down to 4D on a K\"ahler twofold gives rise to only $(\ap)^{\rm even}$ corrections to the scalar potential, starting from $(\ap)^2$ level which might yield at most a correction of the form $V \sim \vo^{-8/3}$ (starting at the disk and projective-plane level, in the string coupling expansion).

Let us finally mention that we focused above only on stacks of D7-branes in isolation whose physics is accurately described by the DBI and WZ actions. However in $\mc{N}=1$ compactifications D7-branes can also intersect in complex codimension-1 loci where the 8D action fails to fully capture the physics due to the possible presence of massless matter at the intersection. Such special loci may be viewed as 6D defects of the 8D theory with their own EFT. Unfortunately, not much is known about the structure of higher derivative corrections to such a theory. However two intersecting stacks of D7-branes\footnote{We assume that the stacks wrap homotopically equivalent 4-cycles. We conjecture the same conclusions to hold in the more general case where however we cannot use the continuity arguments employed here.} can approximately be described as a single stack (of size the sum of the two sizes) with a non-trivial profile for the worldvolume scalars \cite{Beasley:2008dc,Marchesano:2019azf}.\footnote{The smaller the intersection angle, the more accurate this description compared to the defect picture.} Such a profile encodes the information of the wavefunctions of localised fields as can be seen by solving the D-term differential equations.\footnote{As an easy example in affine space, consider two D7-branes intersecting on $z_1=z_2=0$ in $\mathbb{C}^2$. This system can equivalently be described by a stack of two D7-branes on $z_1=0$ with a Higgs field given by $\Phi={\rm diag}(z_2,-z_2)$.} Aside from the details, what this reasoning teaches us is that there cannot be higher derivative corrections on the defects which cannot be continuously extracted from corrections already present in the 8D worldvolume action.

\subsubsection*{Backreaction}

The analysis of the previous paragraphs does not take into account the effect of branes and fluxes on the bulk background. A clever way to capture at least some of them is to regard the bulk fields as functions of the brane worldvolume scalars and Taylor expand them.\footnote{This method has been introduced in \cite{Myers:1999ps} and later used in \cite{Camara:2003ku,Grana:2003ek,Camara:2004jj} to compute soft supersymmetry breaking terms.} The couplings that arise induce new operators on the worldvolume field theory, which softly break the original 16 supercharges. This phenomenon makes also the D3-brane worldvolume theory contribute to the 4D scalar potential. Imaginary-anti-self-dual bulk fluxes indeed generate terms like $g_s (\ap)^2 (*_6 \tilde{G}_3-{\rm i} \tilde{G}_3)_{ijk}\Phi^i\Phi^j\Phi^k$ on a stack of D3-branes,\footnote{They arise by a first order Taylor expansion of the non-Abelian DBI coupling $B[\Phi,\Phi]$ \cite{Camara:2003ku}.} where the power of $g_s$ shows that such effects appear at 1-loop in string perturbation theory. Analogous terms are expected to pop up also on D7-branes and to contribute to the 4D potential after integration on the internal 4-cycle. An interesting example in the case of T-branes is a term which scales as $\vo^{-8/3}$ that has been used to achieve dS vacua \cite{Cicoli:2015ylx}. Given that the Taylor expansion does not require any metric contraction, the structure of all these terms is such that a Lorentz invariant can be constructed only in the presence of an \emph{even} number of net inverse metric factors. Thus $\lambda$ has to be even, and so no $\vo^{-7/3}$ correction can be generated this way.

Another important backreaction effect is the generation of warping in the spacetime metric due to branes and fluxes \cite{Giddings:2005ff}. Thanks to open/closed string duality, by solving for the warp factor the tree-level equations of motion in the closed string sector, we infer a 1-loop correction in the open string (and non-orientable closed string) sector. Following the discussion of \cite{Junghans:2014zla}, the $\vo$ dependence of such a correction to 4D scalar potential depends on the dimension of the D-branes/O-planes involved. In the type of compactifications we are analysing, this dependence is however bounded from below by $\vo^{-8/3}$ (due to graviton exchange between D7-branes).

\subsubsection*{Final remarks}

Altogether, the arguments given above lead us to state with reasonable certainty that $\vo^{-7/3}$ corrections are absent in the 4D scalar potential \emph{at string tree-level}. This is because all 4D corrections at string tree-level must already be present in the higher dimensional (and more supersymmetric) theories whose zero-mode reductions we have analysed in detail. It is starting from the string 1-loop level that new states (such as KK and winding modes) come into play and potentially contribute to amplitudes between low-energy states. Therefore we cannot guarantee that the reduction of supersymmetry down to 4 supercharges caused by compactification does not give birth to novel 4D perturbative corrections. Famous examples of such loop corrections due to exchange of KK and winding modes are those computed in \cite{Berg:2005ja} for both $\mc{N}=2$ and $\mc{N}=1$ toroidal orientifolds. They appear at $(\ap)^{\rm even}$ order but their origin as higher derivative corrections in the D7 worldvolume theory (or equivalently in M/F-theory) is still unclear.

Apart from the backreaction effects discussed above, little is known about the consequences of supersymmetry breaking on the starting bulk and brane actions. A hint in this direction might be obtained by analysing loop amplitudes of 11D supergravity compactified on elliptically fibred CY fourfolds. Such amplitudes, albeit in the case of toroidal reductions only, were shown in \cite{Green:1997as,Green:1999by} to efficiently capture string loop and non-perturbative corrections.\footnote{Examples of that are the bulk couplings $(\ap)^3\mc{L}'$ in \eqref{10dBulk}.} In order to able to make an \emph{exact-in-$g_s$} claim of absence of $\vo^{-7/3}$ corrections, one should find a way to perform a dimensional analysis of the kinematic structures expected from loop amplitudes of 11D supergravity on non-toroidal backgrounds. This will be further motivated in Sec.~\ref{sec:QuantumReductions}. In the remainder of this paper we shall use M/F-theory techniques to infer the form of $\alpha'$ corrections at different orders in the 4D superspace derivative expansion without however being able to shed too much light on the $g_s$ expansion.

\section{$\alpha'$ corrections from dimensional analysis in F/M-theory}
\label{sec:MF}

Here we come to the core of our study. In Sec.~\ref{sec:Flimit} we outline the rules for connecting the 4D EFT to the intermediate 3D theory obtained after CY fourfold reductions of M-theory. In Sec.~\ref{sec:dimAnalysis} we then apply these rules to the reduction of higher derivative structures appearing in 11D supergravity. This will allow us to make general statements about the ensuing $\alpha'$ corrections to the 4D F-theory effective action.

\subsection{The F-theory limit}
\label{sec:Flimit}

In this section we will discuss the rules to extract the 4D F-theory effective action from the 3D one arising from the associated M-theory reduction \cite{Grimm:2010ks,Grimm:2011sk}. To do so, we will exploit F/M-theory duality, which can be roughly summarised as follows \cite{Denef:2008wq}. One compactifies M-theory on an elliptically fibred CY fourfold $Y_4$ and picks up a basis of fundamental 1-cycles, called $A$ and $B$-cycles, on the generic smooth fibre. First of all one reduces the 11D theory on the $A$-cycle and then T-dualises it along the $B$-cycle, ending up with type IIB compactified on the threefold base of the fibration (over which the axio-dilaton has a non-trivial profile) times the circle T-dual to the $B$-cycle. The final step is the limit of vanishing volume of the original fibre, the so-called `F-theory limit', which renders the effective theory Poincar\'e invariant in 4D, by decompactifying the extra circle in IIB.\footnote{See \cite{Grimm:2013gma,Grimm:2013bha} for earlier formulations of the F-theory limit. The present one differs from them only in the powers of the base volume, $\vo$, which is a finite quantity in the F-theory limit. However, since one of the main goals of this paper is to estimate the behaviour of string corrections at large $\vo$, it is crucial to derive how the various terms in the action precisely scale with $\vo$ after the F-theory limit.}

The best strategy is to compare two actions in 3D: one obtained from a fourfold reduction of the 11D supergravity action, and the other from a preliminary threefold reduction of the type IIB supergravity action, followed by a circle reduction. We first make this comparison at the classical level, i.e.~considering all terms at lowest order in $\ell_M$ and $\ap$, in order to derive all the formulas which are relevant to take the F-theory limit. These formulas are then used in Sec.~\ref{sec:dimAnalysis} to discuss the structure of $\alpha'$ corrections. We will not pay attention to the exact derivation of all $2\pi$ factors since our focus will instead be on volume factors and on how the duality between M and F-theory relates $\ell_M$ and $\ap$.

Since fluxes will play a crucial r\^ole in what follows, let us first say a few preliminary words about them. According to the F/M-theory duality, Poincar\'e invariance in 4D forces the internal M-theory $G_4$ flux to have 1 leg along the fibre and 3 legs on the base \cite{Dasgupta:1999ss}.\footnote{We will not discuss the case where $G_4$ has 3 legs on $\mathbb{R}^{1,2}$ which is instead associated to $F_5$ and to the warp factor in type IIB.} Depending on whether the 3 legs on the base are along a 3-cycle or a 3-chain, $G_4$ gives rise respectively to bulk type IIB $F_3$ and $H_3$ fluxes or to the D7-brane $\mc{F}_2$ flux. The integral flux quanta on both sides of the duality must be the same, translating into the following equality of vacuum expectation values:
\be
\label{G4F3H3}
\frac{1}{\ell_{M}^3}\int_{\mc{C}_4} G_4 = \frac{1}{\alpha'}\int_{\mc{C}_3} p F_3 +q H_3\,.
\ee
Here the 4-cycle $\mc{C}_4$ is a circle bundle over the 3-cycle/3-chain $\mathcal{C}_3$ with fibre $pS^1_A+q S^1_B$, where $p,q$ are integers and $S^1_{A,B}$ are the $A$ and $B$-cycle respectively. As a consequence, the flux-induced M2/D3 tadpole reads:
\be
\frac{1}{2\ell_M^6}\int_{Y_4} G_4\wedge G_4=\frac{1}{(\alpha')^2}\int_{B_3} F_3\wedge H_3\,,
\ee
where $Y_4\to B_3$ is the elliptic fibration and the factor of $1/2$ is combinatorial. Hence one gets a match between the Chern-Simons terms of 11D and type IIB supergravity by requiring:
\be
\frac{1}{\ell_M^3}\int_{\mathbb{R}^{1,2}} C_3 = \frac{1}{(\alpha')^2}\int_{\mathbb{R}^{1,2}\times S^1} C_4\,.
\label{C3C4}
\ee
We will find below how to write $G_4$ in terms of $F_3$ and $H_3$, and $C_4$ in terms of $C_3$, in order to satisfy both \eqref{G4F3H3} and \eqref{C3C4}.

On the CP-even side of the 11D action, the kinetic term of the $G_4$ flux combines with the 8-derivative curvature correction $\mc{R}^4$ to give rise, upon using M2-brane tadpole cancellation\cite{Duff:1995wd,Sethi:1996es,Dasgupta:1996yh}, to the 3D scalar potential \cite{Becker:1996gj,Haack:2001jz}:
\be
V_{\rm tree}^{\rm (M)}=\frac{1}{2\ell_M^6}\int_{Y_4}\left(G_4\wedge\star_8 G_4- G_4\wedge G_4\right)
=\frac{1}{\ell_M^6}\int_{Y_4} G_{4-} \wedge \star_8 G_{4-}\,,
\label{VM}
\ee
where $G_{4-}$ denotes the anti-self-dual part of $G_4$. Correspondingly, on the side of type IIB compactified on $B_3$, one has:
\be
V_{\rm tree}^{\rm (F)}=\frac{1}{(\alpha')^2}\int_{B_3}\frac{G_{3-}\wedge \star_6 \bar{G}_{3-}}{{\rm Im}\,\tau}\,,
\ee 
which originates from the kinetic term of the $G_3$ flux, after removing the contribution of its imaginary-self-dual part $G_{3+}$, fixed by the D3 tadpole. The F/M-theory duality in this case amounts to the statement:
\be
V_{\rm tree}^{\rm (M)}= V_{\rm tree}^{\rm (F)}\,.
\label{scalarpotential}
\ee 
Moreover, according to the M/F theory duality, Euclidean M5-branes in M-theory which are `vertical', i.e. wrapped around a 6-cycle in $Y_4$ having the structure of an elliptic fibration over a 4-cycle in $B_3$, descend to Euclidean D3-branes wrapped on the same 4-cycle. This leads us to equating their tension in the respective units, namely:
\be
v^2 v_f = v_b^2\,,
\label{EM5ED3} 
\ee
where $v_f$ and $v$ are respectively the volumes of the fibre and a 2-cycle of the base, computed with the M-theory metric in units of $\ell_M$, whereas $v_b$ is the volume of a 2-cycle of the base, computed with the type IIB Einstein frame metric in units of $\alpha'$.\footnote{Given that we are interested just in volume scalings, we can consider without loss of generality a base manifold with just a single K\"ahler modulus.}

After reducing the 11D classical action down to 3D on an elliptically fibred CY fourfold, one finds:
\be
S_{\rm M}^{(3)} = \frac{1}{\ell_M}\int \cV_4\, \tilde{R}^{(3)}\,\sqrt{-\tilde{g}^{(3)}}\,{\rm d}^3x + \frac{1}{\ell_M^3}\int V_{\rm tree}^{\rm (M)}\,\sqrt{-\tilde{g}^{(3)}}\,{\rm d}^3x\,,
\ee
where $\tilde{R}^{(3)}$ denotes the Ricci scalar of the 3D metric $\tilde{g}_{\mu\nu}^{(3)}$, and $\cV_4$ is the volume of $Y_4$ in units of $\ell_M$ which can be written in terms of the K\"ahler form $J$ as:
\be
\cV_4=\dfrac{1}{\ell_M^8}\int_{Y_4}\, J\wedge J\wedge J\wedge J\,.
\label{eq:DefFourfoldVol} 
\ee
To bring this action to the standard 3D Einstein frame, we have to rescale the metric as:
\be
\tilde{g}_{\mu\nu}^{(3)}=g_{\mu\nu}^{(3)} \left(\frac{\langle \cV_4 \rangle}{\cV_4}\right)^2\,,
\ee
so that we get:
\be
S_{\rm M}^{(3)} = M_3\int R^{(3)}\,\sqrt{-g^{(3)}}\,{\rm d}^3x + M_3^3\int\frac{V_{\rm tree}^{\rm (M)}}{\cV_4^3}\,\sqrt{-g^{(3)}}\,{\rm d}^3x\,,
\label{3dM}
\ee
where we have defined the 3D Planck mass as $M_3\equiv\langle \cV_4\rangle/\ell_M$.

On the other hand, the reduction of the type IIB Einstein frame classical action down to 4D on the base $B_3$ of the elliptic fibration gives:
\be
S_{\rm IIB}^{(4)}\;=\;\frac{1}{\alpha'}\int \vo\, \tilde{R}^{(4)}\,\sqrt{-\tilde{g}^{(4)}}\,{\rm d}^4x + \frac{1}{(\alpha')^2}\int V_{\rm tree}^{\rm (F)}\,\sqrt{-\tilde{g}^{(4)}}\,{\rm d}^4x\,,
\ee
where $\vo$ is the volume of $B_3$ in units of $\alpha'$. To bring this action to the standard 4D Einstein frame, the 4D metric has to be rescaled as:
\be
\tilde{g}_{\mu\nu}^{(4)}=g_{\mu\nu}^{(4)} \,\frac{\langle \vo \rangle}{\vo}\,,
\ee
so that we get:
\be
S_{\rm IIB}^{(4)} = M_p^2\int{R}^{(4)}\,\sqrt{-{g}^{(4)}}\,{\rm d}^4x + M_p^4\int\frac{V_{\rm tree}^{\rm (F)}}{\vo^2}\,\sqrt{-{g}^{(4)}}\,{\rm d}^4x\,,
\ee
where the 4D Planck mass has been defined as $M_p\equiv\sqrt{\langle \vo \rangle/\alpha'}$. Next, we dimensionally reduce this action to 3D on a circle of radius $r \,\sqrt{\alpha'}$, obtaining:
\be
S_{\rm IIB}^{(3)} = M_p^2\sqrt{\alpha'}\int r\,\check{R}^{(3)}\,\sqrt{-\check{g}^{(3)}}\,{\rm d}^3x + M_p^4\sqrt{\alpha'} \int\frac{V_{\rm tree}^{\rm (F)}}{\vo^2}\,r\,\sqrt{-\check{g}^{(3)}}\,{\rm d}^3x\,.
\label{3dIIB}
\ee
In order to match \eqref{3dM} and \eqref{3dIIB}, we first have to write both of them in terms of the same dynamical fields (in this case just the 3D metric). This leads us to perform another Weyl rescaling:
\be
\check{g}_{\mu\nu}^{(3)}=g_{\mu\nu}^{(3)} \left(\frac{\langle r\rangle}{r}\right)^2\,,
\label{eq:WeylRescaling3dmetricR} 
\ee
which turns \eqref{3dIIB} into:
\be
S_{\rm IIB}^{(3)} = M_p^2\sqrt{\alpha'} \langle r\rangle\int{R}^{(3)}\,\sqrt{-{g}^{(3)}}\,{\rm d}^3x
+ M_p^4 \sqrt{\alpha'} \langle r\rangle^3\int\frac{V_{\rm tree}^{\rm (F)}}{\vo^2 r^2}\,\sqrt{-{g}^{(3)}}\,{\rm d}^3x\,.
\label{3dIIBresc}
\ee
Now we are ready to match the 3D Lagrangians \eqref{3dM} and \eqref{3dIIBresc}. Using \eqref{scalarpotential} and the definitions of the 3D and 4D Planck masses, we find:
\be
\cV_4^3 = \langle \vo \rangle \, \vo^2 r^2 \qquad\text{and}\qquad \frac{\ell_M}{\sqrt{\alpha'}}=\frac{1}{\langle r\rangle^{1/3}}\,.
\label{matching}
\ee
Notice that the first of the above equations is consistent with the expected relation between the M-theory (3D) and the type IIB (4D) K\"ahler potentials, up to a constant shift (which does not affect the field space metric):
\be
K^{\rm M} = K^{\rm IIB}-2\ln r-\ln{\langle \vo \rangle}\,.
\ee
The second equation in \eqref{matching} can also be rewritten as:
\be
\frac{\ell_M}{\sqrt{\alpha'}}=\langle v_f \rangle^{1/4} \,,
\label{lMls}
\ee
which can be easily seen from the fact that (\ref{EM5ED3}) implies $\vo =  \cV_4 \sqrt{v_f}$ from $\cV_4 = v^3 v_f$ and $\vo =v_b^3$. Notice that these are the same relations valid in the trivial fibration case \cite{Green:1997di}. Moreover it can be easily shown that \eqref{lMls} holds also for compactifications to 5D and 7D, indicating that these are universal relations imposed by M/F-theory duality. It is also worth pointing out that the dimensionful volumes of any $2p$-cycles ($p$=1,2,3) of the base of the elliptic fibration (measured with the respective metric and fundamental scale) are the same on both sides of the duality:
\be
\left(\langle v\rangle \,\ell_M^2\right)^p = \left(\langle v_b\rangle \,\alpha'\right)^p\,,
\ee
which is easy to verify using \eqref{EM5ED3} and \eqref{lMls}.

To conclude, we observe that, due to \eqref{lMls}, the relation:
\be
\frac{G_4}{\ell_M^3}=\frac{\sqrt{\langle v_f\rangle}}{\ell_M^2}\left(F_3\wedge\frac{{\rm d}X^A}{\ell_M}+H_3\wedge\frac{{\rm d}X^B}{\ell_M}\right)
\label{G4G3}
\ee
between 4-form and 3-form fluxes is compatible with \eqref{G4F3H3}, provided that $X^{A,B}/\ell_M$ are angular variables normalised in such a way that $\ell_M^{-1}\int_{S^1_{A,B}} {\rm d}X^{A,B}=1$. Similarly, we have:
\be
\frac{C_4}{(\alpha')^2}=\frac{\langle r\rangle}{(\alpha')^{3/2}}\,C_3\wedge\frac{{\rm d}Y^B}{\sqrt{\alpha'}}\,,
\label{C4C3r}
\ee
which, again due to \eqref{lMls}, guarantees that \eqref{C3C4} is satisfied, provided that $Y^B/\sqrt{\alpha'}$ is the angular variable of the circle $\tilde{S}^1_B$ T-dual to the $B$-cycle, normalised in such a way that $(\alpha')^{-1/2}\int_{\tilde{S}^1_B} {\rm d}Y^B=1$.

A few comments are now in order. Even though we talked about 3D actions, the actual duality match concerns the \emph{Lagrangians} in 3D, which are quantities with the dimension of \emph{length}$^{-3}$. This is why we obtained a relation between the two fundamental scales $\ell_M$ and $\alpha'$. Internally, in contrast, we take the \emph{dimensionless} coordinates used to parametrise the base of the elliptic fibration to be the \emph{same} on the two sides of the duality. More precisely, we have:
\be
\frac{X^J_{\rm M}}{\ell_M} = \frac{X^J_{\rm IIB}}{\sqrt{\alpha'}}\,,\qquad J=1,\ldots, 6\,.
\label{DimlessCoord}
\ee
The differentials of these coordinates are used to expand the various forms in the respective contexts (e.g.~$G_4$ in terms of d$X^J_{\rm M}$, and $F_3$ and $H_3$ in terms of d$X^J_{\rm IIB}$). For this reason, \eqref{DimlessCoord} combined with \eqref{G4G3} allows us to derive the following formulas connecting the form coefficients:
\be
\ell_M\,(G_4)_{IJKA}=\sqrt{\alpha'}\,(F_3)_{IJK}\,,\qquad \ell_M\,(G_4)_{IJKB}=\sqrt{\alpha'}\,(H_3)_{IJK}\,.
\ee
These relations essentially state that dimensionless, metric-independent quantities are duality invariant, and this is particularly useful when estimating the volume behaviour of the terms in the 11D Lagrangian generating the low-energy scalar potential after compactification. In addition, using this observation, it is easy to prove \eqref{scalarpotential} when the classical scalar potentials are written in terms of the form coefficients. In the following we shall exploit this result to analyse perturbative corrections to the 4D EFT.

\subsection{$\alpha'$ corrections to the 4D scalar potential}
\label{sec:dimAnalysis}

\subsubsection*{General framework}

In this section we propose a scheme to argue for or against the existence of certain $\ap$ corrections in the 4D F-theory effective action. In particular, we shall focus on corrections to the tree-level flux potential due to 8-derivative terms in the 11D M-theory action.\footnote{Our analysis can also be applied to other terms in the 3D/4D action as corrections to the kinetic terms. However the formulae need to be adjusted to account for the Weyl rescaling of the 3D metric.} In full generality, the 3D action obtained from reducing M-theory on a fourfold $Y_4$ contains a scalar potential of the form:
\be
S_{\rm M}^{(3)}\;\supset\; \int\sqrt{-g^{(3)}}\,{\rm d}^3x\,\left(\frac{\langle \cV_4\rangle}{\cV_4}\right)^3\frac{1}{\ell_M^3}\; V^{\rm (M)}\,,
\label{eq:ThreeDimActionPotMTH} 
\ee
where:
\be
V^{\rm (M)}= V_{\rm tree}^{\rm (M)} + V_{\rm corr}^{\rm (M)} = \int_{Y_4} V^{\rm (M)}(G_4,\mc{R})\,\sqrt{g^{(8)}}\,{\rm d}^8x\,.
\ee
Up to $\ell_M^6$ order, we have (schematically):
\be
V^{\rm (M)} (G_4,\mc{R})\;=\;\frac{|G_4|^2}{\ell_M^6}\,+(\nabla G_4)^2 \mc{R}^2+G_4^2 \mc{R}^3+(\nabla G_4)^4 + G_4^4 \mc{R}^2 + G_4^6 \mc{R} + G_4^8\,.
\label{eq:MtheorySalarPot3DCorrected} 
\ee
Strictly speaking, these are only the 8-derivative couplings appearing in the CP-even sector. We ignore CP-odd terms as derived in \cite{Liu:2013dna} which do not contribute to $V^{\rm (M)}$.\footnote{These couplings might however become relevant once our scheme is applied to deriving corrections to other 4D quantities.} Recalling that 4D Poincar\'e invariance requires the 4-form flux to have exactly 1 leg along the fibre, and applying the results of Sec.~\ref{sec:Flimit}, the tree-level contribution in \eqref{eq:ThreeDimActionPotMTH} reads:
\be
\dfrac{\langle \cV_4\rangle^3}{\cV_4^3} \frac{1}{\ell_M^3} = M_p^4 \left(\frac{\langle r\rangle}{r}\right)^3\, r\sqrt{\alpha'}\, \frac{1}{\vo^2}\,,
\label{eq:TreeLevelScalingPotential3D} 
\ee
where $(\langle r\rangle/r)^3$ disappears by undoing the Weyl rescaling \eqref{eq:WeylRescaling3dmetricR} of the 3D metric. Furthermore $\vo^{-2}$ reproduces the correct volume scaling of the tree-level scalar potential in 4D Einstein frame. Finally $r\sqrt{\alpha'}$ generates the 4-th dimension upon taking the F-theory limit: 
$$
r\sqrt{\alpha'}\int_{\mathbb{R}^3} \sqrt{-\check{g}^{(3)}}  d^3x = \int_{\tilde{S}^1_B} \sqrt{g^{(1)}} dy\int_{\mathbb{R}^3} \sqrt{-\check{g}^{(3)}} d^3x  \qquad\xrightarrow{v_f\raw 0} \qquad \int_{\mathbb{R}^4}\sqrt{-g^{(4)}}d^4x\:,
$$
where $y$ is the dimensionful coordinate along the circle. This is just the opposite process to the one in Eq.~\eqref{3dIIB}.
Having isolated these prefactors, the terms in $V^{\rm (M)}(G_4,\mc{R})$ that contribute to the 4D scalar potential are those which are \emph{independent} on $v_f$. All in all, the F-theory limit results in:
\be
M_3^3 \int \frac{{\rm d}^3x}{ \cV_4^3}\,\int_{Y_4} V^{\rm (M)} (G_4,\mc{R})\,\sqrt{g^{(8)}}\,{\rm d}^8x\xrightarrow{v_f\raw 0} 
M_p^4 \int \frac{{\rm d}^4x}{\vo^2}\,\int_{B_3} V^{\rm (F)} (G_3,\mc{R})\,\sqrt{g^{(6)}}\,{\rm d}^6x\,.
\ee
The contribution to the scalar potential of the tree-level term in \eqref{eq:TreeLevelScalingPotential3D} is $v_f$-independent, therefore leading to \eqref{scalarpotential}. In what follows, we shall focus on metric contractions of the 8-derivative terms in \eqref{eq:MtheorySalarPot3DCorrected} and on their behaviour under the F-theory limit to derive the volume scaling of $\alpha'$ corrections to the 4D scalar potential. 

\subsubsection*{A metric ansatz for elliptically fibred CY fourfolds}
\label{sec:MetricAnsatzCYFEFG} 

For the subsequent dimensional analysis and in contrast to \cite{Conlon:2005ki}, we require an ansatz for the internal metric in order to distinguish between the scaling with respect to fibre and base volume. The former determines the behaviour of a given metric contraction in the F-theory limit $v_f\raw 0$, whereas the latter specifies the $\vo$ dependence of the corresponding correction in $V^{\rm (F)}$. In this sense, the upcoming analysis goes beyond the type IIB arguments of \cite{Conlon:2005ki}, by identifying all relevant M-theory structures responsible for $\ap$ effects in F-theory.\footnote{Here we focus solely on the zero-mode KK reduction of the M-theory action which, as we argue below, is not sufficient to generate all $\ap$ effects in F-theory compactifications.}

To start with, we recall that $Y_4$ is a K\"ahler manifold and both base and fibre are K\"ahler submanifolds of $Y_4$. For this reason, the various metric components are obtained from a K\"ahler potential which can be split into two pieces. We denote local complex coordinates on $Y_4$ as $Z^A$ with $A=1,\ldots,4$, and divide them as fibre coordinates $\zeta^a$ with $a=1$, and base coordinates $z^\alpha$ with $\alpha=1,2,3$. Then the K\"ahler potential on $Y_4$ reads:
\be
K(Z,\bar{Z})= K^f(\zeta,\bar{\zeta},z,\bar{z})+ K^b(z,\bar{z})\,,
\label{eq:KPAnsatzG} 
\ee
where $K^b$ and $K^f$ are respectively the K\"ahler potential of the base and the fibre, and the non-triviality of the fibration is encoded in the dependence of $K^f$ on $z$ and $\bar{z}$.

In the following, we assume that all metric components scale with \emph{integer} powers of $v_f$ and $v$, resulting in:\footnote{We use the fact that $v_f$ does not vary as we move over the base since $J$ is closed in a K\"ahler manifold. According to \cite{Grimm:2014xva}, however, deviations from K\"ahlerity are possible due to backreaction effects, starting at order $\ell_M^9$. The present analysis is purely classical and does not take into account such effects.}
\be
K^f(\zeta,\bar{\zeta},z,\bar{z})=v_f\, k^f(\zeta,\bar{\zeta},z,\bar{z})\,,\qquad K^b(z,\bar{z})=v\,  k^b(z,\bar{z})\,,
\ee
where $k^f$ and $k^b$ are two scale-independent functions. Our assumption is justified because the K\"ahler form can be expanded as $J=v_f\,\omega_f+v\,\omega$, where $\omega_f$ and $\omega$ are the harmonic (1,1)-forms Poincar\'e dual to the horizontal and vertical divisor respectively. In other words, there are no divisors wrapping only a 1-cycle of the fibre. Let us denote the metric components as:
\begin{align}
\label{eq:MetricSplitFourfold} 
g_{A\bar{B}}=\left (\begin{array}{cc}
g_{a\bar{b}} & g_{a\bar{\beta}} \\ 
g_{\alpha\bar{b}} &g_{\alpha\bar{\beta}}
\end{array} \right ), \qquad 
g_{\alpha\bar{\beta}}= K^f_{\alpha\bar{\beta}}+ K^b_{\alpha\bar{\beta}}\,,\qquad g_{a\bar{b}}= K^f_{a\bar{b}}\,,\qquad g_{a\bar{\beta}}
=K^f_{a\bar{\beta}}\,.
\end{align}
Given that in the F-theory limit $K^f_{a\bar{b}}\sim K^f_{a\bar{\beta}}\sim K^f_{\alpha\bar{\beta}}\sim v_f$ and $K^b_{\alpha\bar{\beta}}\sim v$, the components of the metric and its inverse scale as:
\bea
g_{\alpha\bar{\beta}}&\sim& v_f+v\, \qquad g_{a\bar{b}}\sim v_f\, \qquad g_{a\bar{\beta}}\sim v_f\,, \nn \\
g^{\alpha\bar{\beta}}&\sim& \frac{1}{v_f+v}\,\qquad g^{a\bar{b}}\sim \frac{1}{v_f}+\frac{1}{v_f+v}\,,\qquad  g^{a\bar{\beta}}\sim \frac{1}{v_f+v}\,,
\label{eq:InverseMetricScaling} 
\eea
implying:
\be
\cV_4= \sqrt{\det(g)}\sim v^3 v_f \left(1+\frac{v_f}{v}+\frac{v_f^2}{v^2}+\frac{v_f^3}{v^3}\right)\,.
\label{eq:ExpansionFourfoldVol} 
\ee
This result may also be obtained directly from \eqref{eq:DefFourfoldVol} using $\omega_f^2=-\omega_f \wedge c_1(B_3)$, with $c_1(B_3)$ the first Chern class of the base. For the same reason, \eqref{EM5ED3} gets corrected as:
\be
v_b^2 \sim v_f \left(v+v_f\right)^2\,,
\ee
which implies the following useful formula:
\be
v^3 \sim \frac{\vo}{v_f^{3/2}} + \vo^{2/3} + \vo^{1/3} v_f^{3/2} + v_f^3\,.
\label{Useful}
\ee
Next we compute the connection coefficients and the Riemann tensor. Up to symmetries and complex conjugation, the non-vanishing components on a generic K\"ahler manifold are:
\be
\Gamma^A_{BC}=g^{A\bar{D}}\p_{B}g_{C\bar{D}}\qquad\text{and}\qquad R^A\,_{B\bar{C}D}=\p_{\bar{C}} \Gamma^A_{DB} \,.
\ee
In App.~\ref{app:TensorsEllFibGen} we give the details of the various components after the fibre/base split $Z^A\raw (\zeta^a,z^\alpha)$. Concentrating only on the parametric volume dependence, we use \eqref{eq:MetricSplitFourfold} and \eqref{eq:InverseMetricScaling} to find:
\be
\Gamma^a_{bc},\Gamma^a_{\mu b},\Gamma^a_{\alpha\beta},\Gamma^\alpha_{bc},\Gamma^\alpha_{\beta\gamma}\sim\cO(1)+\dfrac{v_f}{v} +\ldots \,,\qquad 
\Gamma^\alpha_{\gamma b}\sim{ \dfrac{v_f}{v}} +\ldots\,,
\label{eq:VolScalGamma}
\ee
where $\ldots$ encodes additional terms of higher order in $v_f/v$. Similarly, the non-vanishing components of the curvature tensor satisfy:
\begin{align}
\label{eq:VolScalRiemannT} 
&R^\alpha\,_{\beta\bar{\gamma}\delta},R^a\,_{\beta\bar{\gamma}\delta},R^a\,_{\beta\bar{c}\delta},R^{a}\,_{b\bar{\gamma}\delta},R^{a}\,_{\beta\bar{\gamma} d},R^a\,_{b\bar{c}\delta},R^{a}\,_{b\bar{\gamma} d},R^a\,_{b \bar{c}d}\sim\cO(1)+{ \dfrac{v_f}{v}} +\ldots\, ,\nn\\
& R^\alpha\,_{\beta\bar{c}\delta},R^{\alpha}\,_{\beta\bar{\gamma}d},R^\alpha\,_{a\bar{\gamma}\delta},R^{\alpha}\,_{a\bar{\gamma}d},R^{\alpha}\,_{a\bar{c}\delta}, R^\alpha\,_{a\bar{c}d}  \sim{ \dfrac{v_f}{v}}+\ldots \, .
\end{align}

\subsection{No $(\alpha')^{\rm odd}$ terms from dimensional reduction}
\label{sec:ScalingAnalysisClas} 

\begin{table}[t!]
\centering
{\small
\begin{tabular}{|c|c|}
\hline 
&\\[-1.em]
Parameter & Specification \\ [0.2em]
\hline 
\hline 
&\\[-1.em]
$\lambda_f$ & net number of \textbf{inverse fibre} metrics $g^{a\bar{b}}$ \\ [0.2em]
\hline 
&\\[-1.em]
$\lambda_b$ & net number of \textbf{inverse base} metrics $g^{\alpha\bar{\beta}}$ \\ [0.2em]
\hline 
&\\[-1.em]
$\lambda_{\rm mix}$ & net number of \textbf{inverse mixed} metrics $g^{a\bar{\beta}}$ \\ [0.2em]
\hline 
\hline
&\\[-1.em]
$\lambda$ & net number of inverse metrics \\ [0.2em]
\hline 
&\\[-1.em]
$x$ & number of tensors with \textbf{non-trivial} scaling \\ [0.2em]
\hline 
&\\[-1.em]
$\lambda_{\rm crit}$ & critical value of $\lambda_f$ for \textbf{finite} results as $v_f\raw 0$  \\ [0.2em]
\hline 
\end{tabular}
}
\caption{Summary of parameters used in the dimensional analysis.}
\label{tab:NotationScalingAnalysis} 
\end{table}

We are now ready to perform our scaling analysis. The idea is to look at all 8-derivative Lorentz-invariant contractions allowed to appear in \eqref{eq:MtheorySalarPot3DCorrected}. We will actually be more general and, analogously to (\ref{ScorrEinst}) for 10D type IIB string theory but ignoring dilaton factors, we schematically denote any 11D higher derivative term as:
\be
K^{PRL} \propto \Bigl(  g^{\circ \circ} R^\circ_{\circ \circ\circ} \Bigr)^{P+1}  \Bigl[ (g^{\circ \circ})^4\, G_{\circ\circ\circ\circ} G_{\circ\circ\circ\circ} \Bigr]^R  \Bigl[ (g^{\circ \circ})^5\,  \nabla_{\circ}G_{\circ\circ\circ\circ} \nabla_{\circ}G_{\circ\circ\circ\circ} \Bigr]^L\,, 
\label{eq:KNML} 
\ee
with all indices taken along internal directions. As for the type IIB case, given that we are interested just in scaling considerations, we can set $L=0$ without loss of generality since each power of $(\nabla G_4)^2$ scales as $\mc{R} G_4^2$. We now play a multi-parameter game counting all possible contractions of \eqref{eq:KNML}, using the various parameters summarised in Tab.~\ref{tab:NotationScalingAnalysis}. We want to build contractions using all possible metric components. The total number of inverse metrics $\lambda$ satisfies:
\be
\lambda=4 R + P +1 + 5 L =\lambda_{\rm mix}+\lambda_f+\lambda_b \,,
\label{lambda}
\ee
which we use to eliminate $\lambda_b$ in favour of the other parameters. Contrary to the type IIB discussion in Sec.~\ref{TypeIIBscalings}, we have to take into account the possible non-trivial scaling of the various connection coefficients and Riemann tensor components as in \eqref{eq:VolScalGamma} and \eqref{eq:VolScalRiemannT}. We count these scaling factors with an additional parameter $x$.

A standard KK reduction of the various 8D contractions results in a scaling of the 4D scalar potential $V$ with respect to base and fibre volume given by:
\be
V \sim \frac{M_p^4}{\vo^2}\,\int_{Y_4}\, d^8 x \sqrt{-g^{(8)}}\,K^{PRL}  \sim \frac{M_p^4}{\vo^2}\,\underbrace{(g^{a\bar{\beta}})^{\lambda_{\rm mix}} 
(g^{\alpha\bar{\beta}})^{\lambda-\lambda_{\rm mix}-\lambda_f} (g^{a\bar{b}})^{\lambda_f}}_{\text{metric contractions}} 
\underbrace{\cV_4}_{\int_{Y_4}}\, \underbrace{{ \left (\dfrac{v_f^{3/2}}{\vo^{1/3}} \right )^x}}_{R,\,\Gamma}\,. \nn
\ee
Using (\ref{eq:InverseMetricScaling}), (\ref{eq:ExpansionFourfoldVol}) and (\ref{Useful}), we end up with (setting $M_p=1$):
\be
V \sim \frac{v_f^{\frac12(\lambda+3(x-\lambda_f)-1)}}{\vo^{1+\frac13 (\lambda-\lambda_f+x)}} \left (1+\dfrac{v_f^{3/2}}{\vo^{1/3}}+\ldots \right),
\label{eq:ClassicalScalingAnalysisContractions} 
\ee
where the bracket encodes an expansion in powers of $v_f^{3/2}/\vo^{1/3}$ and $\lambda_{\rm mix}$ drops out since $g^{a\bar{\beta}}\sim g^{\alpha\bar{\beta}}$. Naively from \eqref{eq:ClassicalScalingAnalysisContractions} one may be worried that the reduction of the general 11D term \eqref{eq:KNML} might give rise to divergent contributions in the F-theory limit $v_f\raw 0$. However our scaling analysis does not allow us to determine the coefficients of $\alpha'$ corrections to the 4D scalar potential arising from 11D higher derivative terms. We therefore assume all apparently divergent terms in \eqref{eq:ClassicalScalingAnalysisContractions} come along with vanishing coefficients so that M/F-theory duality holds at all orders in $\ap$. Below we provide concrete evidence for this assumption. Let us stress that the only input in \eqref{eq:KNML} is parity invariance which constrains all terms to be even in powers of $G_4$. Moreover only particular kinematic structures are expected to appear in the 11D action which is however not fully known yet, even at the 8-derivative level. The special nature of the compact geometry also plays a crucial r\^ole in determining what terms survive after reduction. 

For the above reasons, the terms in \eqref{eq:ClassicalScalingAnalysisContractions} that are amenable to give non-trivial contributions to the 4D scalar potential are those independent of $v_f$. This allows us to deduce a \emph{critical} value for the number of inverse fibre metrics $\lambda_f$. If a finite term in the F-theory limit arises at order $o$ in the expansion in $v_f^{3/2}/\vo^{1/3}$, such a critical value is:
\be
\lambda_{\rm crit}=\dfrac{\lambda-1}{3}+x+o\,.
\label{eq:MCritClassicalScaling} 
\ee
This relation implies that $(\lambda-1)$ must be a multiple of 3 in order to have $\lambda_{\rm crit}\in\mathbb{N}$,\footnote{Here we take $\mathbb{N}=\{0,1,2,...\}$.} i.e.~$\lambda = 1+3\mathfrak{n}$ with $\mathfrak{n}\in\mathbb{N}$. Given that the $\ell_M^l$ order of the generic term (\ref{eq:KNML}) is counted by $l = 2(P+R+ 2L)$, using (\ref{lambda}) we can easily infer $l = 6 (\mathfrak{n}-R-L)$ which implies that higher derivative corrections in M-theory should appear only at order $\ell_M^{6\mathfrak{p}}$ with $\mathfrak{p}\in\mathbb{N}$ assuming they contribute in the F-theory limit. Remarkably, this is exactly what follows from the general structure of M-theory higher derivative couplings conjectured by \cite{Damour:2005zb}. 

If we now plug (\ref{eq:MCritClassicalScaling}) back into \eqref{eq:ClassicalScalingAnalysisContractions}, we obtain:
\begin{align}
V  & \xrightarrow{v_f \raw 0} \begin{cases}
0 & \lambda_f<\lambda_{\rm crit}\\
\vo^{-\frac23 (\mathfrak{n}+2)}& \lambda_f=\lambda_{\rm crit} \qquad \mathfrak{n}\in\mathbb{N} \\
\infty & \lambda_f>\lambda_{\rm crit}
\end{cases}
\label{eq:ScalingFTlimit} 
\end{align}
Analogously to the type IIB result (\ref{GenVcorrDimRed}), corrections to the 4D scalar potential do not depend on the detailed choice of contractions, but only on the net number of inverse metric factors $\lambda=1+3\mathfrak{n}$ which is fixed for a given 11D term. Combining (\ref{eq:ScalingFTlimit}) with (\ref{Important}) for $q = \frac23 (\mathfrak{n}+2)$, we realise that the $(\alpha')^m$ order is:
\be
m = 2\left(\mathfrak{n}-r\right)= \frac23 \left(P + R + 5L\right), 
\label{m}
\ee
where we have used (\ref{lambda}) and we have set $r=R$ since the order $F^{2r}$ or $D^{2r}$ of the F- or D-term expansion of the 4D EFT is counted by the number of $G_4$ powers in 11D, given that F-term contributions arise when $G_4$ reduces to $G_3$ while D-term effects emerge when $G_4$ reduces to $\mc{F}_2$. The result (\ref{m}) shows clearly that the classical KK reduction of the M-theory action can only give rise to $(\alpha')^{\rm even}$ corrections to the 4D scalar potential at different F- or D-term order. This is in agreement with \cite{Grimm:2013gma,Grimm:2013bha,Grimm:2017pid,Weissenbacher:2019mef}. Notice the crucial factor of $2/3$ in (\ref{m}) which implies an important difference in the counting of the $(\alpha')^m$ order in comparison with the type IIB analysis performed in Sec.~\ref{TypeIIBscalings} (setting $L=0$):
\be
\text{Type IIB:}\quad m = p + r \qquad\qquad\text{F/M-theory:}\quad m = \frac23 \left(P+R\right)\,,
\label{alphaPcomparison}
\ee
where the only allowed values of $P$ and $R$ are those such that satisfy $4 R +P =3\mathfrak{n}$ with $\mathfrak{n}\in\mathbb{N}$.

Primary examples of contributions to the $\alpha'$ expansion of the 4D scalar potential from the classical KK reduction of generic 11D terms are: ($i$) the tree-level 11D term $G_4^2$ ($P=-1$, $R=1$ and $L=0$) which, according to (\ref{lambda}) and (\ref{m}), gives $\lambda=4$, $m = 0$ and $V\sim \vo^{-2}$, that corresponds to either the classical flux potential at order $F^2$ or to tree-level moduli-dependent FI terms at order $D^2$; ($ii$) the $\ell_M^6$ 11D term $\mc{R}^4$ ($P=3$, $R=0$ and $L=0$) which would yield $\lambda=4$, $m = 2$ and $V\sim \vo^{-2}$, and so a potential $(\alpha')^2$ correction which however would not contribute to the 4D scalar potential since it is cancelled by the self-dual part of the flux kinetic term  \cite{Haack:2001jz,Grimm:2013gma}; ($iii$) the $\ell_M^6$ 11D term $\mc{R}^3 G_4^2$ ($P=2$, $R=1$ and $L=0$) which gives $\lambda = 7$, $m = 2$ and $V\sim \vo^{-8/3}$ corresponding to $(\alpha')^2$ corrections at $\mc{O}(F^2)$ (or potential $(\alpha')^2$ corrections to FI-terms), in agreement with explicit reductions performed in \cite{Grimm:2013bha,Grimm:2017pid,Weissenbacher:2019mef}. Whether these $(\alpha')^2$ effects correct the scalar potential or give rise just to moduli redefinitions is still an open issue. Interestingly (\ref{alphaPcomparison}) implies that the 10D higher derivative term $\mc{R}^3 G_3^2$ is not naively related to the corresponding 11D $\mc{R}^3 G_4^2$ term by a classical reduction since the first corresponds to $(\alpha')^3$ effects while the second would generate $(\alpha')^2$ corrections. Results for different 11D terms are summarised in Tab.~\ref{tab:ClassicalKK}.

\begin{table}[t!]
\centering
\begin{tabular}{|c|c|c|c|c|c|c|c|}
\hline 
&  &  &  &  & & &   \\ [-1.em]
$\ell_M^l$ & $P$ & $R$ & 11D term & $(\lambda-1)$  & $\lambda_{\rm crit}$ & $(\alpha')^m$ &  $V(\vo)$ \\ 
\hline 
\hline 
&  &  &  &  & &  &  \\ [-0.6em]
$\ell_M^0$ & -1 & 1 & $G_4^2$ & 3  &$1$ & 0 & $\vo^{-2}$ \\[0.5em]
\hline 
&  &  &  & & &   &  \\ [-0.8em]
$\ell_M^6$ & 2 & 1 & $\mc{R}^3 G_4^2$ & 6 &  $2$ & 2 & $\vo^{-8/3}$ \\ [0.8em]
\hline  
&  &  &  & & &   &  \\ [-0.8em]
$\ell_M^6$ & 1 & 2 & $\mc{R}^2 G_4^4$ & 9 & $3$ & 2 & $\vo^{-10/3}$ \\ [0.8em]
\hline 
&  &  &  & & & &    \\ [-0.6em]
$\ell_M^6$ & 0 & 3 & $\mc{R} G_4^6$ & 12  & $4$ & 2 & $\vo^{-4}$ \\ [0.5em]
\hline 
&  &  &  & & & &   \\ [-0.8em]
$\ell_M^6$ & -1 & 4 & $G_4^8$ & 15 & $5$ & 2 & $\vo^{-14/3}$ \\ [0.8em]
\hline 
&  &  &  & & &  &   \\ [-0.8em]
$\ell_M^{12}$ & 5 & 1 & $\mc{R}^6 G_4^2$ & 9 & $3$ & 4 & $\vo^{-10/3}$ \\ [0.8em]
\hline 
\end{tabular} 
\caption{Summary of results for some $(\alpha')^{\rm even}$ corrections to the 4D potential at different F- and D-term order from classical reduction of higher derivative M-theory terms in the F-theory limit. For simplicity, we set $x=o=0$ in $\lambda_{\text{crit}}$ because the volume scaling in $V(\vo)$ is independent on both.}
\label{tab:ClassicalKK} 
\end{table}

Let us now comment on our metric ansatz \eqref{eq:KPAnsatzG}. We worked with a general $K^f$ which is not necessarily flat.
One might have assumed instead an ansatz for $K^f$ like:\footnote{This proposal has been used originally in \cite{Greene:1989ya} to describe elliptic K3 manifolds.} 
\be
K^f(\zeta,z)=- \frac{v_f}{2\,{\rm Im}(\tau(z))}\, (\zeta-\bar{\zeta})^2\,,
\label{eq:KPAnsatzGSVY} 
\ee
which would yield a so-called semi-flat fourfold metric which is flat when restricted to the fibre \cite{2000math8018G}. In particular, since $K^f$ is only quadratic in $\zeta$, it satisfies $\p_c g_{a\bar{b}}=0$. Looking at the expressions listed in App.~\ref{app:TensorsEllFibGen}, this implies that $R^\alpha\,_{a\bar{c}d},R^a\,_{b \bar{c}e},R^a\,_{b\bar{c}\lambda},R^a\,_{b\bar{\gamma} e}$ are suppressed by an additional factor of $v_f/v$. Combined with \eqref{eq:VolScalRiemannT}, this suggests that components of $R^a\,_{\circ\circ\circ}$ and $R^\alpha\,_{\circ\circ\circ}$ with more than one fibre index downstairs scale with a positive power of $v_f$. Ultimately, restricting to the ansatz \eqref{eq:KPAnsatzGSVY} causes all corrections to the 3D scalar potential to vanish in the F-theory limit. However (\ref{eq:KPAnsatzGSVY}) is the correct expression for $K^f$ only away from singular fibres. Thus our analysis proves that the classical reduction of 11D terms captures only effects due to 7-branes in F-theory compactifications of M-theory. This is consistent with the discussion in Sec.~\ref{AbsenceOfAPOne} where we argued that $(\alpha')^{\rm even}$ corrections are induced just by higher derivative couplings on D7-brane worldvolumes. On the other hand, the type IIB closed string degrees of freedom are not captured in classical reductions since they would generate $(\alpha')^{\rm odd}$ corrections to the 4D scalar potential. This raises the obvious question how the well-known $(\alpha')^3$ effects in type IIB CY threefold compactifications \cite{Becker:2002nn} and orientifold generalisations thereof \cite{Minasian:2015bxa} can actually be recovered from F/M-theory duality. We will discuss this issue in more detail in Sec.~\ref{sec:QuantumReductions}.

Let us conclude with commenting on the limitations of our procedure. Here we are only able to predict the $\alpha'$ order of a given correction, and not whether this correction actually appears in the 4D EFT or not. Clearly, some of these terms could be washed away by applying field redefinitions \cite{Grimm:2013bha,Junghans:2014zla}. Finally, given that our analysis is only classical, we are unable to account for possible loop effects in type IIB \cite{Conlon:2009kt,Conlon:2010ji} and F-theory \cite{Weissenbacher:2019mef} which could generate $\ln\vo$-type $(\alpha')^2$ corrections in the 4D scalar potential.\footnote{Similar types of corrections at loop-level at order $(\alpha')^3$ have been proposed in \cite{Antoniadis:1998ax,Antoniadis:2018hqy,Antoniadis:2019rkh}.}

\subsubsection*{Absence of divergences in known kinematic structures}
\label{NoDiv}

In the previous analysis we assumed that all terms that naively would diverge in the $v_f\raw 0$ limit, are actually multiplied by vanishing coefficients which our analysis is insensitive to. This is essentially the requirement that the M/F-theory duality makes sense beyond tree-level in $\alpha'$. In this section we give some evidence in this direction, following a logic that works for any K\"ahler metric in the compact space.

The simplest higher derivative tensor structure is $\mc{R}^4$ which separates into two pieces \cite{Green:1997di,Green:1997as,Kiritsis:1997em,Russo:1997mk,Antoniadis:1997eg,Tseytlin:2000sf}:
\be
S_{\mc{R}^4}=\frac{1}{\ell_M^3} \int\left (J_0-\frac12 E_8\right )\star_{11}1\,,
\label{eq:ActionR4} 
\ee
where schematically (for definitions and conventions see App.~\ref{app:TensorsEllFibGen}):
\be
J_0 = t_8 t_8 \mc{R}^4+\frac14 E_8\qquad\text{with}\qquad E_8=\dfrac{1}{3!}\epsilon_{11}\epsilon_{11} \mc{R}^4\,.
\ee
We now argue that, even though our analysis predicts divergent terms as $v_f\raw 0$, they cancel among each other in the 3D scalar potential (effectively due to K\"ahlerity of $Y_4$). First of all $J_0$ can be expressed in terms of the Weyl tensor $C^{M}\,_{NPQ}$ as \cite{Banks:1998nr,Gubser:1998nz}:
\be
J_0 = C^U\,_{MRV}C_{USQ}\,^V\left [C^{MNPQ}C^R\,_{NP}\,^S+ \frac12 C^{MQPN}C^{RS}\,_{PN}\right].
\ee
We decompose $J_0$ into an internal and an external part. As already noted in \cite{Haack:2001jz}, the external part vanishes because $C^M\,_{NPQ}=0$ in 3D.
Furthermore the following integral vanishes for Ricci-flat K\"ahler manifolds (see App.~\ref{app:TensorsEllFibGen} for details) \cite{Gross:1986iv,Freeman:1986zh}:
\be
\int_{Y_4}\, J_0 \star_8 1=0\,.
\label{J0vanishing}
\ee
Nonetheless, within $J_0$ there are contractions of the form:
\be
R_{a\bar{a}\bar{a}a} R_{\bar{a}aa}\,^a R^{\bar{a}}\,_{\bar{a}}\,^a\,_a R^{a\bar{a}\bar{a}a}\sim (g_{a\bar{a}})^2\, (g^{a\bar{a}})^5 (R^{\bar{a}}\,_{\bar{a}a\bar{a}})^2 (R^a\,_{a\bar{a}a})^2\,,
\ee
which would clearly be divergent from (\ref{eq:ScalingFTlimit}) since $\lambda_f = 3 > 1 = \lambda_{\rm crit}$. However the full kinematics proves that this term must be multiplied by a vanishing coefficient. Therefore from (\ref{eq:ActionR4}) we realise that the only contributions to the 3D potential are those associated with the 8D Euler density $E_8$. Putting all legs along the internal directions, we recover \cite{Haack:2001jz,Grimm:2013gma}:
\be
\frac14 E_8(Y_4)\star_8 = 1536\, c_4(Y_4)\,,
\ee
in terms of the $4$th Chern class defined in \eqref{eq:FourthCC}. By definition this quantity is topological and, in particular, finite in the $v_f \raw 0$ limit. In fact, it contributes to the M2-tadpole thereby cancelling the self-dual part of $G_4$ in the tree-level scalar potential~\eqref{VM}~\cite{Duff:1995wd,Becker:1996gj,Sethi:1996es,Dasgupta:1996yh,Haack:2001jz}.

We next investigate the higher derivative term $\mc{R}^3 G_4^2$ whose 11D kinematics was determined in \cite{Hyakutake:2007sm,Liu:2013dna} as (see App.~\ref{app:TensorsEllFibGen} for definitions and conventions):
\be
S_{\mc{R}^3 G_4^2}=\frac{1}{\ell_M^3}\int\, \left (t_8 t_8+\frac{1}{96}\epsilon_{11}\epsilon_{11}\right) \mc{R}^3 G_4^2 \star_{11}1\,.
\label{eq:HDActionG2R3} 
\ee
As before, we compactify both terms on $Y_4$ and argue that there are no divergent terms stemming from \eqref{eq:HDActionG2R3}. The term $\epsilon_{11}\epsilon_{11} \mc{R}^3 G_4^2$ does not contribute to the 3D scalar potential $V^{\rm (M)}$ for dimensional reasons: $\epsilon_D$ vanishes identically when putting all indices along $d<D$ directions. This suggests that only terms within $t_8 t_8 G_4^2 \mc{R}^3$ are potentially dangerous in the F-theory limit. In App.~\ref{app:TensorsEllFibGen} we managed to show that a cancellation of divergent terms does occur in a subset of terms within this kinematic structure. Full absence of divergences is achieved through additional assumptions about the metric ansatz.
For instance, it turns out that imposing:\footnote{This condition is trivially satisfied for the ansatz \eqref{eq:KPAnsatzGSVY}.}
\be
\label{eq:MetCond} 
R^{a}\,_{a\bar{a} a}=0+\cO\left (\dfrac{v_{f}}{v}\right )\quad\Rightarrow\quad \p_{\bar{a}} \p_{a}g_{a\bar{a}}=g^{a\bar{a}}\left (\p_{\bar{a}}g_{\bar{a}a}\right )\left (\p_{a}g_{a\bar{a}}\right ),
\ee
guarantees that there are no divergent contractions stemming from $\mc{R}^3 G_4^2$ or $\mc{R}^2 (\nabla G_4)^2$, although there remain dangerous terms in e.g.~$\mc{R}^4$ and $\mc{R}^2 G_4^4$. While \eqref{eq:MetCond} remains a conjecture, we stress again that the absence of divergences should really hold true due to M/F-theory duality.

\section{$\alpha'$ corrections from 11D loops}
\label{sec:QuantumReductions} 

In the previous section we found convincing evidence that classical KK compactifications on smooth elliptic fourfolds of the $\ell_M$-corrected 11D supergravity action can lead, upon F-theory limit, to only $(\alpha')^{\rm even}$ corrections to the 4D scalar potential. This procedure restricts however to zero-modes only, ignoring loops of non-zero KK and winding modes. This implies that $(\alpha')^{\rm odd}$ effects have to emerge from 11D loops, potentially together with additional $(\alpha')^{\rm even}$ corrections. 

This agrees with the findings of \cite{Green:1997as,Green:1997me,Green:1999by} where it has been shown that closed string degrees of freedom at higher order in the $\alpha'$ expansion are encoded in non-zero winding states on the $T^2$. For instance, the well-known coefficient of the type IIB $\mc{R}^4$ coupling is invisible to a classical $T^2$ reduction of M-theory, while it can be derived by a 1-loop calculation in the 11D superparticle formalism \cite{Green:1997as}.\footnote{Specifically, this is a Schwinger-type computation based on a string-inspired formalism \cite{Schubert:2001he} applied to the 11D superparticle \cite{Green:1997as}. A famous example is the \emph{Brink-Schwarz superparticle} \cite{Brink:1981nb,Brink:1981rt} as a zero-mode approximation of the Green-Schwarz superstring \cite{Green:1983wt,Green:1983sg}. Rather than using covariant quantisation in the pure spinor formulation \cite{Berkovits:2002uc,Anguelova:2004pg}, the calculus is based on light cone quantisation \cite{Green:1999by}. The framework of a Brink-Schwarz-like superparticle was shown to be equivalent to the 11D pure spinor formalism in \cite{Guillen:2017mte} (see also \cite{Berkovits:2019szu,Guillen:2020mmd} for discussions).} A direct comparison to string amplitudes in \cite{Russo:1997mk} led to the observation that such 1-loop amplitudes in 11D supergravity contain complete information about both perturbative and non-perturbative corrections in $g_s$. This derivation takes into account the interplay between field theoretic loop effects and stringy winding modes when compactifying to lower dimensions, as opposed to the standard classical procedure of simply ignoring non-zero KK and winding states.

We will argue that generalising the computation of 11D loops to elliptic CY fourfold compactifications of M-theory is a way to recover $(\alpha')^{\rm odd}$ corrections in 4D upon F-theory limit. Performing this computation explicitly is a difficult task since deriving corrections to the 4D scalar potential would require to investigate purely internal 1-loop amplitudes in a non-trivially curved background.\footnote{See \cite{Schubert:2001he,Green:2016tfs} for brief discussions of backgrounds more intricate than $T^2$.} As usual, this brings along all sorts of complications such as the factor ordering in the Hamiltonian. In addition, the torus is now the elliptic fibre over a base manifold, and so $\tau$ becomes a monodromic function of the base coordinates. A complete evaluation of the amplitude is therefore beyond the scope of this paper.

Nevertheless determining the volume dependence of the 4D scalar potential is in principle possible. For simplicity we shall focus on 8-derivative terms which arise at 1-loop level for the simple case of trivial fibrations where all classical contributions vanish in the F-theory limit, according to our findings in Sec.~\ref{sec:ScalingAnalysisClas}. Recall that the fibre volume is constant as a function of the base since the fibre itself is a K\"ahler submanifold of $Y_4$ (ignoring backreaction effects violating K\"ahlerity) \cite{Denef:2008wq}. From the 3D EFT perspective, $Q$-point effective vertices of the form \eqref{eq:KNML} with $Q=P+1+2(R+L)$ read schematically:
\be
\Gamma^{(Q)}(\mc{R}, G_4,\nabla G_4) = K^{PRL} \left( C + F(\tau,\bar{\tau})\, v_f^{Q-\frac12(11+p_f)}\right)\;+\;\cdots, 
\label{eq:EffVerticesFrom11DOneLoop} 
\ee
where $Q=4+R$ at the 8-derivative level. The first term is the contribution of zero-modes associated with the classical KK reduction, and so generates the $\alpha'$ corrections discussed in Sec.~\ref{sec:ScalingAnalysisClas}. In the decompactification limit $\cV_4\raw \infty$, it leads back to the 11D M-theory action.\footnote{As explained in \cite{Green:1997as}, the numerical coefficient of this term cannot be determined by the 11D loop amplitude, and must be fixed by the UV completion of the theory, which is M-theory itself.} The non-zero modes are instead encoded in the second term in (\ref{eq:EffVerticesFrom11DOneLoop}) where the fibre volume dependence is partially due to the integral over Schwinger time. Moreover $p_f=2R$ is the number of KK momenta along the fibre appearing in the vertex operators, while the functions $F(\tau,\bar{\tau})$ are generalisations of the typical Eisenstein series appearing in type IIB.\footnote{Formula (3.7) in \cite{Minasian:2015bxa} should contain one such generalisation, implicitly given as an integral over the base of the elliptic fibration. In that context, it pops up through a different (12D-inspired) derivation of $(\alpha')^3$ corrections of the 4D EFT which makes no use of M-theory.} 

The dots in formula \eqref{eq:EffVerticesFrom11DOneLoop} mean that in non-trivial fibrations we do not expect such amplitudes to take a factorised form, like the term with $K^{PRL}$ multiplying the factor in brackets (which is instead the full answer for trivial fibrations). Such a factorised structure would also imply that in principle there might be divergent terms from contractions.\footnote{The factorised structure arises from tracing over fermionic zero modes which is independent of the winding sector in $T^2$ compactifications. Since we are considering here \emph{internal} contributions in 8D rather than \emph{external} ones in 9D as in \cite{Green:1997as}, we expect this problem to be alleviated once we examine in more detail the fermion and 11D vertex operators under their decomposition in $\mathrm{SO}(1,10)\raw \mathrm{SO}(1,2)\times \mathrm{SO}(8)$.} This is because we potentially multiply by additional negative powers of $v_f$ at the non-zero winding level. Similarly to Sec.~\ref{sec:ScalingAnalysisClas}, we therefore assume that the coefficient of these terms has to vanish. Moreover the base does not admit non-trivial 1-cycles, and so there is no obvious counterpart for multiple windings of the superparticle worldline around internal directions.

Still, nothing prevents us from applying the techniques to trivial fibrations. The procedure then becomes a 2-stage process where we initially compute the 1-loop amplitude for compactifications on a $T^2$, and subsequently reduce the 9D result on a CY threefold before taking the $v_f\raw 0$ limit. By the duality arguments presented in Fig.~\ref{fig:MFIIB}, the two scenarios:
\be
11D \raw 9D \xrightarrow{{\rm Vol}(T^2) \raw 0} 10D \raw 4D \qquad\text{and}\qquad 11D \raw 9D \raw 3D \xrightarrow{v_f\raw 0} 4D\,,
\ee
are of course equivalent for trivial fibrations. Proceeding as with the classical reduction in Sec.~\ref{sec:ScalingAnalysisClas}, we identify the volume scaling of a generic higher derivative correction to the 4D scalar potential at the non-zero winding level as (setting again $M_p=1$):
\be
V\sim \frac{v_f^{Q-\frac12(11+p_f)}}{\vo^2} \int_{Y_4}\,d^8x\,\sqrt{-g^{(8)}} K^{PRL} 
\sim \frac{v_f^{\frac12(\lambda+3(x-\lambda_f)-12-p_f)+Q}}{\vo^{1+\frac13 (\lambda-\lambda_f+x)}} \left (1+\dfrac{v_f^{3/2}}{\vo^{1/3}}+\ldots \right),
\label{Vscal}
\ee
where we used \eqref{eq:ClassicalScalingAnalysisContractions}. Focusing on $p_f=2 R$, this gives rise to:
\be
\lambda_{\rm crit}=\frac{\lambda+2(Q-R)}{3}-4+x+o  = P+2R+3(L-1)+x+o\,.
\label{eq:ResultsParamQR} 
\ee
Non-zero contributions to the 4D scalar potential arise when $\lambda_f =\lambda_{\rm crit}$, implying from (\ref{Vscal}) that they would scale as:
\be
V \xrightarrow{v_f \raw 0} \vo^{-\frac13 [7+ 2(R+L)]}\,.
\label{Vvol}
\ee
Combining (\ref{Vvol}) with (\ref{Important}) for $q = \frac13 [7+ 2(R+L)]$ and $r=R+L$, we find $(\alpha')^m$ corrections with $m = 3$ at different F-term orders (counted by $r$), as expected for 8-derivative terms. Some examples of these $(\alpha')^3$ corrections to the 4D scalar potential emerging from 11D loops are: ($i$) the $\ell_M^6$ 11D term $\mc{R}^4$ ($P=3$, $R=0$ and $L=0$) which would yield $m = 3$ and $V\sim \vo^{-7/3}$, and so a potential $(\alpha')^3$ correction which however does not contribute to the 4D scalar potential since the contractions of $\mc{R}^4$ are in fact zero, similar to reducing the 10D $\mc{R}^4$ term on a 6D space \cite{Becker:2002nn} (as discussed in Sec.~\ref{AbsenceOfAPOne}, supersymmetry ensures that all moduli remain massless at every order in $\alpha'$ and $g_s$ in the absence of background flux); ($ii$) the $\ell_M^6$ 11D term $\mc{R}^3 G_4^2$ ($P=2$, $R=1$ and $L=0$) which gives $m = 3$ and $V\sim \vo^{-3}$ corresponding to the $(\alpha')^3$ corrections at $\mc{O}(F^2)$ computed in \cite{Becker:2002nn};  ($iii$) the $\ell_M^6$ 11D term $\mc{R}^2 (\nabla G_4)^2$ ($P=1$, $R=0$ and $L=1$) which gives $m = 3$ and $V\sim \vo^{-3}$ corresponding again to $(\alpha')^3$ corrections at $\mc{O}(F^2)$. Results for different 11D terms are summarised in Tab.~\ref{tab:QuantumKK}. 

\begin{table}[t!]
\centering
\begin{tabular}{|c|c|c|c|c|c|c|c|c|}
\hline 
&  &  &  & & &  &  & \\ [-1.em]
$\ell_M^l$ & $P$ & $R$ & $L$ & 11D term   & $\lambda_{\rm crit}$ & $(\alpha')^m$ &  $V(\vo)$ & $F^{2r}$ \\ 
\hline 
\hline 
&  &  &  & & &   & & \\ [-0.8em]
$\ell_M^6$ & 2 & 1 & 0 & $\mc{R}^3 G_4^2$ &   $1$ & 3 & $\vo^{-3}$ & $F^2$ \\ [0.8em]
\hline  
&  &  &  & &   & & & \\ [-0.8em]
$\ell_M^6$ & 1 & 0 & 1 & $\mc{R}^2 (\nabla G_4)^2$ & $1$ & 3 & $\vo^{-3}$ & $F^2$ \\ [0.8em]
\hline 
&  &  &  & &   & & & \\ [-0.8em]
$\ell_M^6$ & 1 & 2 & 0 & $\mc{R}^2 G_4^4$ & $2$ & 3 & $\vo^{-11/3}$ & $F^4$ \\ [0.8em]
\hline 
&  &  &  & &   & & & \\ [-0.8em]
$\ell_M^6$ & -1 & 0 & 2 & $(\nabla G_4)^4$ & $3$ & 3 & $\vo^{-11/3}$ & $F^4$ \\ [0.8em]
\hline
&  &  &  & & &  &  & \\ [-0.6em]
$\ell_M^6$ & 0 & 3 & 0 & $\mc{R} G_4^6$ & $3$ & 3 & $\vo^{-13/3}$ & $F^6$ \\ [0.5em]
\hline 
&  &  &  & & &  & & \\ [-0.8em]
$\ell_M^6$ & -1 & 4 & 0 & $G_4^8$ & $4$ & 3 & $\vo^{-5}$ & $F^8$ \\ [0.8em]
\hline 
\end{tabular} 
\caption{Summary of $(\alpha')^3$ corrections to the 4D potential at different F-term order from quantum reductions of M-theory compactified on trivially-fibred fourfolds. For convenience, we set $x=o=0$ in $\lambda_{\text{crit}}$ because the volume behaviour in $V(\vo)$ is independent on both.}
\label{tab:QuantumKK}
\end{table}

This analysis gives evidence that $(\alpha')^{\rm odd}$ corrections to the 4D scalar potential at different F- or D-term order should arise from the quantum reduction of the M-theory action. Although these results look promising, this is certainly not the full picture since we did not investigate higher loops and non-trivial fibrations. For instance, the results of \cite{GarciaEtxebarria:2012zm} suggest that $(\alpha')^2$ corrections enjoy a non-trivial modular behaviour which can only arise from a proper treatment of KK and winding modes on an elliptically-fibred K3 manifold. We therefore expect that 11D loops, when computed on non-trivial fibrations, should generate also $(\alpha')^{\rm even}$ effects. Moreover we have not yet considered the presence of non-perturbative degrees of freedom from M2/M5-brane instantons which certainly raises new challenges \cite{Green:2006gt}. These are important questions for F-theory compactifications that deserve further scrutiny.

\section{Conclusions}
\label{sec:Con}

This paper  provides a step towards a systematic understanding of the $\alpha'$ expansion in F-theory, with the final goal of classifying the moduli dependence of arbitrary perturbative corrections to the 4D scalar potential of type IIB string theory, where moduli stabilisation is best understood. Understanding at which order in $\alpha'$ and $g_s$ the characteristic no-scale structure of these compactifications gets broken, is fundamental for controlling moduli stabilisation, which is the primary goal to connect string theory to low-energy particle physics and cosmology. 

The first part was concerned with the picture of type IIB CY orientifold compactifications. By exploiting the two approximate scaling symmetries of the underlying 10D theory, combined with supersymmetry and shift symmetry, we managed to infer the dependence on the dilaton and the CY volume of an arbitrary perturbative correction in $\alpha'$ and $g_s$ to the 4D scalar potential at different orders in the low-energy superspace derivative expansion. Due to the absence of $(\alpha')^1$ corrections in 10D and 8D, and the fact that $(\alpha')^2$ corrections enjoy an extended no-scale cancellation \cite{Cicoli:2007xp}, we deduced that the dominant no-scale breaking effects at string tree-level arise from known $(\alpha')^3$ corrections \cite{Becker:2002nn,Minasian:2015bxa} modulo potential logarithmic corrections.

However higher orders in $g_s$ require further scrutiny. This is because we reduce higher dimensional theories with more than 16 supercharges on K\"ahler manifolds to 4D $\cN=1$ supergravity theories with 4 supercharges by retaining only KK zero modes. At string tree-level, all 4D corrections originate from the higher dimensional effective actions. Starting from string 1-loop, however, additional states such as KK or winding states with non-vanishing charge become relevant by participating in amplitudes with low-energy states. Thus at the loop level it remains obscure whether the severe reduction of the number of supercharges yields additional $\alpha'$ corrections. 

Such effects were for instance observed in \cite{Berg:2005ja} via loop corrections due to the exchange of KK and winding modes in $\cN=2$ and $\cN=1$ toroidal orientifold compactifications. Despite many efforts, their origin from the worldvolume theory of D7-branes wrapped on 4-cycles continues to be vague. A hint in this direction might come from analysing loop amplitudes of 11D supergravity compactified on elliptically fibred CY fourfolds briefly introduced in Sec.~\ref{sec:QuantumReductions}. Such an undertaking would allow for an exact-in-$g_s$ statement regarding $(\alpha')^1$ corrections because these amplitudes efficiently capture string loop and non-perturbative corrections. In App.~\ref{AppB} we have however shown that $(\alpha')^1$ loop effects, if present at all, instead of destabilising known LVS vacua, can give rise to new dS minima in a regime where the EFT can be under control. 

In the second part of this paper we addressed instead the issue of $\alpha'$ corrections in the 4D F-theory effective action from compactifications of M-theory on elliptically fibred CY fourfolds $Y_4$. In the context of the F/M-theory duality, we derived scaling relations between the variables in the two duality frames. We utilised a general ansatz for the metric on $Y_4$ depending only on integer powers of the fibre volume $v_{f}$ and of the $2$-cycle volume $v$ on the base. The split of the metric components along base and fibre directions allowed us to define the parametric volume scaling of various tensor components. Subsequently we performed an exhaustive dimensional analysis of a generic higher derivative 11D term constructed from $\mc{R}$, $G_4$ and $\nabla G_4$. This investigation showed that, in conventional KK reductions of M-theory on $Y_4$, only $(\alpha')^{\rm even}$ corrections survive in the 4D F-theory limit. This procedure does not allow to make statements about possible cancellation effects, as some surviving terms may be identically zero. However, we can state which contributions vanish in the limit $v_{f}\raw 0$. In particular, we found that all corrections in 4D necessarily disappear for trivial fibrations and even for the semi-flat ansatz of \cite{Greene:1989ya} since they are killed by the F-theory limit.

Overall these findings provide convincing evidence that our treatment of F-theory to extract the low energy effective action needs to be revised in order to capture $(\alpha')^{\rm odd}$ effects. Historically this might not really come as a surprise given that the $(\alpha')^3$-corrected 10D type IIB action (\ref{AP3Action}) cannot simply be recovered from classical KK reductions of M-theory on a $T^2$, but only when winding modes along the torus are properly integrated out \cite{Green:1997as}. In other words, the type IIB bulk or closed string degrees of freedom are associated with winding states on the $T^2$ in the ${\rm Vol}(T^{2})\raw 0$ limit. Therefore we argued that incorporating KK and winding states on the elliptic fibration is crucial in understanding the full range of $\alpha'$ corrections in F-theory compactifications. We hope to address some of these issues in the future.

\section*{Acknowledgements}

We would like to thank Antonella Grassi, Michael Green, Daniel Junghans, James Liu, Fernando Marchesano, Ruben Minasian and Gary Shiu for useful discussions. The work of FQ has been partially supported by STFC consolidated grants ST/P000681/1, ST/T000694/1.
AS acknowledges support by the German Academic Scholarship Foundation as well as by DAMTP through an STFC studentship. AS also thanks the Cambridge Trust for his Helen Stone Scholarship in support of his studies.

\appendix

\section{Compactifications on elliptically fibred Calabi-Yau manifolds}
\label{app:TensorsEllFibGen}

In this appendix we summarise useful definitions and identities relevant for the bulk of this paper. We compute Riemann tensor components for an elliptic fibration in order to determine their non-trivial volume scaling in Sec.~\ref{sec:dimAnalysis}. Furthermore, we expand on the discussion in Sec.~\ref{sec:ScalingAnalysisClas} about the absence of divergent terms in higher derivative structures $\mc{R}^4$ and $\mc{R}^3 G_4^2$.

\subsection{Definitions and conventions}

We start by giving some definitions and conventions for the various tensor structures encountered in the bulk of the paper. For 11D coordinates, we use capital letters $M,N,P,\ldots$ as indices. We mostly work with quantities along the internal direction of an elliptically fibred CY manifold. We denote $n$-dimensional complex coordinates $Z^{A}$ with capital letters $A,B,\ldots=1,\ldots,n$. Similarly, complex coordinates on the base are defined as $z^{\alpha}$ using Greek indices $\alpha,\beta,\ldots=1,\ldots,n-1$ and on the fibre as $\zeta^{a}$ with small letters $a,b,\ldots=1$.

\subsection*{Hermitian, K\"ahler and Calabi-Yau manifolds}

Let $X$ be a compact Hermitian manifold of complex dimension $n$ with real coordinates $\lbrace x^{1},\ldots ,x^{2n}\rbrace$. We define complex coordinates $Z^{A}$, $A=1,\ldots, n$, as:
\begin{equation}
(Z^{1},\ldots ,Z^{n})=\left (\dfrac{1}{\sqrt{2}}(x^{1}+\I x^{2}),\ldots ,\dfrac{1}{\sqrt{2}}(x^{2n-1}+\I x^{2n})\right )\, .
\end{equation}
Then:
\begin{equation}
\sqrt{g}\dif x^{1}\wedge\ldots\wedge \dif x^{2n}=\sqrt{g}(-1)^{\frac{n(n-1)}{2}}\, \I^{n}\dif Z^{1}\wedge \dif Z^{n}\wedge\dif \bar{Z}^{1}\wedge\ldots\wedge\dif \bar{Z}^{n}=\dfrac{1}{n!}J^{n}
\end{equation}
where $\sqrt{g}=\det(g_{A\bar{B}})$ and $J$ is the K\"ahler form:
\begin{equation}
J=\I g_{A\bar{B}}\dif Z^{A}\wedge\dif \bar{Z}^{\bar{B}}\, .
\end{equation}
The non-vanishing connection coefficients and curvature tensor components are (together with the corresponding complex conjugates):
\begin{align}
\label{eq:ConnRTHM} 
\Gamma^{A}_{BC}=g^{A\bar{D}}\p_{B}g_{C\bar{D}}\kom R^{A}\,_{B\bar{C}D}=\p_{\bar{C}}\Gamma^{A}_{DB} \, .
\end{align}
Furthermore, the curvature 2-form is defined as:
\begin{equation}
\cR^{A}\,_{B}=R^{A}\,_{BC\bar{D}}\dif Z^{C}\wedge\dif\bar{Z}^{\bar{D}}\, .
\end{equation}

A Hermitian manifold $X$ is K\"ahler if its K\"ahler form $J$ is closed, $\dif J=0$. The associated metric $g_{A\bar{B}}$ is referred to as K\"ahler metric.
In local coordinates $Z^{A}$, it is obtained from a K\"ahler potential $K$ via:
\begin{equation}
g_{A\bar{B}}=\p_{A}\p_{\bar{B}}K\, .
\end{equation}
Since $\dif J=0$ implies $\p_{A}g_{B\bar{C}}=\p_{B}g_{A\bar{C}}$, the connection coefficients and Riemann tensor components enjoy the additional symmetries:
\begin{align}
\Gamma^{A}_{BC}=\Gamma^{A}_{CB}\kom  R^{A}\,_{B\bar{C}D}= R^{A}\,_{D\bar{C}B}\, .
\end{align}

Finally, we call $X$ CY if its canonical bundle is trivial. Then, the 4-th Chern class is given in terms of the curvature 2-form by:
\begin{equation}
\label{eq:FourthCC} 
c_{4}(Y_{4})=\dfrac{1}{8}\left (\tr(\cR^{2})^{2}-2\tr(\cR^{4})\right )\,,
\end{equation}
where:
\begin{align}
\tr(\cR^{2})&=R^{A}\,_{BC_{1}\bar{D}_{1}}R^{B}\,_{AC_{2}\bar{D}_{2}}\dif Z^{C_{1}}\wedge\dif\bar{Z}^{\bar{D}_{1}}\wedge\dif Z^{C_{2}}\wedge\dif\bar{Z}^{\bar{D}_{2}}\, ,\\
\tr(\cR^{4})&=R^{A_{1}}\,_{B_{1}C_{1}\bar{D}_{1}}R^{B_{1}}\,_{A_{2}C_{2}\bar{D}_{2}}R^{A_{2}}\,_{B_{2}C_{3}\bar{D}_{3}}R^{B_{2}}\,_{A_{1}C_{4}\bar{D}_{4}} \bigwedge_{i=1}^{4}\dif Z^{C_{i}}\wedge\dif\bar{Z}^{\bar{D}_{i}}\, .
\end{align}

\subsection*{Higher derivative structures}

At the 8-derivative level, the M-theory action contains higher derivative corrections of the schematic form summarised in \eqref{eq:MtheorySalarPot3DCorrected}. In the CP-even sector, the corresponding index structures are nicely encoded in terms of the tensor $t_{8}$ as well as the totally anti-symmetric Levi-Civita symbol $\epsilon_{D}$ in $D$ dimensions. The tensor $t_{8}$ is defined as \cite{Freeman:1986zh,Freeman:1986br}:
\begin{equation}
t_{8}M^{4}=24\left (\tr(M^{4})-\dfrac{1}{4}\tr(M^{2})^{2}\right )
\end{equation}
for an anti-symmetric matrix $M$. It further is symmetric under the exchange of pairs of indices, while anti-symmetric within each pairs of indices, i.e.:
\begin{equation}
t_{M_{1}M_{2}M_{3}M_{4}M_{5}\ldots M_{8}}=-t_{M_{2}M_{1}M_{3}M_{4}M_{5}\ldots M_{8}}=t_{M_{3}M_{4}M_{1}M_{2}M_{5}\ldots M_{8}}\, .
\end{equation}
In Lorentzian space, we use a convention for the totally anti-symmetric tensor in an orthonormal frame where $\epsilon_{0\, 1\, 2\ldots 10}=+1$. 
In terms of the generalised Kronecker-$\delta$, we write:
\begin{equation}
\epsilon_{M_{1}\ldots M_{D}}\epsilon^{N_{1}\ldots N_{D}}=D!\, \delta^{[N_{1}\ldots N_{D}]}_{M_{1}\ldots M_{D}}
\end{equation}
as well as:
\begin{equation}
\epsilon_{M_{1}\ldots M_{D}}\epsilon^{M_{1}\ldots M_{D-n}N_{D-n+1}\ldots N_{D}}=(-1)^{s}n!\, (D-n)!\,  \delta^{[N_{D-n+1}\ldots N_{D}]}_{M_{D-n+1}\ldots M_{D}}
\end{equation}
with $s=1$ ($s=0$) in Lorentzian (Euclidean) signature. 

The higher derivative corrections \eqref{eq:ActionR4} to the Einstein-Hilbert term are encoded in the two quantities \cite{Green:1997di,Green:1997as,Kiritsis:1997em,Russo:1997mk,Antoniadis:1997eg,Tseytlin:2000sf}:
\begin{align}
t_{8}t_{8}R^{4}&=t_{ M_{1}\ldots M_{8}}t^{ N_{1}\ldots N_{8}}R^{ M_{1} M_{2}}\,_{ N_{1} N_{2}} R^{ M_{3} M_{4}}\,_{ N_{3} N_{4}}\ldots R^{ M_{7} M_{8}}\,_{ N_{7} N_{8}}\\
E_{8}&=\dfrac{1}{3!}\epsilon_{M_{1}\ldots M_{11}}\epsilon^{M_{1}M_{2}M_{3}N_{4}\ldots N_{11}} R^{M_{4}M_{5}}\,_{N_{4}N_{5}}\ldots R^{M_{10}M_{11}}\,_{N_{10}N_{11}}\nn\\
&=-8!\, R^{[M_{1}M_{2}}\,_{M_{1}M_{2}}\ldots R^{M_{7}M_{8}]}\,_{M_{7}M_{8}}
\end{align}
Furthermore, formula \eqref{eq:HDActionG2R3} expanded reads \cite{Hyakutake:2007sm,Liu:2013dna}:
\begin{align}
\label{eq:t8t8G2R3} t_{8}t_{8}G_{4}^{2}R^{3}&=t_{8}^{M_{1}\ldots M_{8}}t_{8}^{N_{1}\ldots N_{8}}G_{N_{1}M_{1}PQ}G_{N_{2}M_{2}}\,^{PQ} R_{M_{3}M_{4}N_{3}N_{4}}R_{M_{5}M_{6}N_{5}N_{6}}R_{M_{7}M_{8}N_{7}N_{8}}\\
\epsilon_{11}\epsilon_{11}G_{4}^{2}R^{3}&=\epsilon_{N_{0}N_{1}\ldots N_{10}}\epsilon^{N_{0}M_{1}\ldots M_{10}} G^{N_{1}N_{2}}\,_{M_{1}M_{2}}G^{N_{3}N_{4}}\,_{M_{3}M_{4}}R^{N_{5}N_{6}}\,_{M_{5}M_{6}}R^{N_{7}N_{8}}\,_{M_{7}M_{8}}R^{N_{9}N_{10}}\,_{M_{9}M_{10}}\nn\\
&=-10! G^{[M_{1}M_{2}}\,_{M_{1}M_{2}}G^{M_{3}M_{4}}\,_{M_{3}M_{4}}R^{M_{5}M_{6}}\,_{M_{5}M_{6}}R^{M_{7}M_{8}}\,_{M_{7}M_{8}}R^{M_{9}M_{10}]}\,_{M_{9}M_{10}}\,.
\end{align}

\subsection{Details on the dimensional analysis}

In this appendix we present additional material in support of the analysis in Sec.~\ref{sec:MF}.

\subsection*{Curvature tensors for elliptic fibrations}

We now compute the connection coefficients and curvature components \eqref{eq:ConnRTHM} for the metric ansatz \eqref{eq:KPAnsatzG}. Splitting the indices along base and fibre leads to:
\begin{align}
\Gamma^{a}_{bc}&={\color{red} g^{a\bar{d}}\p_{b}g_{c\bar{d}}}+g^{a\bar{\delta}}\p_{b}g_{c\bar{\delta}}\kom \Gamma^{a}_{b\mu}=g^{a\bar{d}}\p_{b}g_{\mu\bar{d}}+g^{a\bar{\delta}}\p_{b}g_{\mu\bar{\delta}}\, ,\nn\\
\Gamma^{a}_{\alpha\beta}&=g^{a\bar{d}}\p_{\alpha}g_{\beta\bar{d}}+g^{a\bar{\delta}}\p_{\alpha}g_{\beta\bar{\delta}}\kom \Gamma^{\alpha}_{bc}=g^{\alpha\bar{\delta}}\p_{b}g_{c\bar{\delta}}+{\color{red} g^{\alpha\bar{d}}\p_{b}g_{c\bar{d}}}\, ,\nn\\
\Gamma^{\alpha}_{\beta\gamma}&=g^{\alpha\bar{d}}\p_{\beta}g_{\gamma\bar{d}}+g^{\alpha\bar{\delta}}\p_{\beta}g_{\gamma\bar{\delta}}\kom \Gamma^{\alpha}_{b\gamma}=g^{\alpha\bar{d}}\p_{b}g_{\gamma\bar{d}}+g^{\alpha\bar{\delta}}\p_{b}g_{\gamma\bar{\delta}}\, .
\end{align}
Then, we compute:
\begin{align}
R^{\alpha}\,_{\beta\bar{\gamma}\lambda}&=\left (\p_{\bar{\gamma}}g^{\alpha\bar{d}}\right ) \p_{\lambda}g_{\beta\bar{d}}+g^{\alpha\bar{d}}\p_{\bar{\gamma}}\p_{\lambda}g_{\beta\bar{d}}+\left (\p_{\bar{\gamma}}g^{\alpha\bar{\delta}}\right )\p_{\lambda}g_{\beta\bar{\delta}}+g^{\alpha\bar{\delta}}\p_{\bar{\gamma}}\p_{\lambda}g_{\beta\bar{\delta}}\, ,\nn\\
R^{\alpha}\,_{\beta\bar{c}\lambda}&=\left (\p_{\bar{c}}g^{\alpha\bar{d}}\right ) \p_{\lambda}g_{\beta\bar{d}}+g^{\alpha\bar{d}}\p_{\bar{c}}\p_{\lambda}g_{\beta\bar{d}}+\left (\p_{\bar{c}}g^{\alpha\bar{\delta}}\right )\p_{\lambda}g_{\beta\bar{\delta}}+g^{\alpha\bar{\delta}}\p_{\bar{c}}\p_{\lambda}g_{\beta\bar{\delta}}\, ,\nn\\
R^{\alpha}\,_{a\bar{\gamma}\lambda}&=\left (\p_{\bar{\gamma}}g^{\alpha\bar{d}}\right )\p_{\lambda}g_{a\bar{d}}+g^{\alpha\bar{d}}\p_{\bar{\gamma}}\p_{\lambda}g_{a\bar{d}}+\left (\p_{\bar{\gamma}}g^{\alpha\bar{\delta}}\right )\p_{\lambda}g_{a\bar{\delta}}+g^{\alpha\bar{\delta}}\p_{\bar{\gamma}}\p_{\lambda}g_{a\bar{\delta}}\, ,\nn\\
 R^{\alpha}\,_{a\bar{\gamma}b}&=\left (\p_{\bar{\gamma}} g^{\alpha\bar{\delta}}\right )\p_{b}g_{a\bar{\delta}}+g^{\alpha\bar{\delta}}\p_{\bar{\gamma}}\p_{b}g_{a\bar{\delta}}+{\color{red} \left (\p_{\bar{\gamma}}g^{\alpha\bar{d}}\right )\p_{b}g_{a\bar{d}}}+{\color{red} g^{\alpha\bar{d}}\p_{\bar{\gamma}}\p_{b}g_{a\bar{d}}}\, ,\nn\\
 R^{\alpha}\,_{a\bar{c}\lambda}&=\left (\p_{\bar{c}}g^{\alpha\bar{d}}\right )\p_{\lambda}g_{a\bar{d}}+{\color{red}g^{\alpha\bar{d}}\p_{\bar{c}}\p_{\lambda}g_{a\bar{d}}}+\left (\p_{\bar{c}}g^{\alpha\bar{\delta}}\right )\p_{\lambda}g_{a\bar{\delta}}+g^{\alpha\bar{\delta}}\p_{\bar{c}}\p_{\lambda}g_{a\bar{\delta}}\, ,\nn\\
 R^{\alpha}\,_{a\bar{c}b}&=\left (\p_{\bar{c}} g^{\alpha\bar{\delta}}\right )\p_{b}g_{a\bar{\delta}}+{\color{red}g^{\alpha\bar{\delta}}\p_{\bar{c}}\p_{b}g_{a\bar{\delta}}}+{\color{red} \left (\p_{\bar{c}}g^{\alpha\bar{d}}\right )\p_{b}g_{a\bar{d}}}+{\color{red} g^{\alpha\bar{d}}\p_{\bar{c}}\p_{b}g_{a\bar{d}}}\, ,\nn\\
R^{a}\,_{\beta\bar{\gamma}\alpha}&=\left (\p_{\bar{\gamma}}g^{a\bar{d}}\right )\p_{\alpha}g_{\beta\bar{d}}+g^{a\bar{d}}\p_{\bar{\gamma}}\p_{\alpha}g_{\beta\bar{d}}+\left (\p_{\bar{\gamma}}g^{a\bar{\delta}}\right )\p_{\alpha}g_{\beta\bar{\delta}}+g^{a\bar{\delta}}\p_{\bar{\gamma}}\p_{\alpha}g_{\beta\bar{\delta}}\, ,\nn\\
 R^{a}\,_{\beta\bar{c}\alpha}&=\left (\p_{\bar{c}}g^{a\bar{d}}\right )\p_{\alpha}g_{\beta\bar{d}}+g^{a\bar{d}}\p_{\bar{c}}\p_{\alpha}g_{\beta\bar{d}}+\left (\p_{\bar{c}}g^{a\bar{\delta}}\right )\p_{\alpha}g_{\beta\bar{\delta}}+g^{a\bar{\delta}}\p_{\bar{c}}\p_{\alpha}g_{\beta\bar{\delta}}\, ,\nn\\
R^{a}\,_{b\bar{\gamma}\lambda}&= \left ( \p_{\bar{\gamma}} g^{a\bar{d}}\right )\p_{\lambda}g_{b\bar{d}}+g^{a\bar{d}}\p_{\bar{\gamma}}\p_{\lambda}g_{b\bar{d}}+\left (\p_{\bar{\gamma}}g^{a\bar{\delta}}\right )\p_{\lambda}g_{b\bar{\delta}}+g^{a\bar{\delta}}\p_{\bar{\gamma}}\p_{\lambda}g_{b\bar{\delta}} \, ,\nn\\
 R^{a}\,_{b\bar{\gamma} e}&={\color{red}\left ( \p_{\bar{\gamma}} g^{a\bar{d}}\right )\p_{e}g_{b\bar{d}}}+{\color{red} g^{a\bar{d}}\p_{\bar{\gamma}} \p_{e}g_{b\bar{d}}}+\left (\p_{\bar{\gamma}}g^{a\bar{\delta}}\right )\p_{e}g_{b\bar{\delta}}+g^{a\bar{\delta}}\p_{\bar{\gamma}}\p_{e}g_{b\bar{\delta}}\, ,\nn\\
 R^{a}\,_{b\bar{c}\lambda}&=\left (\p_{\bar{c}}g^{a\bar{d}}\right )\p_{\lambda}g_{b\bar{d}}+{\color{red}g^{a\bar{d}} \p_{\bar{c}}\p_{\lambda}g_{b\bar{d}}}+\left (\p_{\bar{c}}g^{a\bar{\delta}}\right )\p_{\lambda}g_{b\bar{\delta}}+g^{a\bar{\delta}}\p_{\bar{c}}\p_{\lambda}g_{b\bar{\delta}}\, ,\nn\\
 R^{a}\,_{b \bar{c}e}& ={\color{red} \left (\p_{\bar{c}} g^{a\bar{d}}\right )\p_{e}g_{b\bar{d}}}+{\color{red} g^{a\bar{d}}\p_{\bar{c}} \p_{e}g_{b\bar{d}}}+\left (\p_{\bar{c}}g^{a\bar{\delta}}\right )\p_{e}g_{b\bar{\delta}}+{\color{red}g^{a\bar{\delta}}\p_{\bar{c}}\p_{e}g_{b\bar{\delta}}} \, .
\end{align}
All terms highlighted in {\color{red}red} vanish for a quadratic K\"ahler potential $K^{f}$ in the fibre coordinates $\zeta$ (such as for the choice \eqref{eq:KPAnsatzGSVY}). One can further simplify the above expressions by using:
\begin{align}
\p_{\bar{\gamma}}g^{A\bar{B}}=-g^{A\bar{C}}\left (\p_{\bar{\gamma}}g_{\bar{C}D}\right )g^{D\bar{B}}\kom \p_{\bar{c}}g^{A\bar{B}}=-g^{A\bar{C}}\left (\p_{\bar{c}}g_{\bar{C}D}\right )g^{D\bar{B}}\, .
\end{align}
From these expressions, one derives the volume scalings:
\begin{align}
R^{\alpha}\,_{\beta\bar{\gamma}\lambda},R^{a}\,_{\beta\bar{\gamma}\alpha}, R^{a}\,_{\beta\bar{\gamma} b},R^{a}\,_{b\bar{\gamma}\lambda}&=\cO\left (1 \right ) \kom R^{\alpha}\,_{\beta\bar{\gamma}e},R^{\alpha}\,_{\beta\bar{c}\lambda},R^{\alpha}\,_{a\bar{\gamma}\lambda}=\cO\left (\dfrac{v_{f}}{v}\right )\, ,\nn\\
 R^{\alpha}\,_{\beta\bar{c}e},R^{\alpha}\,_{a\bar{\gamma}b},R^{\alpha}\,_{a\bar{c}\lambda}&={\color{red} \cO\left (\dfrac{v_{f}}{v}\right )}+\cO\left (\dfrac{v_{f}}{v} \right ) \kom R^{\alpha}\,_{a\bar{c}b}={\color{red} \cO\left (\dfrac{v_{f}}{v}\right )}+\cO\left (\dfrac{v_{f}^{2}}{v^{2}}\right )\, ,\nn\\
R^{a}\,_{\beta\bar{c}b},R^{a}\,_{b\bar{\gamma} e},R^{a}\,_{b\bar{c}\lambda},R^{a}\,_{b \bar{c}e}&={\color{red}\cO\left (1\right )}+\cO\left (\dfrac{v_{f}}{v} \right )\kom  R^{a}\,_{\beta\bar{c}\alpha}={\color{red}\cO\left (1\right )}+\cO\left (1 \right )
\end{align}
as summarised in \eqref{eq:VolScalGamma} and \eqref{eq:VolScalRiemannT}. Again, we indicate scaling behaviours vanishing for the ansatz \eqref{eq:KPAnsatzGSVY} in {\color{red}red}.

\subsection*{Absence of divergences}

We now want to prove formula \eqref{J0vanishing}. This becomes clear when considering the tensor:
\begin{equation}
Z_{UT}=R_{UMRV}R_{TSQ}\,^{V}\left (R^{M}\,_{P}\,^{S}\,_{N}R^{QPRN}-\dfrac{1}{2}R^{MQ}\,_{PN}R^{SRPN}\right )\,,
\end{equation}
which for Ricci-flat (but not necessarily K\"ahler) manifolds is related to $J_{0}$ as $J_{0}=g^{UT}Z_{UT}$. The authors of \cite{Grisaru:1986px} showed that $Z_{UT}\equiv 0$ on K\"ahler spaces. This can easily be seen by switching to complex coordinates where:\footnote{Here, we make use of $g_{MN}=g_{\bar{A}B}+g_{A\bar{B}}$ and $R_{MNPQ}=R_{A\bar{B}C\bar{D}}+R_{\bar{A}BC\bar{D}}+R_{A\bar{B}\bar{C}D}+R_{\bar{A}B\bar{C}D}$, which holds on any K\"ahler manifold. In particular, one finds the useful identities $R^{A}\,_{\bar{B}MN}=R^{\bar{A}}\,_{BMN}=0$.}
\begin{equation}
\label{eq:CompTensorR4} 
Z_{A\bar{B}}=R_{A\bar{C}\bar{D}E}R_{\bar{B}FG}\,^{E}\left (R^{\bar{C}}\,_{\bar{H}}\,^{F}\,_{I}R^{G\bar{H}\bar{D}I}-R^{\bar{C}G}\,_{\bar{H}I}R^{F\bar{D}\bar{H}I}\right )\, .
\end{equation}
For K\"ahler manifolds, we can further use:
\begin{equation}
R^{\bar{C}}\,_{\bar{H}\bar{F}}\,_{I}=R^{\bar{C}}\,_{\bar{F}\bar{H}}\,_{I}\kom R_{\bar{G}{H}{D}\bar{I}}=R_{\bar{G}{D}H\bar{I}}
\end{equation}
to rewrite the first term in \eqref{eq:CompTensorR4} in such a way that:
\begin{equation}\label{eq:CompTensorR4Mod} 
Z_{A\bar{B}}=2R_{A\bar{C}\bar{D}}\,^{\bar{E}}R_{\bar{B}FG\bar{E}}  R^{\bar{C}[F}\,_{\bar{H}I}R^{G]\bar{D}\bar{H}I}\, .
\end{equation}
Since for K\"ahler manifolds $R_{\bar{B}FG\bar{E}}=R_{\bar{B}GF\bar{E}}$ is symmetric under the exchange of labels $G$, $F$, we find:
\begin{equation}
Z_{A\bar{B}}\equiv 0\, .
\end{equation}
This implies that $J_{0}$ vanishes in compactifications of both type IIB on CY threefolds and M-theory on CY fourfolds.In the former case, $J_{0}$ encodes the full $\mc{R}^4$ dependence of the 10D action which is why there is no contribution to the scalar potential from $\mc{R}^4$.

Now we turn our attention to the kinematic structure $t_{8}t_{8} \mc{R}^3 G_4^2$. We are going to show that at least a specific subset of terms contained within this structure are free of divergences when reduced on $Y_4$. Indeed, starting from the definition \eqref{eq:t8t8G2R3} and switching to complex coordinates as above one can show that on K\"ahler spaces:
\begin{align}
\label{eq:t8t8G2R3CC} 
\dfrac{t_{8}t_{8} \mc{R}^3 G_4^2}{12}&=4G^{2}_{\bar{A}_{4}A_{5}A_{6}\bar{A}_{7}}\biggl \{-4R^{\bar{A}_{8}A_{1}A_{2}\bar{A}_{7}}\biggl [R_{\bar{A}_{8}}\,^{\bar{A}_{4}}\,_{A_{2}}\,^{A_{3}}R_{A_{1}}\,^{A_{6}}\,_{A_{3}}\,^{A_{5}}+R_{\bar{A}_{8}A_{1}}\,^{\bar{A}_{3}A_{5}}R_{A_{2}\bar{A}_{3}}\,^{A_{6}\bar{A}_{4}}\nn\\
&\quad+R_{A_{1}}\,^{A_{6}}\,_{A_{2}}\,^{A_{3}}R_{\bar{A}_{8}}\,^{\bar{A}_{4}}\,_{A_{3}}\,^{A_{5}}\biggl ]-8R_{A_{8}\bar{A}_{1}\bar{A}_{2}}\,^{\bar{A}_{4}} R_{A_{3}}\,^{A_{6}A_{5}\bar{A}_{7}}R^{A_{8}\bar{A}_{1}\bar{A}_{2}A_{3}}\nn\\
&\quad+R^{\bar{A}_{8}A_{1}A_{2}\bar{A}_{3}}\biggl [R^{\bar{A}_{4}A_{6}A_{5}\bar{A}_{7}}R_{\bar{A}_{8}A_{1}A_{2}\bar{A}_{3}}+2R_{\bar{A}_{8}}\,^{\bar{A}_{4}}\,_{\bar{A}_{3}}\,^{\bar{A}_{7}}R_{A_{2}}\,^{A_{6}}\,_{A_{1}}\,^{A_{5}}\\
&\quad+2R_{\bar{A}_{8}}\,^{\bar{A}_{4}}\,_{A_{2}}\,^{A_{6}}R_{\bar{A}_{3}}\,^{\bar{A}_{7}}\,_{A_{1}}\,^{A_{5}}+2R_{\bar{A}_{8}}\,^{\bar{A}_{4}}\,_{A_{2}}\,^{A_{5}}R_{A_{1}}\,^{A_{6}}\,_{\bar{A}_{3}}\,^{\bar{A}_{7}} \biggl ]\biggl \}\nn\\
&\quad+16G^{2}_{\bar{A}_{4}\bar{A}_{5}A_{6}A_{7}}\,  R^{\bar{A}_{8}A_{1}\bar{A}_{2}A_{7}}\biggl [-R_{\bar{A}_{8}}\,^{\bar{A}_{4}}\,_{\bar{A}_{2}}\,^{\bar{A}_{3}}R_{A_{1}}\,^{A_{6}}\,_{\bar{A}_{3}}\,^{\bar{A}_{5}}+R_{\bar{A}_{8}A_{1}}\,^{A_{3}\bar{A}_{4}}R_{\bar{A}_{2}A_{3}}\,^{A_{6}\bar{A}_{5}}\biggl ]\nn\\
&\quad+\text{c.c.}\,.\nn
\end{align}
Here, we defined the following two types of flux contractions:
\begin{align} 
\label{eq:G2Contractions1}G^{2}_{\bar{A}_{4}A_{5}A_{6}\bar{A}_{7}}&=G_{\bar{A}_{4}A_{5}}\,^{A_{9}A_{10}} G_{A_{6}\bar{A}_{7} A_{9}A_{10}}+2G_{\bar{A}_{4}A_{5}}\,^{\bar{A}_{9}A_{10}} G_{A_{6}\bar{A}_{7}  \bar{A}_{9}A_{10}}+G_{\bar{A}_{4}A_{5}}\,^{\bar{A}_{9}\bar{A}_{10}} G_{A_{6}\bar{A}_{7}  \bar{A}_{9}\bar{A}_{10}}\nn\\
&=(1,3)\, (3,1)+ (2,2)\, (2,2)+(3,1)\, (1,3)\, ,\\
\label{eq:G2Contractions2}G^{2}_{\bar{A}_{4}\bar{A}_{5}A_{6}A_{7}}&=G_{\bar{A}_{4}\bar{A}_{5}}\,^{A_{9}A_{10}} G_{A_{6}A_{7} A_{9}A_{10}}+2G_{\bar{A}_{4}\bar{A}_{5}}\,^{\bar{A}_{9}A_{10}} G_{A_{6}A_{7}  \bar{A}_{9}A_{10}}+G_{\bar{A}_{4}\bar{A}_{5}}\,^{\bar{A}_{9}\bar{A}_{10}} G_{A_{6}A_{7}  \bar{A}_{9}\bar{A}_{10}}\nn\\
&=(0,4)\, (4,0)+(1,3)\, (3,1)+(2,2)\, (2,2)\, ,
\end{align}
where the second and fourth line indicate the $(p,q)$-type of the various components.

According to our scaling analysis, there are 14 types of contractions that diverge upon taking the limit $v_{f}\raw 0$ involving the following combinations of Riemann tensors:
\begin{align}
&(R^{a}\,_{b\bar{c}d})^{2}R^{\alpha}\,_{\beta\bar{\gamma}\delta}\kom (R^{a}\,_{b\bar{c}d})^{2}R^{\alpha}\,_{b\bar{c}\delta}\kom R^{a}\,_{b\bar{c}d} \left (R^{\alpha}\,_{b\bar{c}d}\right )^{2}\kom R^{a}\,_{b\bar{c}d}R^{a}\,_{b\bar{c}\delta}R^{\alpha}\,_{b\bar{c}d} \, ,\nn\\
& (R^{a}\,_{b\bar{c}d})^{2}R^{a}\,_{b\bar{\gamma}\delta}\kom R^{a}\,_{b\bar{c}d} \left (R^{a}\,_{b\bar{c}\delta}\right )^{2}\kom (R^{a}\,_{b\bar{c}d})^{3}\, .
\end{align}
The divergent contractions must have $\lambda_{f}>2$, recall Tab.~\ref{tab:ClassicalKK}. We claim that the absence of divergences in \eqref{eq:t8t8G2R3CC} is mainly due to the following two reasons: a) the K\"ahlerity of the underlying CY fourfold and b) the dimension of the fibre itself.

Although proving this claim in full generality without making any further assumptions about the metric seems out of reach, we may be able to provide clear evidence. We assume that cancellation of divergences must be manifest for all $(p,q)$-types of fluxes independently. Further, we observe that only \eqref{eq:G2Contractions2} contains contributions from $(4,0)$-flux for which \eqref{eq:t8t8G2R3CC} reduces to:
\begin{align}
\dfrac{t_{8}t_{8} \mc{R}^3 G_4^2}{12}\biggl |_{(4,0)}&=32G^{2}_{\bar{A}_{4}\bar{A}_{5}A_{6}A_{7}}\,  R^{\bar{A}_{8}A_{1}\bar{A}_{2}A_{7}}\biggl [-R_{\bar{A}_{8}}\,^{\bar{A}_{4}}\,_{\bar{A}_{2}}\,^{\bar{A}_{3}}R_{A_{1}}\,^{A_{6}}\,_{\bar{A}_{3}}\,^{\bar{A}_{5}}\nn\\
&\quad+R_{\bar{A}_{8}A_{1}}\,^{A_{3}\bar{A}_{4}}R_{\bar{A}_{2}A_{3}}\,^{A_{6}\bar{A}_{5}}\biggl ]\,.
\end{align}
Each $G_4$ has at most one index along the fibre such that:
\begin{align}
\label{eq:DivFZFluxH2R3} 
\dfrac{t_{8}t_{8} \mc{R}^3 G_4^2}{12}\biggl |_{(4,0)}&= 32 \biggl \{4G^{2}_{\bar{a} \bar{\alpha} b\beta}\, g^{\bar{a}[A_{14}}g^{A_{15}]\bar{\alpha}}\, g^{b[\bar{A}_{16}}g^{\bar{A}_{17}]\beta} +2G^{2}_{\bar{a} \bar{\alpha} \beta\gamma}\, g^{\bar{a}[A_{14}}g^{A_{15}]\bar{\alpha}}\,  g^{\beta\bar{A}_{16}}g^{\gamma\bar{A}_{17}} \nn\\
&\quad+2G^{2}_{\bar{\alpha} \bar{\beta} b\gamma}\,  g^{\bar{\alpha}A_{14}}g^{\bar{\beta}A_{15}} \, g^{b[\bar{A}_{16}}g^{\bar{A}_{17}]\gamma} +G^{2}_{\bar{\gamma} \bar{\alpha} \delta\beta}\,  g^{\bar{\gamma}A_{14}}g^{\bar{\alpha}A_{15}}\, g^{\delta\bar{A}_{16}}g^{\beta\bar{A}_{17}}\biggl \} \\
&\quad\biggl [-g_{\bar{A}_{8}A_{13}}g^{\bar{A}_{3}A_{12}}g^{\bar{A}_{2}A_{11}}R^{\bar{A}_{8}}\,_{\bar{A}_{1}A_{11}\bar{A}_{17}} R^{A_{13}}\,_{A_{14}\bar{A}_{2}A_{12}}R^{\bar{A}_{1}}\,_{\bar{A}_{16}\bar{A}_{3}A_{15}}\nn\\
&\quad+g_{\bar{A}_{8}A_{13}}g^{A_{3}\bar{A}_{12}}g^{A_{1}\bar{A}_{11}} R^{\bar{A}_{8}}\,_{\bar{A}_{11}A_{2}\bar{A}_{17}}R^{A_{13}}\,_{A_{1}\bar{A}_{12}A_{14}}R^{A_{2}}\,_{A_{3}\bar{A}_{16}A_{15}}\biggl ]\nn
\end{align}
where for the $(4,0)$/$(0,4)$-components:
\begin{align}
G^{2}_{\bar{a} \bar{\alpha} b\beta}&=G_{\bar{a} \bar{\alpha}}\,^{\sigma\rho} G_{ b\beta \sigma\rho}=g^{\sigma\bar{\lambda}}g^{\rho\bar{\mu}} G_{\bar{a} \bar{\alpha}\bar{\lambda}\bar{\mu}}G_{ b\beta \sigma\rho}\\
G^{2}_{\bar{a} \bar{\alpha} \beta\gamma}&=2G_{\bar{a} \bar{\alpha}}\,^{\lambda b} G_{\beta\gamma \lambda b}=2g^{\lambda\bar{\mu}}g^{b\bar{\nu}} G_{\bar{a} \bar{\alpha}\bar{\mu}\bar{\nu}} G_{\beta\gamma \lambda b}\\
G^{2}_{\bar{\gamma} \bar{\alpha} \delta\beta}&=G_{\bar{\gamma} \bar{\alpha}}\,^{a\lambda} G_{ \delta\beta a\lambda}=g^{a\bar{b}}g^{\lambda\bar{\mu}} G_{\bar{\gamma} \bar{\alpha}\bar{b}\bar{\mu}} G_{ \delta\beta a\lambda}+g^{a\bar{\mu}}g^{\lambda\bar{b}} G_{\bar{\gamma} \bar{\alpha}\bar{b}\bar{\mu}} G_{ \delta\beta a\lambda} \, .
\end{align}
Critically, only the last line involves at most a single inverse fibre metric. Hence, looking at the way inverse metrics appear in the above expression we deduce that terms built from $G^{2}_{\bar{a} \bar{\alpha} b\beta}$ have a total number of inverse fibre metrics $\lambda_{f}^{\text{tot}}\leq 4$, whereas those built using $G^{2}_{\bar{a} \bar{\alpha} \beta\gamma}$ or $G^{2}_{\bar{\gamma} \bar{\alpha} \delta\beta}$ have $\lambda_{f}^{\text{tot}}\leq 3$.
This eliminates $4$ divergent contractions with $\lambda_{f}^{\text{tot}}=5$ leaving us with 10 dangerous index structures. In the case $(R^{a}\,_{b\bar{c}d})^{3}$, $g_{\bar{A}_{8}A_{13}}\raw g_{\bar{a}a}$ contributes positive powers of $v_{f}$. In addition, $g^{\bar{a}[A_{14}}g^{A_{15}]\bar{\alpha}}\raw g^{\bar{a}[a}g^{a]\bar{\alpha}}=0$ vanishes. Both arguments imply the absence of 2 further divergent contractions.

Next, we have to distinguish between the different types of flux contractions. In fact, for $G^{2}_{\bar{a} \bar{\alpha} \beta\gamma}$ or  $G^{2}_{\bar{\gamma} \bar{\alpha} \delta\beta}$ all inverse metrics in the last two lines of \eqref{eq:DivFZFluxH2R3} have to be on the fibre to find a divergent term in the limit $v_{f}\raw 0$. Those terms are summarised as:
\begin{align}
\dfrac{t_{8}t_{8} \mc{R}^3 G_4^2}{12}\biggl |_{(4,0)}&\supset 32 g_{\bar{a} a}(g^{\bar{a}a})^{2}\biggl \{+2G^{2}_{\bar{a} \bar{\alpha} \beta\gamma}\, g^{\bar{a}[A_{14}}g^{A_{15}]\bar{\alpha}}\,  g^{\beta\bar{A}_{16}}g^{\gamma\bar{A}_{17}}+2G^{2}_{\bar{\alpha} \bar{\beta} b\gamma}\,  g^{\bar{\alpha}A_{14}}g^{\bar{\beta}A_{15}} \, g^{b[\bar{A}_{16}}g^{\bar{A}_{17}]\gamma} \nn\\
&\quad+G^{2}_{\bar{\gamma} \bar{\alpha} \delta\beta}\,  g^{\bar{\gamma}A_{14}}g^{\bar{\alpha}A_{15}}\, g^{\delta\bar{A}_{16}}g^{\beta\bar{A}_{17}}\biggl \} \biggl [-R^{\bar{a}}\,_{\bar{a} a\bar{A}_{17}} R^{a}\,_{A_{14} \bar{a}a}R^{\bar{a}}\,_{\bar{A}_{16}\bar{a} A_{15}}\nn\\
&\quad+R^{\bar{a}}\,_{\bar{a} a\bar{A}_{17}}R^{a}\,_{a\bar{a} A_{14}}R^{a}\,_{a\bar{A}_{16}A_{15}}\biggl ]\, .
\end{align}
Given that there is a direct factor of the fibre metric and $\lbrace\ldots\rbrace$ involves at most a single inverse fibre metric, this means that $\lambda_{f}\leq 2$ and there are no dangerous contractions.

It remains to show the same for the flux contractions $G^{2}_{\bar{a} \bar{\alpha} b\beta}$. Among the 8 potentially dangerous terms, most can be excluded directly. Going through similar arguments as above and working at leading order in the $\vo$ expansion, one finds the following combination of two divergent contractions:
\begin{align}
\dfrac{t_{8}t_{8} \mc{R}^3 G_4^2}{12}\biggl |_{(4,0)}&\supset-64 G^{2}_{\bar{a} \bar{\alpha} b\beta}\,  g_{\bar{a}a}(g^{\bar{a}a})^{4}g^{\bar{\alpha}\delta} g^{\beta\bar{\gamma}} R^{a}\,_{a\bar{a} a}\biggl \{ 3R^{a}\,_{a \bar{a}a} R^{a}\,_{a\bar{\gamma}\delta}-  R^{\bar{a}}\,_{\bar{a} a\bar{\gamma}}R^{\bar{a}}\,_{\bar{a} \delta\bar{a}} \biggl \}\, .
\end{align}
This expression vanishes at leading order in the volume if either:
\begin{equation}
\label{eq:Condition2} 
R^{a}\,_{a\bar{a} a}=0+\cO\left (\dfrac{v_{f}}{v}\right )\quad\Rightarrow\quad \p_{\bar{a}} \p_{a}g_{a\bar{a}}=g^{a\bar{a}}\left (\p_{\bar{a}}g_{\bar{a}a}\right )\left (\p_{a}g_{a\bar{a}}\right )
\end{equation}
or:
\begin{align}
\label{eq:Condition1} 
R^{a}\,_{a\bar{\gamma}\delta}&=0+\cO\left (\dfrac{v_{f}}{v}\right )\quad\Rightarrow\quad\p_{\bar{\gamma}}\p_{\delta}g_{a\bar{a}}=g^{a\bar{a}}\left (\p_{\bar{\gamma}}g_{\bar{a}a}\right )\left (\p_{\delta}g_{a\bar{a}}\right )\nn\\
 R^{\bar{a}}\,_{\bar{a} \delta\bar{a}} &=0+\cO\left (\dfrac{v_{f}}{v}\right )\quad\Rightarrow\quad\p_{\bar{a}}\p_{\delta}g_{a\bar{a}}=g^{a\bar{a}} \left (\p_{\bar{a}}g_{\bar{a}a}\right )\left (\p_{\delta}g_{a\bar{a}}\right )\nn\\
R^{\bar{a}}\,_{\bar{a} a\bar{\gamma}}&=0+\cO\left (\dfrac{v_{f}}{v}\right )\quad\Rightarrow\quad \p_{\bar{\gamma}} \p_{a}g_{a\bar{a}}=g^{a\bar{a}}\left (\p_{\bar{\gamma}}g_{\bar{a}a}\right )\left (\p_{a}g_{a\bar{a}}\right )\,.
\end{align}
These are second order PDEs in the metric components which can in principle be solved for. Notice that \eqref{eq:Condition2} and the last two conditions of \eqref{eq:Condition1} are trivially satisfied for the ansatz \eqref{eq:KPAnsatzGSVY}. More generally, imposing \eqref{eq:Condition2} implies that there are no divergent terms coming from $\mc{R}^3 G_4^2$ and $\mc{R}^2 (\nabla G_{4})^2$, although there are divergent contractions of $\mc{R}^4$ and $\mc{R}^2 G_4^4$.
If we add \eqref{eq:Condition1}, then there are no divergent terms from $\mc{R}^4$, $\mc{R}^3 G_4^2$ and $\mc{R}^2 (\nabla G_{4})^2$, but there remains one divergent term in $\mc{R}^2 G_4^4$.

\section{Vacua from potential $(\ap)^1$ loop effects}
\label{AppB}

In Sec.~\ref{AbsenceOfAPOne} we have seen that the leading no-scale breaking effects at tree-level in $g_s$ should arise from $(\alpha')^3$ corrections. These effects scale as $\vo^{-3}$ and are crucial to give rise to LVS vacua \cite{Balasubramanian:2005zx,Cicoli:2008va}. Interestingly, $\vo^{-8/3}$ corrections cannot come from $(\alpha')^2$ 10D effects at any order in $g_s$ due to the extended no-scale structure \cite{vonGersdorff:2005bf,Berg:2007wt,Cicoli:2007xp,Cicoli:2008va} (see however \cite{Conlon:2009kt,Conlon:2010ji}), while they could emerge at tree-level from non-zero F-terms of matter fields, corresponding to T-brane uplifting contributions \cite{Cicoli:2015ylx}. Potentially dangerous $\vo^{-7/3}$ corrections can instead arise just at $\mc{O}((\alpha')^1)$ at string loop level $g_s^n$ with $n>0$. In this appendix we show that, if present an any order $n>0$, these corrections would not destroy LVS vacua but would lead to a new class of vacua with potentially interesting phenomenological properties.

\subsubsection*{Scalar potentials with $(\ap)^1$ loop corrections}

We focus on the simple model $X_3=\bC\bP^{4}_{[1,1,1,6,9]}[18]$ with 2 K\"ahler moduli and volume form:
\be
\vo = \frac{1}{9\sqrt{2}}\left(\tau_b^{3/2}-\tau_s^{3/2}\right).
\ee
The K\"ahler potential including $(\alpha')^k\, F^2$ corrections with $k=1,2,3$ reads:
\be
K = -2\ln\vo-\frac{\hat\alpha}{\vo^{\frac13}}-\frac{\hat\beta}{\vo^{\frac23}}-\frac{2\hat\xi}{\vo}-\ln\left(\frac{2}{g_s}\right)\,,
\label{eq:KP1} 
\ee
where, according to our previous discussion, $\hat\alpha=\alpha\,g_s^n/\sqrt{g_s}$ with $n>0$ (the other powers of $g_s$ can be identified from the scaling arguments of Sec.~\ref{TypeIIBscalings}), $\hat\beta=\beta/g_s$ and $\hat\xi=\xi/g_s^{3/2}$ with:\footnote{As discussed in \cite{Minasian:2015bxa}, the value of $\xi$ is corrected by contributions from O7-planes/D7-branes so that in the weak coupling limit $\chi(X_3)\raw \chi(X_3)+2\int_{X_3}\, D_{\text{O7}}^3$. Since one typically works with Fano bases in F-theory which have ample anti-canonical bundle, the integral contributes with a positive sign, see e.g. footnote 5 in \cite{Braun:2015pza}. Crucially, LVS requires $\xi>0$ and hence $\chi(X_3)<0$ which could be spoilt here, although this has not been observed in most examples discussed in the literature \cite{Cicoli:2011qg,Cicoli:2012vw,Louis:2012nb,Cicoli:2013mpa,Cicoli:2013zha,Cicoli:2013cha}.}
\be
\xi=-\frac{\zeta(3)\,\chi(X_3)}{4(2\pi)^3} = 0.654\qquad\text{for}\quad \chi(X_3)=-540\,.
\ee
The superpotential receives instead non-perturbative corrections associated with the blow-up mode:
\be
W= W_0 + A_s\,e^{- a_s T_s}\qquad\text{with}\quad a_s= 2\pi/N_s\,.
\ee
After setting the axion to its VEV, the scalar potential obtained from \eqref{eq:IIB:ScalarPotN1} and (\ref{eq:KP1}) in the limit where $\hat\alpha/\vo^{1/3}\ll 1$ and $a_s\tau_s\gg 1$ reads:
\be
V=\lambda_1 \frac{\sqrt{\tau_s}\,e^{-2a_s\tau_s}}{\vo}
-\lambda_2 W_0 \frac{\tau_s\, e^{-a_s\tau_s}}{\vo^2} 
-\lambda_3\,\frac{\hat\alpha W_0^2}{\vo^{7/3}} 
+\lambda_4\,\frac{\tilde\xi W_0^2}{\vo^3}\,,
\label{eq:ScalPotFullAP} 
\ee
with (setting $e^{K_{\rm cs}}=1$):
\be
\lambda_1=\frac{3\sqrt{2}}{g_s}\,\lambda_2^2 \,,\qquad \lambda_2=2 a_s A_s g_s \,,\qquad 
\lambda_3=\frac{g_s}{8}\,,\qquad \lambda_4=6\lambda_3\,,
\label{eq:CoeffLVSScalarPotential} 
\ee
and:
\be
\tilde\xi= \hat\xi -\frac{3}{32}\hat\alpha^3+\frac{5}{18} \hat\alpha  \hat\beta\,.
\label{eq:AssumptionMinimaAP1} 
\ee
Notice that the term $\sim \hat\beta/\cV^{8/3}$ is absent due to the extended no-scale structure, which is why $\hat\beta$ appears at leading order only inside $\tilde\xi$. Clearly, if $\hat\alpha\neq 0$, the usual balance of terms in LVS is destroyed. Of course, this does not mean that all hope is lost as we now discuss.

\subsubsection*{Minimisation}

We can derive simple conditions for the existence of minima of \eqref{eq:ScalPotFullAP} by requiring $\partial V/\partial \vo = \partial V/\partial\tau_s=0$ which leads to (in the $a_s\tau_s\gg 1$ limit):
\bea
\label{eq:EqAPMinV} 
&&\lambda_1  \vo^2 \sqrt{\tau_s}\,e^{- 2a_s\tau_s}-2\lambda_2 W_0  \vo\,\tau_s\,e^{- a_s\tau_s}
-\frac{\lambda_3 W_0^2}{3} \left (7\hat\alpha \vo^{2/3}-\tilde\xi \right) = 0\,, \\
&&\lambda_1 \vo^{4/3}  4 e^{- a_s\tau_s}-  2\lambda_2 W_0 \vo^{1/3} \sqrt{\tau_s}=0\,. 
\eea
The second equation is identical to the LVS condition:
\be
\vo =\frac{\lambda_2 W_0 \sqrt{\tau_s}}{2\lambda_1}\, e^{a_s\tau_s}\,.
\label{eq:MinVolAp1} 
\ee
Plugging this back into \eqref{eq:EqAPMinV} gives rise to:
\be
7\hat\alpha \vo^{2/3}- 54\tilde\xi + 3\sqrt{2}\,\tau_s^{3/2}  = 0\,.
\label{eq:SolMinWithAddAPCor} 
\ee
In the $\alpha\raw 0$ limit this relation reproduces the standard LVS result $\tau_s = (9\sqrt{2} \hat\xi)^{2/3}\sim 1/g_s$ \cite{Balasubramanian:2005zx}, while for $\alpha\neq 0$ we obtain:
\be
\vo = \left[\frac{1}{7\hat\alpha}\left(54\tilde\xi - 3\sqrt{2}\,\tau_s^{3/2}\right)\right]^{3/2}\,,
\label{eq:SolValVolMinAPCor}
\ee
showing that the volume at the minimum is not exponentially large anymore, unless $\hat\alpha\ll 1$. The stationary points of the full potential \eqref{eq:ScalPotFullAP} can be obtained by looking at the intersection between (\ref{eq:MinVolAp1}) and (\ref{eq:SolValVolMinAPCor}). In the remainder of this appendix, we discuss two classes of minima depending on the sign of $\alpha$.

\subsubsection*{AdS vacua for $\alpha>0$}

\begin{figure}[t!]
\centering
\includegraphics[scale=0.22]{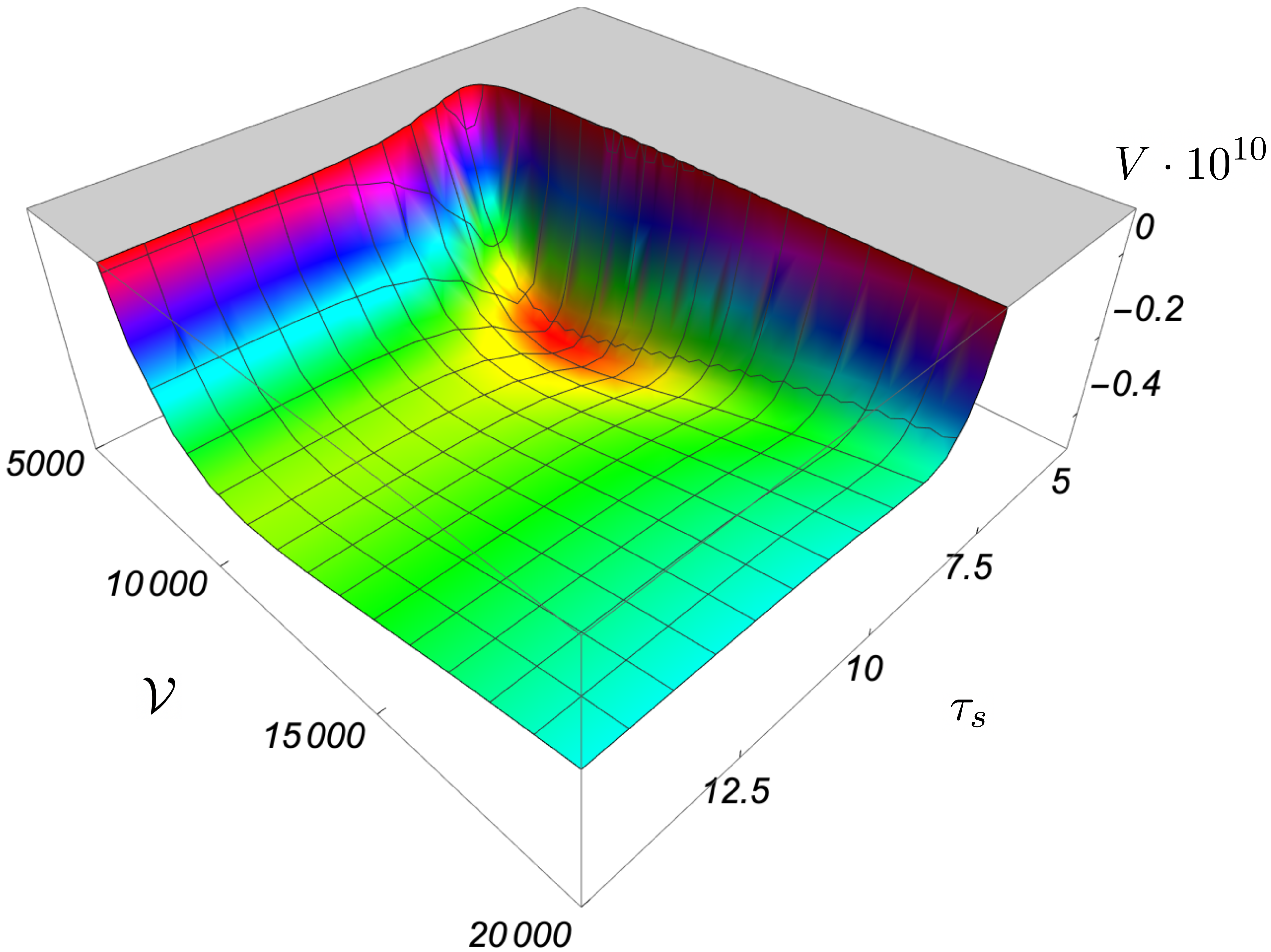}
\caption{New AdS minimum from $(\alpha')^1$ loop effects with $\alpha>0$ and parameters given by \eqref{eq:ParamsAP123Basic1}.}
\label{fig:AP123ApproxVolExample}
\end{figure}

We begin our analysis with explicit examples for $\alpha>0$. To find the values of $\cV$ and $\tau_s$ at the minimum, we compute the intersection of \eqref{eq:MinVolAp1} and \eqref{eq:SolValVolMinAPCor} numerically. For illustrative purposes, we focus on the following choice of underlying parameters: 
\be
g_s=0.1 \qquad W_0=1 \qquad A_s=1 \qquad N_s=5 \qquad n=1 \qquad \alpha=1 \qquad \beta=0 \,,
\label{eq:ParamsAP123Basic1} 
\ee
which yield the non-supersymmetric AdS minimum:
\be
\langle\tau_s\rangle=8.95 \qquad \langle\vo\rangle=1.0\cdot 10^4 \qquad V_{\rm AdS}=-4.09\cdot 10^{-11}\,.
\label{AdSMin}
\ee
The potential is shown in Fig.~\ref{fig:AP123ApproxVolExample}. The fact that the vacuum energy has to be negative can be easily inferred from the fact that the first 3 terms in (\ref{eq:ScalPotFullAP}) scale as $\vo^{-3}$ after substituting (\ref{eq:MinVolAp1}), while the last term scales as $\vo^{-7/3}$. Hence for $\vo\to \infty$ the potential approaches zero from below since the $\vo^{-7/3}$-term has a negative coefficient. Notice that the minimum (\ref{AdSMin}) satisfies our approximations since $\hat\alpha/\vo^{1/3}\simeq 0.015$ and $(a_s\tau_s)^{-1}\simeq 0.09$.

\subsubsection*{dS vacua for $\alpha<0$}

\begin{figure}[t!]
\centering
\includegraphics[scale=0.22]{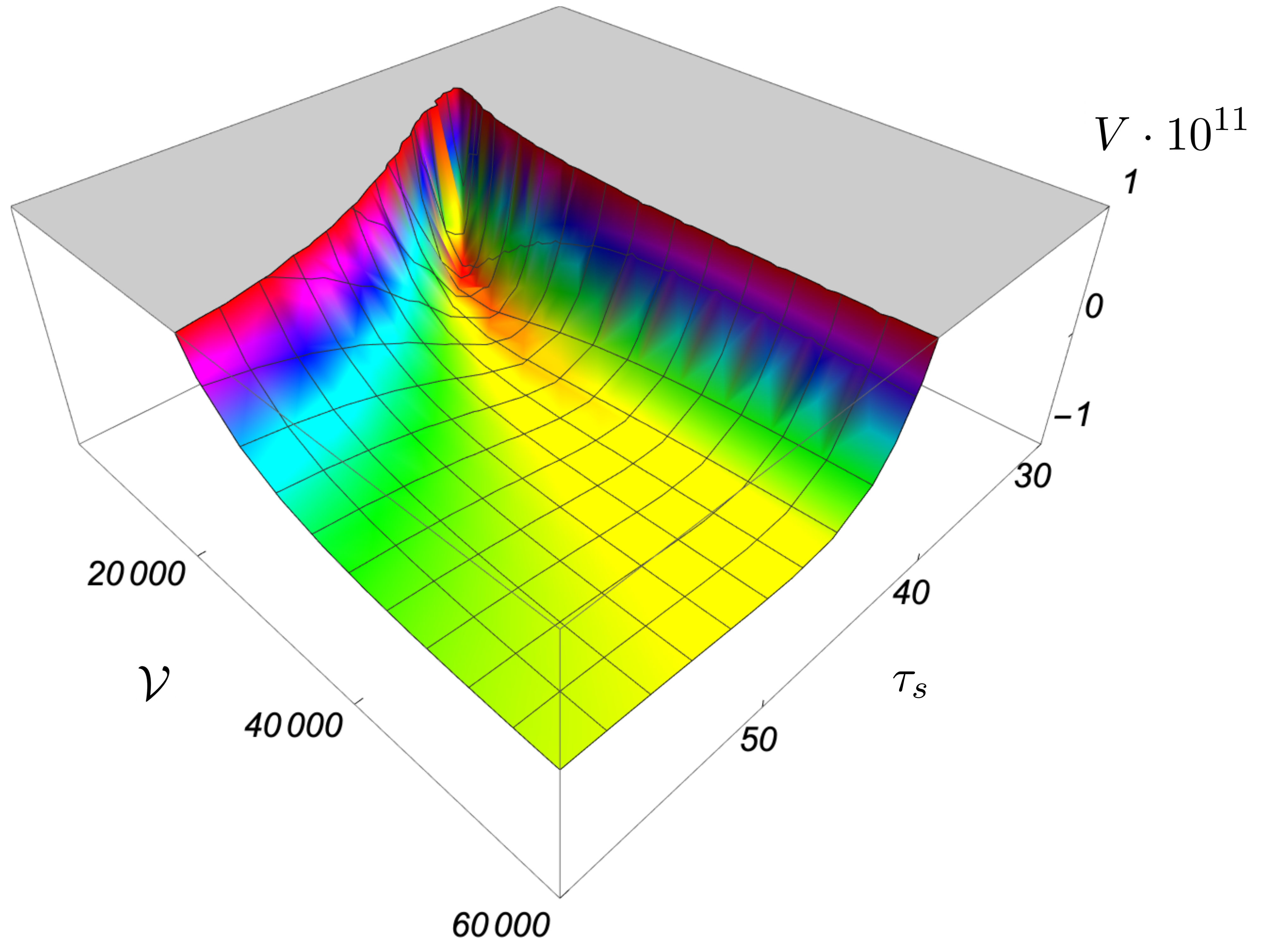}
\caption{New AdS minimum from $(\alpha')^1$ loop effects with $\alpha<0$ and parameter choice \eqref{eq:MinParChoiceNegAl1}.}
\label{Fig3}
\end{figure}

\begin{figure}[t!]
\centering
\includegraphics[scale=0.22]{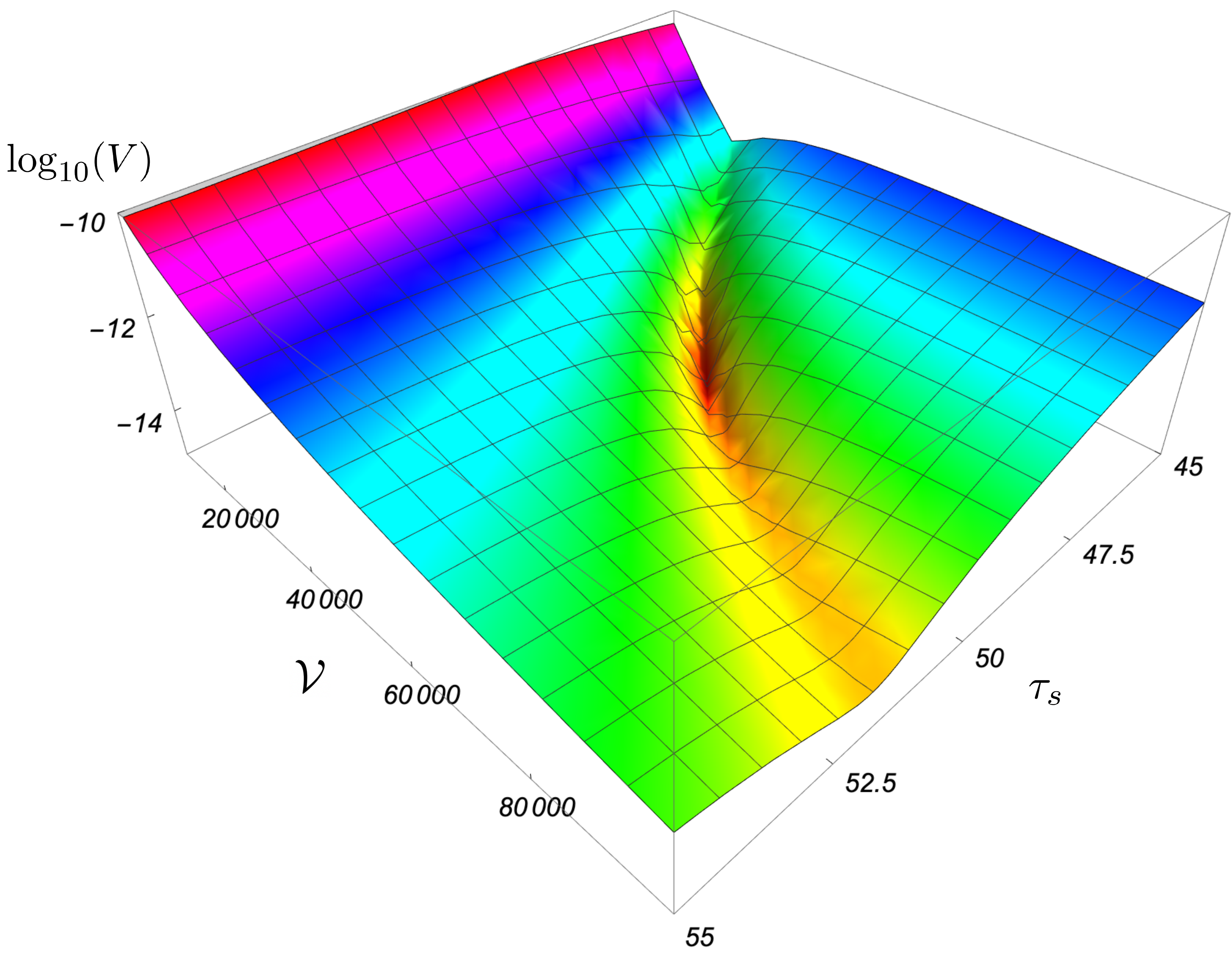}
\caption{New dS minimum from $(\alpha')^1$ loop effects with $\alpha<0$ and parameter choice \eqref{eq:MinParChoiceNegAl2}.}
\label{fig:AP123NegAlDI}
\end{figure}

Let us now analyse the parameter regime $\alpha<0$. In this case the minimum can be dS since the potential (\ref{eq:ScalPotFullAP}) approaches zero from above for $\vo\to\infty$ given that the coefficient of the $\vo^{-7/3}$-term is now positive. Hence different choices of the microscopic parameters can give rise to either an AdS or a dS minimum followed by a maximum (or better a saddle point from the 2-field perspective) at larger $\vo$-values. The presence of two stationary points can be verified numerically by the existence of two intersections between (\ref{eq:MinVolAp1}) and (\ref{eq:SolValVolMinAPCor}) for $\alpha<0$. Let us illustrate this situation with two choices of underlying parameters. In the first case we choose:
\be
g_s=0.1 \qquad W_0=1 \qquad A_s=1 \qquad N_s=30 \qquad n=3 \qquad \alpha=-2 \qquad \beta=0 \,,
\label{eq:MinParChoiceNegAl1} 
\ee
which yield:
\bea
\label{AdSMin2}
\text{AdS minimum:}\quad \langle\tau_s\rangle&=&43.47 \qquad \langle\vo\rangle=1.53\cdot 10^4 \qquad V_{\rm AdS}=-2.16\cdot 10^{-12} \\
\text{Saddle point:}\quad \langle\tau_s\rangle&=&69.64 \qquad \langle\vo\rangle=4.8\cdot 10^6\,.
\eea
In the second case we instead set:
\be
g_s=0.1 \qquad W_0=1 \qquad A_s=1 \qquad N_s=30 \qquad n=3 \qquad \alpha=-8.38 \qquad \beta=0 \,,
\label{eq:MinParChoiceNegAl2} 
\ee
which give a dS minimum at:
\be
\langle\tau_s\rangle =47.80 \qquad \langle\vo\rangle=4.0\cdot 10^4 \qquad V_{\rm dS}=3.06\cdot 10^{-15}\,.
\label{dSMin}
\ee
The two minima are shown respectively in Fig. \ref{Fig3} and \ref{fig:AP123NegAlDI} and are both in the regime where our approximations are trustable.

Finally it is worth stressing that this dS minimum exists only in a finely tuned regime of values for $\alpha$. Nonetheless, given the current debate about the existence of metastable dS vacua in string compactifications \cite{Obied:2018sgi,Cicoli:2018kdo}, it is important to find new examples of dS vacua which do not rely on additional sources as anti-branes. We have therefore found that potential $(\alpha')^1$ loop effects, rather than being a danger for moduli stabilisation, can provide new ways to achieve dS vacua at values of the CY volume which is not exponentially large, but still large enough to keep the EFT under control. In fact, even if we are balancing $(\alpha')^1$ against $(\alpha')^3$ corrections, the $\alpha'$ expansion is still under control due to the fact that $(\alpha')^1$ effects arise at loop-level while $(\alpha')^3$ terms are at tree-level in $g_s$.

\bibliographystyle{utphys}
\bibliography{Literatur}

\providecommand{\href}[2]{#2}\begingroup\raggedright\begin{thebibliography}{100}

\bibitem{Burgess:2020tbq}
C.~P. Burgess, \href{http://dx.doi.org/10.1017/9781139048040}{{\em
  {Introduction to Effective Field Theory}}}.
\newblock Cambridge University Press, 12, 2020.

\bibitem{Vafa:2005ui}
C.~Vafa, ``{The String landscape and the swampland},''
  \href{http://arxiv.org/abs/hep-th/0509212}{{\ttfamily arXiv:hep-th/0509212}}.

\bibitem{Dine:1985he}
M.~Dine and N.~Seiberg, ``{Is the Superstring Weakly Coupled?},''
  \href{http://dx.doi.org/10.1016/0370-2693(85)90927-X}{{\em Phys. Lett. B}
  {\bfseries 162} (1985) 299--302}.

\bibitem{Balasubramanian:2005zx}
V.~Balasubramanian, P.~Berglund, J.~P. Conlon, and F.~Quevedo, ``{Systematics
  of moduli stabilisation in Calabi-Yau flux compactifications},''
  \href{http://dx.doi.org/10.1088/1126-6708/2005/03/007}{{\em JHEP} {\bfseries
  03} (2005) 007}, \href{http://arxiv.org/abs/hep-th/0502058}{{\ttfamily
  arXiv:hep-th/0502058}}.

\bibitem{Conlon:2005ki}
J.~P. Conlon, F.~Quevedo, and K.~Suruliz, ``{Large-volume flux
  compactifications: Moduli spectrum and D3/D7 soft supersymmetry breaking},''
  \href{http://dx.doi.org/10.1088/1126-6708/2005/08/007}{{\em JHEP} {\bfseries
  08} (2005) 007}, \href{http://arxiv.org/abs/hep-th/0505076}{{\ttfamily
  arXiv:hep-th/0505076}}.

\bibitem{Burgess:2001vr}
C.~P. Burgess, P.~Martineau, F.~Quevedo, G.~Rajesh, and R.~J. Zhang, ``{Brane -
  anti-brane inflation in orbifold and orientifold models},''
  \href{http://dx.doi.org/10.1088/1126-6708/2002/03/052}{{\em JHEP} {\bfseries
  03} (2002) 052}, \href{http://arxiv.org/abs/hep-th/0111025}{{\ttfamily
  arXiv:hep-th/0111025}}.

\bibitem{Conlon:2005jm}
J.~P. Conlon and F.~Quevedo, ``{Kahler moduli inflation},''
  \href{http://dx.doi.org/10.1088/1126-6708/2006/01/146}{{\em JHEP} {\bfseries
  01} (2006) 146}, \href{http://arxiv.org/abs/hep-th/0509012}{{\ttfamily
  arXiv:hep-th/0509012}}.

\bibitem{Cicoli:2008gp}
M.~Cicoli, C.~P. Burgess, and F.~Quevedo, ``{Fibre Inflation: Observable
  Gravity Waves from IIB String Compactifications},''
  \href{http://dx.doi.org/10.1088/1475-7516/2009/03/013}{{\em JCAP} {\bfseries
  03} (2009) 013}, \href{http://arxiv.org/abs/0808.0691}{{\ttfamily
  arXiv:0808.0691 [hep-th]}}.

\bibitem{Cicoli:2011ct}
M.~Cicoli, F.~G. Pedro, and G.~Tasinato, ``{Poly-instanton Inflation},''
  \href{http://dx.doi.org/10.1088/1475-7516/2011/12/022}{{\em JCAP} {\bfseries
  12} (2011) 022}, \href{http://arxiv.org/abs/1110.6182}{{\ttfamily
  arXiv:1110.6182 [hep-th]}}.

\bibitem{Burgess:2013sla}
C.~P. Burgess, M.~Cicoli, and F.~Quevedo, ``{String Inflation After Planck
  2013},'' \href{http://dx.doi.org/10.1088/1475-7516/2013/11/003}{{\em JCAP}
  {\bfseries 11} (2013) 003}, \href{http://arxiv.org/abs/1306.3512}{{\ttfamily
  arXiv:1306.3512 [hep-th]}}.

\bibitem{Broy:2015zba}
B.~J. Broy, D.~Ciupke, F.~G. Pedro, and A.~Westphal, ``{Starobinsky-Type
  Inflation from $\alpha'$-Corrections},''
  \href{http://dx.doi.org/10.1088/1475-7516/2016/01/001}{{\em JCAP} {\bfseries
  01} (2016) 001}, \href{http://arxiv.org/abs/1509.00024}{{\ttfamily
  arXiv:1509.00024 [hep-th]}}.

\bibitem{Cicoli:2016chb}
M.~Cicoli, D.~Ciupke, S.~de~Alwis, and F.~Muia, ``{$\alpha'$ Inflation: moduli
  stabilisation and observable tensors from higher derivatives},''
  \href{http://dx.doi.org/10.1007/JHEP09(2016)026}{{\em JHEP} {\bfseries 09}
  (2016) 026}, \href{http://arxiv.org/abs/1607.01395}{{\ttfamily
  arXiv:1607.01395 [hep-th]}}.

\bibitem{Burgess:2014tja}
C.~P. Burgess, M.~Cicoli, F.~Quevedo, and M.~Williams, ``{Inflating with Large
  Effective Fields},''
  \href{http://dx.doi.org/10.1088/1475-7516/2014/11/045}{{\em JCAP} {\bfseries
  11} (2014) 045}, \href{http://arxiv.org/abs/1404.6236}{{\ttfamily
  arXiv:1404.6236 [hep-th]}}.

\bibitem{Burgess:2016owb}
C.~P. Burgess, M.~Cicoli, S.~de~Alwis, and F.~Quevedo, ``{Robust Inflation from
  Fibrous Strings},''
  \href{http://dx.doi.org/10.1088/1475-7516/2016/05/032}{{\em JCAP} {\bfseries
  05} (2016) 032}, \href{http://arxiv.org/abs/1603.06789}{{\ttfamily
  arXiv:1603.06789 [hep-th]}}.

\bibitem{Burgess:2020qsc}
C.~P. Burgess, M.~Cicoli, D.~Ciupke, S.~Krippendorf, and F.~Quevedo, ``{UV
  Shadows in EFTs: Accidental Symmetries, Robustness and No-Scale
  Supergravity},'' \href{http://dx.doi.org/10.1002/prop.202000076}{{\em
  Fortsch. Phys.} {\bfseries 68} no.~10, (2020) 2000076},
  \href{http://arxiv.org/abs/2006.06694}{{\ttfamily arXiv:2006.06694
  [hep-th]}}.

\bibitem{Aparicio:2014wxa}
L.~Aparicio, M.~Cicoli, S.~Krippendorf, A.~Maharana, F.~Muia, and F.~Quevedo,
  ``{Sequestered de Sitter String Scenarios: Soft-terms},''
  \href{http://dx.doi.org/10.1007/JHEP11(2014)071}{{\em JHEP} {\bfseries 11}
  (2014) 071}, \href{http://arxiv.org/abs/1409.1931}{{\ttfamily arXiv:1409.1931
  [hep-th]}}.

\bibitem{Cicoli:2012sz}
M.~Cicoli, M.~Goodsell, and A.~Ringwald, ``{The type IIB string axiverse and
  its low-energy phenomenology},''
  \href{http://dx.doi.org/10.1007/JHEP10(2012)146}{{\em JHEP} {\bfseries 10}
  (2012) 146}, \href{http://arxiv.org/abs/1206.0819}{{\ttfamily arXiv:1206.0819
  [hep-th]}}.

\bibitem{Reece:2015qbf}
M.~Reece and W.~Xue, ``{SUSY\textquoteright{}s Ladder: reframing sequestering
  at Large Volume},'' \href{http://dx.doi.org/10.1007/JHEP04(2016)045}{{\em
  JHEP} {\bfseries 04} (2016) 045},
  \href{http://arxiv.org/abs/1512.04941}{{\ttfamily arXiv:1512.04941
  [hep-ph]}}.

\bibitem{Becker:2002nn}
K.~Becker, M.~Becker, M.~Haack, and J.~Louis, ``{Supersymmetry breaking and
  alpha-prime corrections to flux induced potentials},''
  \href{http://dx.doi.org/10.1088/1126-6708/2002/06/060}{{\em JHEP} {\bfseries
  06} (2002) 060}, \href{http://arxiv.org/abs/hep-th/0204254}{{\ttfamily
  arXiv:hep-th/0204254}}.

\bibitem{Berg:2005ja}
M.~Berg, M.~Haack, and B.~Kors, ``{String loop corrections to Kahler potentials
  in orientifolds},''
  \href{http://dx.doi.org/10.1088/1126-6708/2005/11/030}{{\em JHEP} {\bfseries
  11} (2005) 030}, \href{http://arxiv.org/abs/hep-th/0508043}{{\ttfamily
  arXiv:hep-th/0508043}}.

\bibitem{Berg:2007wt}
M.~Berg, M.~Haack, and E.~Pajer, ``{Jumping Through Loops: On Soft Terms from
  Large Volume Compactifications},''
  \href{http://dx.doi.org/10.1088/1126-6708/2007/09/031}{{\em JHEP} {\bfseries
  09} (2007) 031}, \href{http://arxiv.org/abs/0704.0737}{{\ttfamily
  arXiv:0704.0737 [hep-th]}}.

\bibitem{vonGersdorff:2005bf}
G.~von Gersdorff and A.~Hebecker, ``{Kahler corrections for the volume modulus
  of flux compactifications},''
  \href{http://dx.doi.org/10.1016/j.physletb.2005.08.024}{{\em Phys. Lett. B}
  {\bfseries 624} (2005) 270--274},
  \href{http://arxiv.org/abs/hep-th/0507131}{{\ttfamily arXiv:hep-th/0507131}}.

\bibitem{GarciaEtxebarria:2012zm}
I.~Garcia-Etxebarria, H.~Hayashi, R.~Savelli, and G.~Shiu, ``{On quantum
  corrected Kahler potentials in F-theory},''
  \href{http://dx.doi.org/10.1007/JHEP03(2013)005}{{\em JHEP} {\bfseries 03}
  (2013) 005}, \href{http://arxiv.org/abs/1212.4831}{{\ttfamily arXiv:1212.4831
  [hep-th]}}.

\bibitem{Cicoli:2007xp}
M.~Cicoli, J.~P. Conlon, and F.~Quevedo, ``{Systematics of String Loop
  Corrections in Type IIB Calabi-Yau Flux Compactifications},''
  \href{http://dx.doi.org/10.1088/1126-6708/2008/01/052}{{\em JHEP} {\bfseries
  01} (2008) 052}, \href{http://arxiv.org/abs/0708.1873}{{\ttfamily
  arXiv:0708.1873 [hep-th]}}.

\bibitem{Bonetti:2016dqh}
F.~Bonetti and M.~Weissenbacher, ``{The Euler characteristic correction to the
  K\"ahler potential \textemdash{} revisited},''
  \href{http://dx.doi.org/10.1007/JHEP01(2017)003}{{\em JHEP} {\bfseries 01}
  (2017) 003}, \href{http://arxiv.org/abs/1608.01300}{{\ttfamily
  arXiv:1608.01300 [hep-th]}}.

\bibitem{Ciupke:2015msa}
D.~Ciupke, J.~Louis, and A.~Westphal, ``{Higher-Derivative Supergravity and
  Moduli Stabilization},''
  \href{http://dx.doi.org/10.1007/JHEP10(2015)094}{{\em JHEP} {\bfseries 10}
  (2015) 094}, \href{http://arxiv.org/abs/1505.03092}{{\ttfamily
  arXiv:1505.03092 [hep-th]}}.

\bibitem{Grimm:2017okk}
T.~W. Grimm, K.~Mayer, and M.~Weissenbacher, ``{Higher derivatives in Type II
  and M-theory on Calabi-Yau threefolds},''
  \href{http://dx.doi.org/10.1007/JHEP02(2018)127}{{\em JHEP} {\bfseries 02}
  (2018) 127}, \href{http://arxiv.org/abs/1702.08404}{{\ttfamily
  arXiv:1702.08404 [hep-th]}}.

\bibitem{Minasian:2015bxa}
R.~Minasian, T.~G. Pugh, and R.~Savelli, ``{F-theory at order $\alpha'^3$},''
  \href{http://dx.doi.org/10.1007/JHEP10(2015)050}{{\em JHEP} {\bfseries 10}
  (2015) 050}, \href{http://arxiv.org/abs/1506.06756}{{\ttfamily
  arXiv:1506.06756 [hep-th]}}.

\bibitem{Berg:2014ama}
M.~Berg, M.~Haack, J.~U. Kang, and S.~Sj\"ors, ``{Towards the one-loop K\"ahler
  metric of Calabi-Yau orientifolds},''
  \href{http://dx.doi.org/10.1007/JHEP12(2014)077}{{\em JHEP} {\bfseries 12}
  (2014) 077}, \href{http://arxiv.org/abs/1407.0027}{{\ttfamily arXiv:1407.0027
  [hep-th]}}.

\bibitem{Haack:2015pbv}
M.~Haack and J.~U. Kang, ``{One-loop Einstein-Hilbert term in minimally
  supersymmetric type IIB orientifolds},''
  \href{http://dx.doi.org/10.1007/JHEP02(2016)160}{{\em JHEP} {\bfseries 02}
  (2016) 160}, \href{http://arxiv.org/abs/1511.03957}{{\ttfamily
  arXiv:1511.03957 [hep-th]}}.

\bibitem{Haack:2018ufg}
M.~Haack and J.~U. Kang, ``{Field redefinitions and K\"ahler potential in
  string theory at 1-loop},''
  \href{http://dx.doi.org/10.1007/JHEP08(2018)019}{{\em JHEP} {\bfseries 08}
  (2018) 019}, \href{http://arxiv.org/abs/1805.00817}{{\ttfamily
  arXiv:1805.00817 [hep-th]}}.

\bibitem{Antoniadis:2018hqy}
I.~Antoniadis, Y.~Chen, and G.~K. Leontaris, ``{Perturbative moduli
  stabilisation in type IIB/F-theory framework},''
  \href{http://dx.doi.org/10.1140/epjc/s10052-018-6248-4}{{\em Eur. Phys. J. C}
  {\bfseries 78} no.~9, (2018) 766},
  \href{http://arxiv.org/abs/1803.08941}{{\ttfamily arXiv:1803.08941
  [hep-th]}}.

\bibitem{Grimm:2013gma}
T.~W. Grimm, R.~Savelli, and M.~Weissenbacher, ``{On \textbackslash{}alpha'
  corrections in N=1 F-theory compactifications},''
  \href{http://dx.doi.org/10.1016/j.physletb.2013.07.024}{{\em Phys. Lett. B}
  {\bfseries 725} (2013) 431--436},
  \href{http://arxiv.org/abs/1303.3317}{{\ttfamily arXiv:1303.3317 [hep-th]}}.

\bibitem{Grimm:2013bha}
T.~W. Grimm, J.~Keitel, R.~Savelli, and M.~Weissenbacher, ``{From M-theory
  higher curvature terms to $\alpha'$ corrections in F-theory},''
  \href{http://dx.doi.org/10.1016/j.nuclphysb.2015.12.011}{{\em Nucl. Phys. B}
  {\bfseries 903} (2016) 325--359},
  \href{http://arxiv.org/abs/1312.1376}{{\ttfamily arXiv:1312.1376 [hep-th]}}.

\bibitem{Junghans:2014zla}
D.~Junghans and G.~Shiu, ``{Brane curvature corrections to the $ \mathcal{N} =$
  1 type II/F-theory effective action},''
  \href{http://dx.doi.org/10.1007/JHEP03(2015)107}{{\em JHEP} {\bfseries 03}
  (2015) 107}, \href{http://arxiv.org/abs/1407.0019}{{\ttfamily arXiv:1407.0019
  [hep-th]}}.

\bibitem{Weissenbacher:2019mef}
M.~Weissenbacher, ``{F-theory vacua and $\alpha'$-corrections},''
  \href{http://dx.doi.org/10.1007/JHEP04(2020)032}{{\em JHEP} {\bfseries 04}
  (2020) 032}, \href{http://arxiv.org/abs/1901.04758}{{\ttfamily
  arXiv:1901.04758 [hep-th]}}.

\bibitem{Klaewer:2020lfg}
D.~Klaewer, S.-J. Lee, T.~Weigand, and M.~Wiesner, ``{Quantum corrections in 4d
  $N$ = 1 infinite distance limits and the weak gravity conjecture},''
  \href{http://dx.doi.org/10.1007/JHEP03(2021)252}{{\em JHEP} {\bfseries 03}
  (2021) 252}, \href{http://arxiv.org/abs/2011.00024}{{\ttfamily
  arXiv:2011.00024 [hep-th]}}.

\bibitem{Kachru:2003aw}
S.~Kachru, R.~Kallosh, A.~D. Linde, and S.~P. Trivedi, ``{De Sitter vacua in
  string theory},'' \href{http://dx.doi.org/10.1103/PhysRevD.68.046005}{{\em
  Phys. Rev. D} {\bfseries 68} (2003) 046005},
  \href{http://arxiv.org/abs/hep-th/0301240}{{\ttfamily arXiv:hep-th/0301240}}.

\bibitem{Cicoli:2008va}
M.~Cicoli, J.~P. Conlon, and F.~Quevedo, ``{General Analysis of LARGE Volume
  Scenarios with String Loop Moduli Stabilisation},''
  \href{http://dx.doi.org/10.1088/1126-6708/2008/10/105}{{\em JHEP} {\bfseries
  10} (2008) 105}, \href{http://arxiv.org/abs/0805.1029}{{\ttfamily
  arXiv:0805.1029 [hep-th]}}.

\bibitem{AbdusSalam:2020ywo}
S.~AbdusSalam, S.~Abel, M.~Cicoli, F.~Quevedo, and P.~Shukla, ``{A systematic
  approach to K\"ahler moduli stabilisation},''
  \href{http://dx.doi.org/10.1007/JHEP08(2020)047}{{\em JHEP} {\bfseries 08}
  no.~08, (2020) 047}, \href{http://arxiv.org/abs/2005.11329}{{\ttfamily
  arXiv:2005.11329 [hep-th]}}.

\bibitem{Haack:1999zv}
M.~Haack and J.~Louis, ``{Duality in heterotic vacua with four supercharges},''
  \href{http://dx.doi.org/10.1016/S0550-3213(00)00091-2}{{\em Nucl. Phys. B}
  {\bfseries 575} (2000) 107--133},
  \href{http://arxiv.org/abs/hep-th/9912181}{{\ttfamily arXiv:hep-th/9912181}}.

\bibitem{Haack:2001jz}
M.~Haack and J.~Louis, ``{M theory compactified on Calabi-Yau fourfolds with
  background flux},''
  \href{http://dx.doi.org/10.1016/S0370-2693(01)00464-6}{{\em Phys. Lett. B}
  {\bfseries 507} (2001) 296--304},
  \href{http://arxiv.org/abs/hep-th/0103068}{{\ttfamily arXiv:hep-th/0103068}}.

\bibitem{Berg:2002es}
M.~Berg, M.~Haack, and H.~Samtleben, ``{Calabi-Yau fourfolds with flux and
  supersymmetry breaking},''
  \href{http://dx.doi.org/10.1088/1126-6708/2003/04/046}{{\em JHEP} {\bfseries
  04} (2003) 046}, \href{http://arxiv.org/abs/hep-th/0212255}{{\ttfamily
  arXiv:hep-th/0212255}}.

\bibitem{deWit:2004yr}
B.~de~Wit, H.~Nicolai, and H.~Samtleben, ``{Gauged supergravities in
  three-dimensions: A Panoramic overview},''
  \href{http://dx.doi.org/10.22323/1.011.0016}{{\em PoS} {\bfseries jhw2003}
  (2003) 016}, \href{http://arxiv.org/abs/hep-th/0403014}{{\ttfamily
  arXiv:hep-th/0403014}}.

\bibitem{Grimm:2011tb}
T.~W. Grimm, M.~Kerstan, E.~Palti, and T.~Weigand, ``{Massive Abelian Gauge
  Symmetries and Fluxes in F-theory},''
  \href{http://dx.doi.org/10.1007/JHEP12(2011)004}{{\em JHEP} {\bfseries 12}
  (2011) 004}, \href{http://arxiv.org/abs/1107.3842}{{\ttfamily arXiv:1107.3842
  [hep-th]}}.

\bibitem{Vafa:1996xn}
C.~Vafa, ``{Evidence for F theory},''
  \href{http://dx.doi.org/10.1016/0550-3213(96)00172-1}{{\em Nucl. Phys. B}
  {\bfseries 469} (1996) 403--418},
  \href{http://arxiv.org/abs/hep-th/9602022}{{\ttfamily arXiv:hep-th/9602022}}.

\bibitem{Morrison:1996na}
D.~R. Morrison and C.~Vafa, ``{Compactifications of F theory on Calabi-Yau
  threefolds. 1},'' \href{http://dx.doi.org/10.1016/0550-3213(96)00242-8}{{\em
  Nucl. Phys. B} {\bfseries 473} (1996) 74--92},
  \href{http://arxiv.org/abs/hep-th/9602114}{{\ttfamily arXiv:hep-th/9602114}}.

\bibitem{Morrison:1996pp}
D.~R. Morrison and C.~Vafa, ``{Compactifications of F theory on Calabi-Yau
  threefolds. 2.},'' \href{http://dx.doi.org/10.1016/0550-3213(96)00369-0}{{\em
  Nucl. Phys. B} {\bfseries 476} (1996) 437--469},
  \href{http://arxiv.org/abs/hep-th/9603161}{{\ttfamily arXiv:hep-th/9603161}}.

\bibitem{Denef:2008wq}
F.~Denef, ``{Les Houches Lectures on Constructing String Vacua},'' {\em Les
  Houches} {\bfseries 87} (2008) 483--610,
  \href{http://arxiv.org/abs/0803.1194}{{\ttfamily arXiv:0803.1194 [hep-th]}}.

\bibitem{Grimm:2010ks}
T.~W. Grimm, ``{The N=1 effective action of F-theory compactifications},''
  \href{http://dx.doi.org/10.1016/j.nuclphysb.2010.11.018}{{\em Nucl. Phys. B}
  {\bfseries 845} (2011) 48--92},
  \href{http://arxiv.org/abs/1008.4133}{{\ttfamily arXiv:1008.4133 [hep-th]}}.

\bibitem{Weigand:2018rez}
T.~Weigand, ``{F-theory},'' {\em PoS} {\bfseries TASI2017} (2018) 016,
  \href{http://arxiv.org/abs/1806.01854}{{\ttfamily arXiv:1806.01854
  [hep-th]}}.

\bibitem{Sen:1997gv}
A.~Sen, ``{Orientifold limit of F theory vacua},''
  \href{http://dx.doi.org/10.1103/PhysRevD.55.R7345}{{\em Phys. Rev. D}
  {\bfseries 55} (1997) R7345--R7349},
  \href{http://arxiv.org/abs/hep-th/9702165}{{\ttfamily arXiv:hep-th/9702165}}.

\bibitem{Giveon:1994fu}
A.~Giveon, M.~Porrati, and E.~Rabinovici, ``{Target space duality in string
  theory},'' \href{http://dx.doi.org/10.1016/0370-1573(94)90070-1}{{\em Phys.
  Rept.} {\bfseries 244} (1994) 77--202},
  \href{http://arxiv.org/abs/hep-th/9401139}{{\ttfamily arXiv:hep-th/9401139}}.

\bibitem{Sen:2001di}
A.~Sen, ``{An introduction to duality symmetries in string theory},'' in {\em
  {Les Houches Summer School: Session 76: Euro Summer School on Unity of
  Fundamental Physics: Gravity, Gauge Theory and Strings}}.
\newblock 7, 2001.

\bibitem{Dijkgraaf:1997vv}
R.~Dijkgraaf, E.~P. Verlinde, and H.~L. Verlinde, ``{Matrix string theory},''
  \href{http://dx.doi.org/10.1016/S0550-3213(97)00326-X}{{\em Nucl. Phys. B}
  {\bfseries 500} (1997) 43--61},
  \href{http://arxiv.org/abs/hep-th/9703030}{{\ttfamily arXiv:hep-th/9703030}}.

\bibitem{Cicoli:2013swa}
M.~Cicoli, J.~P. Conlon, A.~Maharana, and F.~Quevedo, ``{A Note on the
  Magnitude of the Flux Superpotential},''
  \href{http://dx.doi.org/10.1007/JHEP01(2014)027}{{\em JHEP} {\bfseries 01}
  (2014) 027}, \href{http://arxiv.org/abs/1310.6694}{{\ttfamily arXiv:1310.6694
  [hep-th]}}.

\bibitem{Damour:2005zb}
T.~Damour and H.~Nicolai, ``{Higher order M theory corrections and the
  Kac-Moody algebra E(10)},''
  \href{http://dx.doi.org/10.1088/0264-9381/22/14/003}{{\em Class. Quant.
  Grav.} {\bfseries 22} (2005) 2849--2880},
  \href{http://arxiv.org/abs/hep-th/0504153}{{\ttfamily arXiv:hep-th/0504153}}.

\bibitem{Polchinski:1998rr}
J.~Polchinski, \href{http://dx.doi.org/10.1017/CBO9780511618123}{{\em {String
  theory. Vol. 2: Superstring theory and beyond}}}.
\newblock Cambridge Monographs on Mathematical Physics. Cambridge University
  Press, 12, 2007.

\bibitem{Peeters:2001ub}
K.~Peeters, P.~Vanhove, and A.~Westerberg, ``{Chiral splitting and world sheet
  gravitinos in higher derivative string amplitudes},''
  \href{http://dx.doi.org/10.1088/0264-9381/19/10/312}{{\em Class. Quant.
  Grav.} {\bfseries 19} (2002) 2699--2716},
  \href{http://arxiv.org/abs/hep-th/0112157}{{\ttfamily arXiv:hep-th/0112157}}.

\bibitem{Richards:2008jg}
D.~M. Richards, ``{The One-Loop Five-Graviton Amplitude and the Effective
  Action},'' \href{http://dx.doi.org/10.1088/1126-6708/2008/10/042}{{\em JHEP}
  {\bfseries 10} (2008) 042}, \href{http://arxiv.org/abs/0807.2421}{{\ttfamily
  arXiv:0807.2421 [hep-th]}}.

\bibitem{Richards:2008sa}
D.~M. Richards, ``{The One-Loop H**2 R**3 and H**2(Delta H)R-2 Terms in the
  Effective Action},''
  \href{http://dx.doi.org/10.1088/1126-6708/2008/10/043}{{\em JHEP} {\bfseries
  10} (2008) 043}, \href{http://arxiv.org/abs/0807.3453}{{\ttfamily
  arXiv:0807.3453 [hep-th]}}.

\bibitem{Liu:2013dna}
J.~T. Liu and R.~Minasian, ``{Higher-derivative couplings in string theory:
  dualities and the B-field},''
  \href{http://dx.doi.org/10.1016/j.nuclphysb.2013.06.002}{{\em Nucl. Phys. B}
  {\bfseries 874} (2013) 413--470},
  \href{http://arxiv.org/abs/1304.3137}{{\ttfamily arXiv:1304.3137 [hep-th]}}.

\bibitem{Liu:2019ses}
J.~T. Liu and R.~Minasian, ``{Higher-derivative couplings in string theory:
  five-point contact terms},''
  \href{http://dx.doi.org/10.1016/j.nuclphysb.2021.115386}{{\em Nucl. Phys. B}
  {\bfseries 967} (2021) 115386},
  \href{http://arxiv.org/abs/1912.10974}{{\ttfamily arXiv:1912.10974
  [hep-th]}}.

\bibitem{Policastro:2006vt}
G.~Policastro and D.~Tsimpis, ``{R**4, purified},''
  \href{http://dx.doi.org/10.1088/0264-9381/23/14/012}{{\em Class. Quant.
  Grav.} {\bfseries 23} (2006) 4753--4780},
  \href{http://arxiv.org/abs/hep-th/0603165}{{\ttfamily arXiv:hep-th/0603165}}.

\bibitem{Policastro:2008hg}
G.~Policastro and D.~Tsimpis, ``{A Note on the quartic effective action of type
  IIB superstring},''
  \href{http://dx.doi.org/10.1088/0264-9381/26/12/125001}{{\em Class. Quant.
  Grav.} {\bfseries 26} (2009) 125001},
  \href{http://arxiv.org/abs/0812.3138}{{\ttfamily arXiv:0812.3138 [hep-th]}}.

\bibitem{Green:1997tv}
M.~B. Green and M.~Gutperle, ``{Effects of D instantons},''
  \href{http://dx.doi.org/10.1016/S0550-3213(97)00269-1}{{\em Nucl. Phys. B}
  {\bfseries 498} (1997) 195--227},
  \href{http://arxiv.org/abs/hep-th/9701093}{{\ttfamily arXiv:hep-th/9701093}}.

\bibitem{Green:1997di}
M.~B. Green and P.~Vanhove, ``{D instantons, strings and M theory},''
  \href{http://dx.doi.org/10.1016/S0370-2693(97)00785-5}{{\em Phys. Lett. B}
  {\bfseries 408} (1997) 122--134},
  \href{http://arxiv.org/abs/hep-th/9704145}{{\ttfamily arXiv:hep-th/9704145}}.

\bibitem{Green:1997as}
M.~B. Green, M.~Gutperle, and P.~Vanhove, ``{One loop in eleven-dimensions},''
  \href{http://dx.doi.org/10.1016/S0370-2693(97)00931-3}{{\em Phys. Lett. B}
  {\bfseries 409} (1997) 177--184},
  \href{http://arxiv.org/abs/hep-th/9706175}{{\ttfamily arXiv:hep-th/9706175}}.

\bibitem{Green:1997tn}
M.~B. Green and M.~Gutperle, ``{D Particle bound states and the D instanton
  measure},'' \href{http://dx.doi.org/10.1088/1126-6708/1998/01/005}{{\em JHEP}
  {\bfseries 01} (1998) 005},
  \href{http://arxiv.org/abs/hep-th/9711107}{{\ttfamily arXiv:hep-th/9711107}}.

\bibitem{Green:1998yf}
M.~B. Green and M.~Gutperle, ``{D instanton partition functions},''
  \href{http://dx.doi.org/10.1103/PhysRevD.58.046007}{{\em Phys. Rev. D}
  {\bfseries 58} (1998) 046007},
  \href{http://arxiv.org/abs/hep-th/9804123}{{\ttfamily arXiv:hep-th/9804123}}.

\bibitem{Green:1999by}
M.~B. Green, M.~Gutperle, and H.~H. Kwon, ``{Light cone quantum mechanics of
  the eleven-dimensional superparticle},''
  \href{http://dx.doi.org/10.1088/1126-6708/1999/08/012}{{\em JHEP} {\bfseries
  08} (1999) 012}, \href{http://arxiv.org/abs/hep-th/9907155}{{\ttfamily
  arXiv:hep-th/9907155}}.

\bibitem{Green:1998by}
M.~B. Green and S.~Sethi, ``{Supersymmetry constraints on type IIB
  supergravity},'' \href{http://dx.doi.org/10.1103/PhysRevD.59.046006}{{\em
  Phys. Rev. D} {\bfseries 59} (1999) 046006},
  \href{http://arxiv.org/abs/hep-th/9808061}{{\ttfamily arXiv:hep-th/9808061}}.

\bibitem{Peeters:2000qj}
K.~Peeters, P.~Vanhove, and A.~Westerberg, ``{Supersymmetric higher derivative
  actions in ten-dimensions and eleven-dimensions, the associated superalgebras
  and their formulation in superspace},''
  \href{http://dx.doi.org/10.1088/0264-9381/18/5/307}{{\em Class. Quant. Grav.}
  {\bfseries 18} (2001) 843--890},
  \href{http://arxiv.org/abs/hep-th/0010167}{{\ttfamily arXiv:hep-th/0010167}}.

\bibitem{Howe:1983sra}
P.~S. Howe and P.~C. West, ``{The Complete N=2, D=10 Supergravity},''
  \href{http://dx.doi.org/10.1016/0550-3213(84)90472-3}{{\em Nucl. Phys. B}
  {\bfseries 238} (1984) 181--220}.

\bibitem{Green:1999qt}
M.~B. Green, ``{Interconnections between type II superstrings, M theory and N=4
  supersymmetric Yang-Mills},''
  \href{http://dx.doi.org/10.1007/BFb0104240}{{\em Lect. Notes Phys.}
  {\bfseries 525} (1999) 22},
  \href{http://arxiv.org/abs/hep-th/9903124}{{\ttfamily arXiv:hep-th/9903124}}.

\bibitem{deHaro:2002vk}
S.~de~Haro, A.~Sinkovics, and K.~Skenderis, ``{On a supersymmetric completion
  of the R4 term in 2B supergravity},''
  \href{http://dx.doi.org/10.1103/PhysRevD.67.084010}{{\em Phys. Rev. D}
  {\bfseries 67} (2003) 084010},
  \href{http://arxiv.org/abs/hep-th/0210080}{{\ttfamily arXiv:hep-th/0210080}}.

\bibitem{Rajaraman:2005up}
A.~Rajaraman, ``{On a supersymmetric completion of the R**4 term in type IIB
  supergravity},'' \href{http://dx.doi.org/10.1103/PhysRevD.74.085018}{{\em
  Phys. Rev. D} {\bfseries 72} (2005) 125008},
  \href{http://arxiv.org/abs/hep-th/0505155}{{\ttfamily arXiv:hep-th/0505155}}.

\bibitem{Green:2003an}
M.~B. Green and C.~Stahn, ``{D3-branes on the Coulomb branch and instantons},''
  \href{http://dx.doi.org/10.1088/1126-6708/2003/09/052}{{\em JHEP} {\bfseries
  09} (2003) 052}, \href{http://arxiv.org/abs/hep-th/0308061}{{\ttfamily
  arXiv:hep-th/0308061}}.

\bibitem{Paulos:2008tn}
M.~F. Paulos, ``{Higher derivative terms including the Ramond-Ramond
  five-form},'' \href{http://dx.doi.org/10.1088/1126-6708/2008/10/047}{{\em
  JHEP} {\bfseries 10} (2008) 047},
  \href{http://arxiv.org/abs/0804.0763}{{\ttfamily arXiv:0804.0763 [hep-th]}}.

\bibitem{Bakhtiarizadeh:2017bpl}
H.~R. Bakhtiarizadeh, ``{Two Ramond\textendash{}Ramond corrections to type II
  supergravity via field-theory amplitude},''
  \href{http://dx.doi.org/10.1140/epjc/s10052-017-5391-7}{{\em Eur. Phys. J. C}
  {\bfseries 77} no.~12, (2017) 823},
  \href{http://arxiv.org/abs/1708.02805}{{\ttfamily arXiv:1708.02805
  [hep-th]}}.

\bibitem{Blaback:2019zig}
J.~Bl\r{a}b\"ack, U.~Danielsson, G.~Dibitetto, and S.~Giri, ``{Constructing
  stable de Sitter in M-theory from higher curvature corrections},''
  \href{http://dx.doi.org/10.1007/JHEP09(2019)042}{{\em JHEP} {\bfseries 09}
  (2019) 042}, \href{http://arxiv.org/abs/1902.04053}{{\ttfamily
  arXiv:1902.04053 [hep-th]}}.

\bibitem{Garousi:2020mqn}
M.~R. Garousi, ``{Minimal gauge invariant couplings at order $\alpha '^3$:
  NS\textendash{}NS fields},''
  \href{http://dx.doi.org/10.1140/epjc/s10052-020-08662-9}{{\em Eur. Phys. J.
  C} {\bfseries 80} no.~11, (2020) 1086},
  \href{http://arxiv.org/abs/2006.09193}{{\ttfamily arXiv:2006.09193
  [hep-th]}}.

\bibitem{Codina:2020yma}
T.~Codina and D.~Marques, ``{Generalized Dualities and Higher Derivatives},''
  \href{http://dx.doi.org/10.1007/JHEP10(2020)002}{{\em JHEP} {\bfseries 10}
  (2020) 002}, \href{http://arxiv.org/abs/2007.09494}{{\ttfamily
  arXiv:2007.09494 [hep-th]}}.

\bibitem{Mayer:2020lpa}
K.~Mayer, \href{http://dx.doi.org/10.33540/136}{{\em {On Quantum Corrections in
  String Compactifications: Effective Actions and Black Holes}}}.
\newblock PhD thesis, University Utrecht, 2020.

\bibitem{Grimm:2017pid}
T.~W. Grimm, K.~Mayer, and M.~Weissenbacher, ``{One-modulus Calabi-Yau fourfold
  reductions with higher-derivative terms},''
  \href{http://dx.doi.org/10.1007/JHEP04(2018)021}{{\em JHEP} {\bfseries 04}
  (2018) 021}, \href{http://arxiv.org/abs/1712.07074}{{\ttfamily
  arXiv:1712.07074 [hep-th]}}.

\bibitem{Weissenbacher:2020cyf}
M.~Weissenbacher, ``{On $\alpha'$-effects from $D$-branes in $4d \; \mathcal{N}
  = 1$},'' \href{http://dx.doi.org/10.1007/JHEP11(2020)076}{{\em JHEP}
  {\bfseries 11} (2020) 076}, \href{http://arxiv.org/abs/2006.15552}{{\ttfamily
  arXiv:2006.15552 [hep-th]}}.

\bibitem{Cicoli:2015ylx}
M.~Cicoli, F.~Quevedo, and R.~Valandro, ``{De Sitter from T-branes},''
  \href{http://dx.doi.org/10.1007/JHEP03(2016)141}{{\em JHEP} {\bfseries 03}
  (2016) 141}, \href{http://arxiv.org/abs/1512.04558}{{\ttfamily
  arXiv:1512.04558 [hep-th]}}.

\bibitem{Grisaru:1986px}
M.~T. Grisaru, A.~E.~M. van~de Ven, and D.~Zanon, ``{Four Loop beta Function
  for the N=1 and N=2 Supersymmetric Nonlinear Sigma Model in
  Two-Dimensions},'' \href{http://dx.doi.org/10.1016/0370-2693(86)90408-9}{{\em
  Phys. Lett. B} {\bfseries 173} (1986) 423--428}.

\bibitem{Grimm:2015mua}
T.~W. Grimm, T.~G. Pugh, and M.~Weissenbacher, ``{The effective action of
  warped M-theory reductions with higher-derivative terms - Part II},''
  \href{http://dx.doi.org/10.1007/JHEP12(2015)117}{{\em JHEP} {\bfseries 12}
  (2015) 117}, \href{http://arxiv.org/abs/1507.00343}{{\ttfamily
  arXiv:1507.00343 [hep-th]}}.

\bibitem{Myers:1999ps}
R.~C. Myers, ``{Dielectric branes},''
  \href{http://dx.doi.org/10.1088/1126-6708/1999/12/022}{{\em JHEP} {\bfseries
  12} (1999) 022}, \href{http://arxiv.org/abs/hep-th/9910053}{{\ttfamily
  arXiv:hep-th/9910053}}.

\bibitem{Haack:2006cy}
M.~Haack, D.~Krefl, D.~Lust, A.~Van~Proeyen, and M.~Zagermann, ``{Gaugino
  Condensates and D-terms from D7-branes},''
  \href{http://dx.doi.org/10.1088/1126-6708/2007/01/078}{{\em JHEP} {\bfseries
  01} (2007) 078}, \href{http://arxiv.org/abs/hep-th/0609211}{{\ttfamily
  arXiv:hep-th/0609211}}.

\bibitem{Beasley:2008dc}
C.~Beasley, J.~J. Heckman, and C.~Vafa, ``{GUTs and Exceptional Branes in
  F-theory - I},'' \href{http://dx.doi.org/10.1088/1126-6708/2009/01/058}{{\em
  JHEP} {\bfseries 01} (2009) 058},
  \href{http://arxiv.org/abs/0802.3391}{{\ttfamily arXiv:0802.3391 [hep-th]}}.

\bibitem{Marchesano:2019azf}
F.~Marchesano, R.~Savelli, and S.~Schwieger, ``{T-branes and defects},''
  \href{http://dx.doi.org/10.1007/JHEP04(2019)110}{{\em JHEP} {\bfseries 04}
  (2019) 110}, \href{http://arxiv.org/abs/1902.04108}{{\ttfamily
  arXiv:1902.04108 [hep-th]}}.

\bibitem{Camara:2003ku}
P.~G. Camara, L.~E. Ibanez, and A.~M. Uranga, ``{Flux induced SUSY breaking
  soft terms},'' \href{http://dx.doi.org/10.1016/j.nuclphysb.2004.04.013}{{\em
  Nucl. Phys. B} {\bfseries 689} (2004) 195--242},
  \href{http://arxiv.org/abs/hep-th/0311241}{{\ttfamily arXiv:hep-th/0311241}}.

\bibitem{Grana:2003ek}
M.~Grana, T.~W. Grimm, H.~Jockers, and J.~Louis, ``{Soft supersymmetry breaking
  in Calabi-Yau orientifolds with D-branes and fluxes},''
  \href{http://dx.doi.org/10.1016/j.nuclphysb.2004.04.021}{{\em Nucl. Phys. B}
  {\bfseries 690} (2004) 21--61},
  \href{http://arxiv.org/abs/hep-th/0312232}{{\ttfamily arXiv:hep-th/0312232}}.

\bibitem{Camara:2004jj}
P.~G. Camara, L.~E. Ibanez, and A.~M. Uranga, ``{Flux-induced SUSY-breaking
  soft terms on D7-D3 brane systems},''
  \href{http://dx.doi.org/10.1016/j.nuclphysb.2004.11.035}{{\em Nucl. Phys. B}
  {\bfseries 708} (2005) 268--316},
  \href{http://arxiv.org/abs/hep-th/0408036}{{\ttfamily arXiv:hep-th/0408036}}.

\bibitem{Giddings:2005ff}
S.~B. Giddings and A.~Maharana, ``{Dynamics of warped compactifications and the
  shape of the warped landscape},''
  \href{http://dx.doi.org/10.1103/PhysRevD.73.126003}{{\em Phys. Rev. D}
  {\bfseries 73} (2006) 126003},
  \href{http://arxiv.org/abs/hep-th/0507158}{{\ttfamily arXiv:hep-th/0507158}}.

\bibitem{Grimm:2011sk}
T.~W. Grimm and R.~Savelli, ``{Gravitational Instantons and Fluxes from
  M/F-theory on Calabi-Yau fourfolds},''
  \href{http://dx.doi.org/10.1103/PhysRevD.85.026003}{{\em Phys. Rev. D}
  {\bfseries 85} (2012) 026003},
  \href{http://arxiv.org/abs/1109.3191}{{\ttfamily arXiv:1109.3191 [hep-th]}}.

\bibitem{Dasgupta:1999ss}
K.~Dasgupta, G.~Rajesh, and S.~Sethi, ``{M theory, orientifolds and G -
  flux},'' \href{http://dx.doi.org/10.1088/1126-6708/1999/08/023}{{\em JHEP}
  {\bfseries 08} (1999) 023},
  \href{http://arxiv.org/abs/hep-th/9908088}{{\ttfamily arXiv:hep-th/9908088}}.

\bibitem{Duff:1995wd}
M.~J. Duff, J.~T. Liu, and R.~Minasian, ``{Eleven-dimensional origin of
  string-string duality: A One loop test},''
  \href{http://dx.doi.org/10.1016/0550-3213(95)00368-3}{{\em Nucl. Phys. B}
  {\bfseries 452} (1995) 261--282},
  \href{http://arxiv.org/abs/hep-th/9506126}{{\ttfamily arXiv:hep-th/9506126}}.

\bibitem{Sethi:1996es}
S.~Sethi, C.~Vafa, and E.~Witten, ``{Constraints on low dimensional string
  compactifications},''
  \href{http://dx.doi.org/10.1016/S0550-3213(96)00483-X}{{\em Nucl. Phys. B}
  {\bfseries 480} (1996) 213--224},
  \href{http://arxiv.org/abs/hep-th/9606122}{{\ttfamily arXiv:hep-th/9606122}}.

\bibitem{Dasgupta:1996yh}
K.~Dasgupta and S.~Mukhi, ``{A Note on low dimensional string
  compactifications},''
  \href{http://dx.doi.org/10.1016/S0370-2693(97)00216-5}{{\em Phys. Lett. B}
  {\bfseries 398} (1997) 285--290},
  \href{http://arxiv.org/abs/hep-th/9612188}{{\ttfamily arXiv:hep-th/9612188}}.

\bibitem{Becker:1996gj}
K.~Becker and M.~Becker, ``{M theory on eight manifolds},''
  \href{http://dx.doi.org/10.1016/0550-3213(96)00367-7}{{\em Nucl. Phys. B}
  {\bfseries 477} (1996) 155--167},
  \href{http://arxiv.org/abs/hep-th/9605053}{{\ttfamily arXiv:hep-th/9605053}}.

\bibitem{Grimm:2014xva}
T.~W. Grimm, T.~G. Pugh, and M.~Weissenbacher, ``{On M-theory fourfold vacua
  with higher curvature terms},''
  \href{http://dx.doi.org/10.1016/j.physletb.2015.02.047}{{\em Phys. Lett. B}
  {\bfseries 743} (2015) 284--289},
  \href{http://arxiv.org/abs/1408.5136}{{\ttfamily arXiv:1408.5136 [hep-th]}}.

\bibitem{Greene:1989ya}
B.~R. Greene, A.~D. Shapere, C.~Vafa, and S.-T. Yau, ``{Stringy Cosmic Strings
  and Noncompact Calabi-Yau Manifolds},''
  \href{http://dx.doi.org/10.1016/0550-3213(90)90248-C}{{\em Nucl. Phys. B}
  {\bfseries 337} (1990) 1--36}.

\bibitem{2000math8018G}
M.~{Gross} and P.~M.~H. {Wilson}, ``{Large Complex Structure Limits of K3
  Surfaces},'' {\em arXiv Mathematics e-prints} (Aug., 2000) math/0008018,
  \href{http://arxiv.org/abs/math/0008018}{{\ttfamily arXiv:math/0008018
  [math.DG]}}.

\bibitem{Conlon:2009kt}
J.~P. Conlon and E.~Palti, ``{Gauge Threshold Corrections for Local
  Orientifolds},'' \href{http://dx.doi.org/10.1088/1126-6708/2009/09/019}{{\em
  JHEP} {\bfseries 09} (2009) 019},
  \href{http://arxiv.org/abs/0906.1920}{{\ttfamily arXiv:0906.1920 [hep-th]}}.

\bibitem{Conlon:2010ji}
J.~P. Conlon and F.~G. Pedro, ``{Moduli Redefinitions and Moduli
  Stabilisation},'' \href{http://dx.doi.org/10.1007/JHEP06(2010)082}{{\em JHEP}
  {\bfseries 06} (2010) 082}, \href{http://arxiv.org/abs/1003.0388}{{\ttfamily
  arXiv:1003.0388 [hep-th]}}.

\bibitem{Antoniadis:1998ax}
I.~Antoniadis and C.~Bachas, ``{Branes and the gauge hierarchy},''
  \href{http://dx.doi.org/10.1016/S0370-2693(99)00102-1}{{\em Phys. Lett. B}
  {\bfseries 450} (1999) 83--91},
  \href{http://arxiv.org/abs/hep-th/9812093}{{\ttfamily arXiv:hep-th/9812093}}.

\bibitem{Antoniadis:2019rkh}
I.~Antoniadis, Y.~Chen, and G.~K. Leontaris, ``{Logarithmic loop corrections,
  moduli stabilisation and de Sitter vacua in string theory},''
  \href{http://dx.doi.org/10.1007/JHEP01(2020)149}{{\em JHEP} {\bfseries 01}
  (2020) 149}, \href{http://arxiv.org/abs/1909.10525}{{\ttfamily
  arXiv:1909.10525 [hep-th]}}.

\bibitem{Kiritsis:1997em}
E.~Kiritsis and B.~Pioline, ``{On R**4 threshold corrections in IIb string
  theory and (p, q) string instantons},''
  \href{http://dx.doi.org/10.1016/S0550-3213(97)00645-7}{{\em Nucl. Phys. B}
  {\bfseries 508} (1997) 509--534},
  \href{http://arxiv.org/abs/hep-th/9707018}{{\ttfamily arXiv:hep-th/9707018}}.

\bibitem{Russo:1997mk}
J.~G. Russo and A.~A. Tseytlin, ``{One loop four graviton amplitude in
  eleven-dimensional supergravity},''
  \href{http://dx.doi.org/10.1016/S0550-3213(97)00631-7}{{\em Nucl. Phys. B}
  {\bfseries 508} (1997) 245--259},
  \href{http://arxiv.org/abs/hep-th/9707134}{{\ttfamily arXiv:hep-th/9707134}}.

\bibitem{Antoniadis:1997eg}
I.~Antoniadis, S.~Ferrara, R.~Minasian, and K.~S. Narain, ``{R**4 couplings in
  M and type II theories on Calabi-Yau spaces},''
  \href{http://dx.doi.org/10.1016/S0550-3213(97)00572-5}{{\em Nucl. Phys. B}
  {\bfseries 507} (1997) 571--588},
  \href{http://arxiv.org/abs/hep-th/9707013}{{\ttfamily arXiv:hep-th/9707013}}.

\bibitem{Tseytlin:2000sf}
A.~A. Tseytlin, ``{R**4 terms in 11 dimensions and conformal anomaly of (2,0)
  theory},'' \href{http://dx.doi.org/10.1016/S0550-3213(00)00380-1}{{\em Nucl.
  Phys. B} {\bfseries 584} (2000) 233--250},
  \href{http://arxiv.org/abs/hep-th/0005072}{{\ttfamily arXiv:hep-th/0005072}}.

\bibitem{Banks:1998nr}
T.~Banks and M.~B. Green, ``{Nonperturbative effects in AdS in five-dimensions
  x S**5 string theory and d = 4 SUSY Yang-Mills},''
  \href{http://dx.doi.org/10.1088/1126-6708/1998/05/002}{{\em JHEP} {\bfseries
  05} (1998) 002}, \href{http://arxiv.org/abs/hep-th/9804170}{{\ttfamily
  arXiv:hep-th/9804170}}.

\bibitem{Gubser:1998nz}
S.~S. Gubser, I.~R. Klebanov, and A.~A. Tseytlin, ``{Coupling constant
  dependence in the thermodynamics of N=4 supersymmetric Yang-Mills theory},''
  \href{http://dx.doi.org/10.1016/S0550-3213(98)00514-8}{{\em Nucl. Phys. B}
  {\bfseries 534} (1998) 202--222},
  \href{http://arxiv.org/abs/hep-th/9805156}{{\ttfamily arXiv:hep-th/9805156}}.

\bibitem{Gross:1986iv}
D.~J. Gross and E.~Witten, ``{Superstring Modifications of Einstein's
  Equations},'' \href{http://dx.doi.org/10.1016/0550-3213(86)90429-3}{{\em
  Nucl. Phys. B} {\bfseries 277} (1986) 1}.

\bibitem{Freeman:1986zh}
M.~D. Freeman, C.~N. Pope, M.~F. Sohnius, and K.~S. Stelle, ``{Higher Order
  $\sigma$ Model Counterterms and the Effective Action for Superstrings},''
  \href{http://dx.doi.org/10.1016/0370-2693(86)91495-4}{{\em Phys. Lett. B}
  {\bfseries 178} (1986) 199--204}.

\bibitem{Hyakutake:2007sm}
Y.~Hyakutake, ``{Toward the Determination of R**3 F**2 Terms in M-theory},''
  \href{http://dx.doi.org/10.1143/PTP.118.109}{{\em Prog. Theor. Phys.}
  {\bfseries 118} (2007) 109},
  \href{http://arxiv.org/abs/hep-th/0703154}{{\ttfamily arXiv:hep-th/0703154}}.

\bibitem{Green:1997me}
M.~B. Green, M.~Gutperle, and H.-h. Kwon, ``{Sixteen fermion and related terms
  in M theory on T**2},''
  \href{http://dx.doi.org/10.1016/S0370-2693(97)01551-7}{{\em Phys. Lett. B}
  {\bfseries 421} (1998) 149--161},
  \href{http://arxiv.org/abs/hep-th/9710151}{{\ttfamily arXiv:hep-th/9710151}}.

\bibitem{Schubert:2001he}
C.~Schubert, ``{Perturbative quantum field theory in the string inspired
  formalism},'' \href{http://dx.doi.org/10.1016/S0370-1573(01)00013-8}{{\em
  Phys. Rept.} {\bfseries 355} (2001) 73--234},
  \href{http://arxiv.org/abs/hep-th/0101036}{{\ttfamily arXiv:hep-th/0101036}}.

\bibitem{Brink:1981nb}
L.~Brink and J.~H. Schwarz, ``{Quantum Superspace},''
  \href{http://dx.doi.org/10.1016/0370-2693(81)90093-9}{{\em Phys. Lett. B}
  {\bfseries 100} (1981) 310--312}.

\bibitem{Brink:1981rt}
L.~Brink and M.~B. Green, ``{Point - Like Particles and Off-shell Supersymmetry
  Algebras},'' \href{http://dx.doi.org/10.1016/0370-2693(81)90649-3}{{\em Phys.
  Lett. B} {\bfseries 106} (1981) 393--398}.

\bibitem{Green:1983wt}
M.~B. Green and J.~H. Schwarz, ``{Covariant Description of Superstrings},''
  \href{http://dx.doi.org/10.1016/0370-2693(84)92021-5}{{\em Phys. Lett. B}
  {\bfseries 136} (1984) 367--370}.

\bibitem{Green:1983sg}
M.~B. Green and J.~H. Schwarz, ``{Properties of the Covariant Formulation of
  Superstring Theories},''
  \href{http://dx.doi.org/10.1016/0550-3213(84)90030-0}{{\em Nucl. Phys. B}
  {\bfseries 243} (1984) 285--306}.

\bibitem{Berkovits:2002uc}
N.~Berkovits, ``{Towards covariant quantization of the supermembrane},''
  \href{http://dx.doi.org/10.1088/1126-6708/2002/09/051}{{\em JHEP} {\bfseries
  09} (2002) 051}, \href{http://arxiv.org/abs/hep-th/0201151}{{\ttfamily
  arXiv:hep-th/0201151}}.

\bibitem{Anguelova:2004pg}
L.~Anguelova, P.~A. Grassi, and P.~Vanhove, ``{Covariant one-loop amplitudes in
  D=11},'' \href{http://dx.doi.org/10.1016/j.nuclphysb.2004.09.024}{{\em Nucl.
  Phys. B} {\bfseries 702} (2004) 269--306},
  \href{http://arxiv.org/abs/hep-th/0408171}{{\ttfamily arXiv:hep-th/0408171}}.

\bibitem{Guillen:2017mte}
M.~Guillen, ``{Equivalence of the 11D pure spinor and Brink-Schwarz-like
  superparticle cohomologies},''
  \href{http://dx.doi.org/10.1103/PhysRevD.97.066002}{{\em Phys. Rev. D}
  {\bfseries 97} no.~6, (2018) 066002},
  \href{http://arxiv.org/abs/1705.06316}{{\ttfamily arXiv:1705.06316
  [hep-th]}}.

\bibitem{Berkovits:2019szu}
N.~Berkovits, E.~Casali, M.~Guillen, and L.~Mason, ``{Notes on the $D=11$ pure
  spinor superparticle},''
  \href{http://dx.doi.org/10.1007/JHEP08(2019)178}{{\em JHEP} {\bfseries 08}
  (2019) 178}, \href{http://arxiv.org/abs/1905.03737}{{\ttfamily
  arXiv:1905.03737 [hep-th]}}.

\bibitem{Guillen:2020mmd}
M.~Guillen, ``{Notes on the 11D pure spinor wordline vertex operators},''
  \href{http://dx.doi.org/10.1007/JHEP08(2020)122}{{\em JHEP} {\bfseries 08}
  (2020) 122}, \href{http://arxiv.org/abs/2006.06022}{{\ttfamily
  arXiv:2006.06022 [hep-th]}}.

\bibitem{Green:2016tfs}
M.~B. Green and A.~Rudra, ``{Type I/heterotic duality and M-theory
  amplitudes},'' \href{http://dx.doi.org/10.1007/JHEP12(2016)060}{{\em JHEP}
  {\bfseries 12} (2016) 060}, \href{http://arxiv.org/abs/1604.00324}{{\ttfamily
  arXiv:1604.00324 [hep-th]}}.

\bibitem{Green:2006gt}
M.~B. Green, J.~G. Russo, and P.~Vanhove, ``{Non-renormalisation conditions in
  type II string theory and maximal supergravity},''
  \href{http://dx.doi.org/10.1088/1126-6708/2007/02/099}{{\em JHEP} {\bfseries
  02} (2007) 099}, \href{http://arxiv.org/abs/hep-th/0610299}{{\ttfamily
  arXiv:hep-th/0610299}}.

\bibitem{Freeman:1986br}
M.~D. Freeman and C.~N. Pope, ``{Beta Functions and Superstring
  Compactifications},''
  \href{http://dx.doi.org/10.1016/0370-2693(86)91127-5}{{\em Phys. Lett. B}
  {\bfseries 174} (1986) 48--50}.

\bibitem{Braun:2015pza}
A.~P. Braun, M.~Rummel, Y.~Sumitomo, and R.~Valandro, ``{De Sitter vacua from a
  D-term generated racetrack potential in hypersurface Calabi-Yau
  compactifications},'' \href{http://dx.doi.org/10.1007/JHEP12(2015)033}{{\em
  JHEP} {\bfseries 12} (2015) 033},
  \href{http://arxiv.org/abs/1509.06918}{{\ttfamily arXiv:1509.06918
  [hep-th]}}.

\bibitem{Cicoli:2011qg}
M.~Cicoli, C.~Mayrhofer, and R.~Valandro, ``{Moduli Stabilisation for Chiral
  Global Models},'' \href{http://dx.doi.org/10.1007/JHEP02(2012)062}{{\em JHEP}
  {\bfseries 02} (2012) 062}, \href{http://arxiv.org/abs/1110.3333}{{\ttfamily
  arXiv:1110.3333 [hep-th]}}.

\bibitem{Cicoli:2012vw}
M.~Cicoli, S.~Krippendorf, C.~Mayrhofer, F.~Quevedo, and R.~Valandro,
  ``{D-Branes at del Pezzo Singularities: Global Embedding and Moduli
  Stabilisation},'' \href{http://dx.doi.org/10.1007/JHEP09(2012)019}{{\em JHEP}
  {\bfseries 09} (2012) 019}, \href{http://arxiv.org/abs/1206.5237}{{\ttfamily
  arXiv:1206.5237 [hep-th]}}.

\bibitem{Louis:2012nb}
J.~Louis, M.~Rummel, R.~Valandro, and A.~Westphal, ``{Building an explicit de
  Sitter},'' \href{http://dx.doi.org/10.1007/JHEP10(2012)163}{{\em JHEP}
  {\bfseries 10} (2012) 163}, \href{http://arxiv.org/abs/1208.3208}{{\ttfamily
  arXiv:1208.3208 [hep-th]}}.

\bibitem{Cicoli:2013mpa}
M.~Cicoli, S.~Krippendorf, C.~Mayrhofer, F.~Quevedo, and R.~Valandro, ``{D3/D7
  Branes at Singularities: Constraints from Global Embedding and Moduli
  Stabilisation},'' \href{http://dx.doi.org/10.1007/JHEP07(2013)150}{{\em JHEP}
  {\bfseries 07} (2013) 150}, \href{http://arxiv.org/abs/1304.0022}{{\ttfamily
  arXiv:1304.0022 [hep-th]}}.

\bibitem{Cicoli:2013zha}
M.~Cicoli, S.~Krippendorf, C.~Mayrhofer, F.~Quevedo, and R.~Valandro, ``{The
  Web of D-branes at Singularities in Compact Calabi-Yau Manifolds},''
  \href{http://dx.doi.org/10.1007/JHEP05(2013)114}{{\em JHEP} {\bfseries 05}
  (2013) 114}, \href{http://arxiv.org/abs/1304.2771}{{\ttfamily arXiv:1304.2771
  [hep-th]}}.

\bibitem{Cicoli:2013cha}
M.~Cicoli, D.~Klevers, S.~Krippendorf, C.~Mayrhofer, F.~Quevedo, and
  R.~Valandro, ``{Explicit de Sitter Flux Vacua for Global String Models with
  Chiral Matter},'' \href{http://dx.doi.org/10.1007/JHEP05(2014)001}{{\em JHEP}
  {\bfseries 05} (2014) 001}, \href{http://arxiv.org/abs/1312.0014}{{\ttfamily
  arXiv:1312.0014 [hep-th]}}.

\bibitem{Obied:2018sgi}
G.~Obied, H.~Ooguri, L.~Spodyneiko, and C.~Vafa, ``{De Sitter Space and the
  Swampland},'' \href{http://arxiv.org/abs/1806.08362}{{\ttfamily
  arXiv:1806.08362 [hep-th]}}.

\bibitem{Cicoli:2018kdo}
M.~Cicoli, S.~De~Alwis, A.~Maharana, F.~Muia, and F.~Quevedo, ``{De Sitter vs
  Quintessence in String Theory},''
  \href{http://dx.doi.org/10.1002/prop.201800079}{{\em Fortsch. Phys.}
  {\bfseries 67} no.~1-2, (2019) 1800079},
  \href{http://arxiv.org/abs/1808.08967}{{\ttfamily arXiv:1808.08967
  [hep-th]}}.

\end{thebibliography}\endgroup

\end{document}